\documentclass[12pt,preprint]{aastex}
\usepackage{amsmath}

\shorttitle{WD\&SNeIa}
\shortauthors{Ablimit at el.}

\begin{document}


\title{Monte Carlo population synthesis of post-common-envelope white dwarf binaries
and type Ia supernova rate}
\author{Iminhaji Ablimit\altaffilmark{1},
Keiichi Maeda\altaffilmark{2,3} and Xiang-Dong Li\altaffilmark{4,5}}
\altaffiltext{1}{Key Laboratory for Optical Astronomy, National Astronomical Observatories,
Chinese Academy of Sciences, Beijing 100012, China}
\altaffiltext{2}{Department of Astronomy, Kyoto University, Kitashirakawa-Oiwake-cho, Sakyo-ku, Kyoto 606-8502}
\altaffiltext{3}{Kavli Institute for the Physics and Mathematics of the
Universe (WPI), The University of Tokyo, 5-1-5 Kashiwanoha, Kashiwa, Chiba 277-8583, Japan}
\altaffiltext{4}{Department of Astronomy, Nanjing University, Nanjing 210046, China}
\altaffiltext{5}{Key Laboratory of Modern Astronomy and Astrophysics,
Ministry of Education, Nanjing 210046, China}


\begin{abstract}
Binary population synthesis (BPS) study provides a comprehensive way to understand evolutions
of binaries and their end products. Close white dwarf (WD) binaries have crucial characteristics
in examining influence of yet-unresolved physical parameters on the binary evolution. In this paper,
we perform Monte Carlo BPS simulations, investigating the population of WD/main sequence (WD/MS) binaries
and double WD binaries, with a publicly available binary star evolution code under 37 different assumptions
on key physical processes and binary initial conditions. We considered different combinations of the binding
energy parameter ($\lambda_{\rm g}$:considering gravitational energy only, $\lambda_{\rm b}$:
considering both gravitational energy and internal energy, and $\lambda_{\rm e}$:considering gravitational energy, internal energy, and entropy of the envelope, the values
of them derived with the MESA code), CE efficiency, critical mass ratio, initial primary mass function and
metallicity. We find that a larger number of post-CE WD/MS binaries in tight orbits are formed when the binding
energy parameters is set by $\lambda_{\rm e}$ than the cases adopting the other prescriptions. We also find
effects of the other input parameters on orbital period and mass distributions of post-CE WD/MS binaries as well.
Containing at least one CO WD, the double WD system evolved from WD/MS binaries may explode as type
Ia supernovae (SNe Ia) by merging. In this work, we also investigate a frequency of two WD mergers and compare
it to the SNe Ia rate. The calculated Galactic SNe Ia rate with $\lambda = \lambda_{\rm e}$ is comparable
with observed SNe Ia rate, $\sim 8.2\times10^{-5}\,\rm{yr}^{-1}$ -- $\sim  4\times10^{-3}\,\rm{yr}^{-1}$ depending
on the other BPS parameters, if a DD system does not require the mass ratio higher than $\sim 0.8$ to become an SNe Ia.
On the other hand, a scenario like a violent merger scenario, which requires a combined mass of two CO WDs
$\geq$ 1.6$M_\sun$ and mass ratio $> 0.8$, results in a much lower SNe Ia rate than observed.
\end{abstract}

\keywords{ binaries: close -- stars: evolution -- white dwarfs
-- stars: supernovae -- X-rays: binaries}

\section{Introduction}
The mass transfer process and the common envelope (CE) phase are crucial for producing all kinds of compact
star binaries. Among the compact star binaries, white dwarf $+$ main sequence
(WD/MS) binaries or double WD systems serve as excellent laboratory to understand yet-unclarified physics
of binary evolution during a common envelope phase, and of type Ia supernovae (SNe Ia). MS/MS close binaries
generate close compact star binaries mainly by going through a CE
phase at least once. In the MS/MS binary (the orbital separation between $\sim$10 and 1000 $R_\sun$), once the
 more massive one (primary) evolves into the first giant branch (FGB) or the asymptotic giant branch (AGB),
 it fills its Roche lobe and the mass transfer may be dynamically unstable, thus the envelope of the primary
 engulfs the less massive one (secondary) \citep{ib1993,we08}.
\cite{pa76} proposed that the orbital energy and orbital angular momentum are removed by the CE ejection.
The stellar parameters of detached WD/MS binaries are easily observed among all compact star binaries, thus
post-CE WD/MS binaries (PCEBs) are ideal systems to test yet-unresolved physical processes involved in the CE phase.

Binary population synthesis (BPS) is a very useful tool to study the physical
parameters and the evolution processes involved in the formation of various types of close
binaries. After \cite{de1993}, a bunch of BPS works have been performed on PCEBs with different CE phase models.
The CE evolution is crucial for producing compact star close binaries. However, the CE evolution is still not
well understood. The $\alpha$-formalism and $\gamma$-formalism have been widely used to study the CE evolution,
the former one considers energy conservation \citep{we84,de00} while the later one considers angular momentum
conservation \citep{ne05}. In the $\alpha$-formalism, the outcome of the CE phase is related
with the efficiency parameter $\alpha_{\rm CE}$,
\begin{equation}
E_{\rm{bind}} = {\alpha_{\rm CE}} \Delta E_{\rm orb}
\end{equation}
where $E_{\rm{bind}}$ and $\Delta E_{\rm orb}$ are the binding energy of the envelope and the change in
the orbital energy during the CE phase, respectively. The two stars coalesce if the system satisfies the
equation 75 of \cite{hu02} and the CE evolution is longer than dynamical timescale (for details of the
criterion for surviving or merging during the CE phase see \cite{hu02}). The binding energy of the envelope
is expressed as the following:
\begin{equation}
E_{\rm{bind}} = - \frac{GM_1 M_{\rm{en}}}{\lambda {\rm{R}}_1}
\end{equation}
where $M_1$, $M_{\rm{en}}$ and ${\rm{R}}_1$ are the total mass, envelope
mass and radius of the primary star, respectively. It is commonly believed that there is no mass
accretion during the CE phase. The binding energy parameter, $\lambda$, depends on the mass and evolutionary
stage of the primary star. In previous BPS studies of PCEBs \citep{de1993,po06,po07,wan15}, the binding energy
parameter has been treated merely as a constant (0.5 or 1.0), or it is fixed with ${\alpha_{\rm CE}}$ \citep{to13}.
However, a number of papers
\citep{han94,de00,we08} claimed that assuming a constant value for $\lambda$ is not a promising way to address all
types of primary stars and their evolution phases. They found that the value of $\lambda$ changes as the star evolves.
According to \cite{de10} and \cite{xu10}, the values of $\lambda$ for evolving stars can be calculated by considering
gravitational energy only (hereafter $\lambda_{\rm g}$), adding internal energy ($\lambda_{\rm b}$), or adding entropy of
the envelope ($\lambda_{\rm e}$). For a $6 M_\sun$ star, the values of $\lambda_{\rm g}$, $\lambda_{\rm b}$ and
$\lambda_{\rm e}$ range from 0.11 and 0.64, 0.4 and 1.5, 0.6 and 9.98, respectively.  We emphasize that all these
parameters (e.g. ${\alpha_{\rm CE}}$, $\lambda$, the critical mass ratios~${q_{\rm cr}}$) (see \S2 for details)
are poorly determined from the first principle, therefore studying binary systems that experienced a CE phase at least
once is crucial: the distributions of orbital periods, primary and secondary masses can be used to constrain these
uncertain values and/or the prescriptions to describe these unresolved physical processes.

Recent observational studies of PCEBs have been providing statistical properties of these systems. The information
serves as crucial observational input to improve the theory of CE evolution \citep{N11,R12,P13}. \cite{z10} adopted
both of the CE parameters in a range of 0.2-0.3, in order to reproduce properties of the observed PCEBs. \cite{to13}
studied the PCEB populations in a similar way. \cite{c14} considered selection effects in the observational samples.
They also performed BPS of PCEBs by considering the relative contributions of recombination energy and orbital energy
to expel the CE \citep{z14}, while this is yet unclear \citep{R12}. Besides, $\lambda$ depends on the structure of the
primary star, ${\alpha_{\rm CE}}$ may change with the secondary mass \citep{de11}.
Therefore, we need further simulations of PCEBs and additional observational data to clarify the nature of the CE evolution.

PCEBs with a relatively massive MS secondary may produce type Ia supernovae (SNe Ia) in the two most popular
scenarios proposed so far: one is the single degenerate model (SD), and the other is the double degenerate model (DD)
\citep[for a review]{wa12}. Most of PCEBs can evolve (either with an unstable or a stable mass transfer) into double WD
binaries. The DD scenario considers that a part of the double WD systems merge via gravitational wave radiation on
timescales shorter than the age of the universe as a potentially significant population leading to SNe Ia \citep{we84}.
The explosion mechanism involves complicated interaction between hydrodynamics evolution and nuclear reactions. Several
 models exist that vary in the physical processes leading to the explosion: for example, the thermonuclear flame can be
 either detonation or deflagration, the mass of the immediate progenitor WD can either be the Chandrasekhar mass or
 sub-Chandrasekhar mass \citep{hi00}. Moreover, there is a serious concern about the capability of the merging product
 to explode as an SN Ia, since the remnant could lead to a formation of a neutron star rather than an SN Ia
 \citep{no85,pi03,pa13,da14,a15}. In sum, both observationally and theoretically, the exact nature of the SNe Ia
 progenitors remains unclear. Studying the birth rate and delay time distribution can help understand the binary
 evolution channel toward SNe Ia. Indeed, the birth rate and delay time distribution of the observed SNe Ia are not
 redily satisfied by either of the SD or DD model \citep[for a recent review]{ma14}. In most theoretical works
 including CE evolution, the standard merger model for SNe Ia, i.e., merging two CO WDs, is not able to reproduce
 the observed SN Ia rate \citep{ru09,me10,to12,ch12,bo13}.

The discrepancy between the theoretical and observational SN Ia rates is an open question.
\cite{cl14} considered effects of various physical parameters on the binary evolution on the rate of
SNe Ia in the SD and DD models. Their BPS results show that even for their optimistic models the predicted
SN Ia rate is a factor of three less than the galaxy-cluster SN Ia rate. \cite{ch12} used different mass transfer
models to compute the SNe Ia rate, with a constraint from a sample of double CO WDs which can merge within Hubble time.
Even though they arbitrarily relaxed the critical mass ratio leading to an SN Ia in a violent merger scenario from the
value obtained by hydrodynamic simulations \citep{pa13}, their BPS models also could not get the SN Ia rate as large as
what is observed. Recently, He WD donors have been considered as a possible important channel for SNe Ia, given that
the number of He WDs can be much higher than that of CO WDs. If mergers of a CO WD with a He WD are hypothesized as a
SN Ia progenitor in the DD scenario, it could be a major contributor to the SN Ia population \citep{na07,ba12,ru11, ru14}.

In this paper, we investigate the influence that a series of binary evolution prescriptions (especially $\lambda$
and $q_{\rm cr}$) has on the formation of WD binaries (WD/MS and double WD systems). We consider both mergers of
CO WDs with CO WDs and those of CO WDs with He WDs to compute  the SNe Ia rate in our BPS study. In \S 2, we describe
our BPS code and our models to treat the binary physical processes including the CE evolution. The properties of WD/MS
binaries produced by our models are presented in \S 3. The resulting double WD systems and SNe Ia rate are discussed
in \S 4. Finally, the paper is closed in \S 5 with conclusions and discussion.

\section{Monte Carlo BPS simulations}
\label{sec:model}
We use the BPS code developed by \cite{hu02} and modified by \cite{ki06} to generate ${10}^7$ initial
MS/MS binaries for each model. For the distribution of the masses of the primary stars, we adopted the initial
 mass function (IMF) of \cite{kr93},

\begin{equation}
f(M_1) = \left\{ \begin{array}{ll}
0 & \textrm{$M_1/M_\odot < 0.1$}\\
0.29056{M_1}^{-1.3} & \textrm{$0.1\leq M_1/M_\odot < 0.5$}\\
0.1557{M_1}^{-2.2} & \textrm{$0.5\leq M_1/M_\odot < 1.0$}\\
0.1557{M_1}^{-2.7} & \textrm{$1.0\leq M_1/M_\odot$},
\end{array} \right.
\end{equation}

The masses of the secondary stars are determined by the distribution of the initial mass ratio,

\begin{equation}
n(q) = \left\{ \begin{array}{ll}
0 & \textrm{$q>1$}\\
\mu q^{\nu} & \textrm{$0\leq q < 1$},
\end{array} \right.
\end{equation}
where $q=M_2/M_1$, $\mu$ is the normalization factor for the assumed power law distribution with
the index $\nu$. We consider two cases for the initial mass ratio distribution (IMRD): a flat IMRD
($\nu = 0$ and $n(q)=$constant) and
an IMRD proportional to $q$ ($\nu = 1$). For the distribution of the initial orbital separation,
$a_{\rm i}$, we adopt the following formalism \citep{da08}:

\begin{equation}
n(a) = \left\{ \begin{array}{ll}
0 & \textrm{$a_{\rm i}/R_\odot < 3$ or $a_{\rm i}/R_\odot > 10^6$}\\
0.078636{a_{\rm i}}^{-1} & \textrm{$3\leq a_{\rm i}/R_\odot \leq 10^6$}
\end{array} \right.
\end{equation}

We assume that all binaries are in circle orbits. We calculate Pop. I and II
binaries with the metallicity given as $Z=0.02$ and 0.001, respectively.

Regarding the key physical processes, our simulations have three main tunable parameters, i.e.,
$\alpha_{\rm CE}$, $\lambda$ and $q_{\rm cr}$. The main aim of this paper is to investigate the
effects that these prescriptions have on the evolution toward WD/MS binaries and double WD binaries.
As introduced in \S 1, the CE evolution is an important but unsolved phase in the binary evolutionary
process. In most works treating the CE evolution, the two main parameters describing the CE evolution
($\alpha_{\rm CE}$ and $\lambda$) have been set as constants. However, in reality, they should change
with the evolutionary process. \cite{de11} and \cite{de12} proposed that $\alpha_{\rm CE}$ may
change with the WD mass, the secondary mass, the mass ratio, or the orbital period. Here we consider the
following formula,
\begin{equation}
{\rm log_{10}}{\alpha_{\rm CE}} = {\epsilon}_0 + {{\epsilon}_1}{\rm log_{10}}{q}
\end{equation}
in \cite{de12} with the values of ${\epsilon}_0$ and ${\epsilon}_1$ taken from their Table 6.
We calculate the CE evolution either by adopting this equation or by fixing ${\alpha_{\rm CE}=1}$.

There have been three prescriptions proposed so far to describe the binding energy parameters
($\lambda_{\rm g}$, $\lambda_{\rm b}$, and $\lambda_{\rm e}$: see \S 1). We test all the three ones
independently in our BPS study, adopting the calculated results with the MESA code by \cite{wa15a,wa15b}.

The critical mass ratio ($q_{\rm cr}$) is another physical key parameter that determines whether the
mass transfer is stable or not. \cite{sh14} computed the critical mass ratio considering the
possible response of the accreting star (i.e., spin-up and rejuvenation) under three different assumptions: (1) Half
of the transferred mass is accreted by the secondary, and the
other half is lost from the system, also taking the specific orbital
angular momentum of the accretor (Also see de Mink et al.
2007). (2) The transferred mass is assumed to be accreted
by the secondary unless its thermal timescale (${\tau_{\rm KH_2}}$) becomes much
shorter than the mass transfer timescale ($\tau_{\dot{M}}$). The accretion
rate is limited by--$[\rm{min}(10(\tau_{\dot{M}} /{\tau_{\rm KH_2}}), 1)]\dot{M}_1$
\citep{hu02}. Rapid mass accretion may drive the accretor
out of thermal equilibrium, which will expand and become
overluminous. \cite{sh14} found the values of ${\tau_{\rm KH_2}}$ are usually much lower
than that of the same star in thermal
equilibrium. Therefore, it is always as ${\tau_{\rm KH_2}} < 10{\tau_{\dot{M}}}$, and the
mass transfer is generally conservative. (3) The accretion
rate onto a rotating star is reduced by a factor of ($1 - {\Omega}/{\Omega_{\rm cr}}$),
where $\Omega$ is the angular velocity of the star and ${\Omega_{\rm cr}}$ is its critical
value. In this prescription, a star cannot accrete
mass when it rotating at ${\Omega_{\rm cr}}$. The remaining material is ejected out of
the binary by the isotropic wind, and it takes away the
specific orbital angular momentum from the accretor. The critical mass ratios are
denoted as $q_{\rm cr1}$, $q_{\rm cr2}$ and $q_{\rm cr3}$, respectively.
For each model as described above, we run the BPS simulations with these different treatment of the
critical mass ratio (i.e., adopting $q_{\rm cr} = q_{\rm cr1}$, $q_{\rm cr2}$, or $q_{\rm cr3}$).
For models 2--13, there are thus 36 different simulations. Including the standard model (model 1),
we cover 37 different models in our numerical calculations. In the Table 1, we summarize our models.
In our standard model, we assume ${\alpha_{\rm CE}=1}$, ${\lambda=1}$ for the CE phase, $Z=0.02$,
the default $q_{\rm cr}$ prescription in the BSE code and a flat
IMRD for the initial conditions. For other initial physical inputs, we adopt the default values
in the BPS code given by \citet{hu02}.


\section{Results }

\subsection{Post common envelope
WD/MS binaries (PCEBs)}

Figure 1 shows the orbital period distributions of PCEBs in our standard model (model 1)
and other twelve models (from model 2 to 13) with three different prescriptions for $q_{\rm cr}$.
The dashed black line, shown in all panels of Figure 1, shows the result of our standard model. In
the upper left panel, the solid, dotted, dashed colored lines show results of model 2, 3, and 4,
respectively. In each model, there are three lines corresponding to three different prescriptions
for $q_{\rm cr}$. Hereafter, different line-styles are used to describe results adopting different
prescriptions for $\lambda$, and different colors are used to describe results adopting different
prescriptions for $q_{\rm cr}$.
It is hard to construct reliable observational distributions, but it is still interesting to
give some observational information. We use the solid black line to demonstrate the observed PCEBs sample
from Tables 1 and 2 of \cite{z11}.

From the upper left panel we see that the different prescriptions for $q_{\rm cr}$ do not have
clear effects on the orbital period distribution of WD binaries.
However, the orbital distribution is sensitive to the prescription for $\lambda$.
The orbital separation is the smallest for $\lambda = \lambda_{\rm e}$, while the largest
for $\lambda=\lambda_{\rm g}$, and the difference in the typical orbital period is more than an order of magnitude.
There are a larger number of PCEBs with short orbital periods when adopting $\lambda=\lambda_{\rm e}$,
where the shortest orbital period is $\sim 0.008$ day. This means that the more PCEBs
can survive CE evolution when $\lambda=\lambda_{\rm e}$ (or more precisely ${\alpha_{\rm CE}}*{\lambda}$ is higher. Most of observed PCEB samples have short orbital periods,
and their orbital period range seems to be covered by the calculated results with $\lambda = \lambda_{\rm e}$ and model 1.
The upper right panel shows the results in the models 5--7, where $\alpha_{\rm CE}$ changes with $q$,
while other parameters are same as the models in the upper left panel. The prescription of $\alpha_{\rm CE}$
affects the distribution clearly. If $\lambda=\lambda_{\rm g}$ or $\lambda=\lambda_{\rm b}$, as seen by
comparing the right panel with the left panel, the distribution becomes narrower and more sharply peaked.
The distribution is however not sensitive to the prescription for $\alpha_{\rm CE}$ if $\lambda=\lambda_{\rm e}$.
In the lower left panel of Figure 1, the results in models 8-10 are shown where the initial mass ratio distribution
is given as $\propto q$, while the other parameters are same with models 2--4. The orbital period distribution is
similar to the corresponding models 2--4, showing that it is not sensitive to the initial mass distribution.
In the lower right panel, the results in models 11--13 with low metallicity ($\rm Z=0.001$) are shown. These
models tend to have shorter orbital periods than those of the solar metallicity models. Comparing our results
with those of \cite{z14}, the PCEBs have orbital periods shorter in our simulations. In the binary population,
the CE evolution with $\lambda_{\rm e}$ leads to the large number of short orbital period systems.

The distributions of the secondary masses are given in Figure 2. The results from models 2, 3, and 4 are
similar to those from the standard model (in the upper left panel of the Figure 2). Nearly all PCEBs
(in the solid black line) have low mass secondaries. We can also see in the upper right
panel that the different prescriptions for $\alpha_{\rm CE}$ (models 5--7) do not lead to clear difference,
as compared to the upper left one (models 2--4 and the standard model). For models 8--10 (in the lower left panel),
the distribution of the secondary mass moves toward a larger value. That is, more massive secondary stars are produced
and can dispel the CE when the IMRD is given by $n(q)\propto q$ rather than adopting the flat IMRD distribution.
For models with low metallicity (models 11--13, in the lower right panel), the low mass secondaries become more
abundant than in the standard model. While the prescriptions for $q_{\rm cr}$ hardly affect the secondary mass
for the solar metallicity, it has some influence on the distribution when $\rm Z=0.001$, The reason might
be that, a star on red giant branch or asymptotic giant branch has a heavier core and a less massive envelope if its metallicity is lower, and this makes the less massive secondary to be able to expel the envelope during the CE phase.

Figure 3 shows the WD mass distributions. There is a gap in the WD mass distribution in all the models,
which separates the systems with a He WD and with a C/O WD. The gap is caused by the stellar radius at
the tip of the FGB being larger than the radius at the beginning of the AGB when the core mass of the
primary is in the gap area. In this range of core masses, the primary star cannot fill its Roche lobe
because it would have done so before on the FGB. In the order of $\lambda = \lambda_{\rm g}$, $\lambda_{\rm b}$
and $\lambda_{\rm e}$ respectively, the peak in the WD mass distribution shifts to a lower value and low-mass
WDs become more abundant. From the distributions of used PCEB samples and our
results, it is seen that most PCEB samples contain low mass WDs as well. The prescription for $q_{\rm cr}$ has no clear
effect. From the results of model 2-10 and our standard model we can see that massive WDs are more abundant
when $\lambda=\lambda_{\rm g}$ and $\lambda=\lambda_{\rm b}$ (especially in model 7), while low mass WDs are
more abundant when $\lambda=\lambda_{\rm e}$ (a larger number of WD binaries are produced when $\lambda = \lambda_{\rm e}$
is adopted rather than $\lambda_{\rm b}$ or $\lambda_{\rm g}$). For models with low metallicity (models 11--13)
the distribution moves to the right (toward more massive WD) than the solar metallicity models (models 1--3).
This shift in the WD mass is likely caused by different evolutionary age and the RLOF moment of the different
metallicity star.

Figure 4 shows the probability distributions in the orbital period--WD mass (left), the orbital period--secondary
mass (middle), and the secondary mass--WD mass (right) planes. Shown here are for the standard model and
models 2--4. As we take into account three different prescriptions for $\lambda$ ($\lambda_{\rm g,\,b,\, e}$)
and three different prescriptions for $q_{\rm qr}$ ($q_{\rm cr1,\,2,\,3}$), there are 10 models shown in Figure 4.
From the left panel we see that systems with short orbital periods are most abundant when $\lambda=\lambda_{\rm e}$.
Also, low-mass WDs become most abundant when $\lambda=\lambda_{\rm e}$. From the middle panel, we see that the
relation between the period and the secondary mass is quite universal, which is not affected substantially by the
treatment of these unresolved physical processes. In the right panel, the relations of the masses of the two binary
members are different for different parameters, and more binaries survive CE evolution with $\lambda=\lambda_{\rm e}$.

\subsection{Double degenerate systems and SNe Ia rate}

\subsubsection{Double degenerate systems}

We let all WD/MS binaries continue their evolution to double WD binaries.
The close double WD binaries which can merge within the Hubble time are produced by going through the
CE evolution at least once. They may evolve into the CE phase during the first or second mass
transfer or may have CE evolution twice.  There is possibility that some systems
may indeed explode as an SN Ia within the SD scenario e.g., \citep{li97,wa12,a14}, but we only consider
double degenerate systems. Figure 5 shows the orbital distribution of binaries containing a
CO WD primary and either a CO or He WD secondary. These binaries have experienced the CE phase
at least once. Therefore, the orbital periods of these systems are less than 40 days as seen in
Figure 5. It is seen that a larger number of double WD systems have a short orbital period when
the value of $\lambda$ is larger (i.e., when $\lambda=\lambda_{\rm e}$ is adopted). The number
of WD binaries with a short orbital period also increases for low metallicity. We donot know
the full observational distribution of double WD systems yet, however, we have a number of confirmed samples.
The solid black lines (in Figs.5, 6, 7) show the distributions of the observed double WD binaries which can merge
within the Hubble time (See tables of Marsh 2011 \& Kaplan 2010), and they have similar orbital period, first WD mass
and mass ratio distribution ranges as our results (see Fig.6 and 7 for WD mass and mass ratio distributions).

Figure 6 shows the distribution of the primary CO WD masses. For all the models, most of the primary
CO WD masses are in the range of $0.5 - 1.0 M_\sun$. As described above, the double WD may go through first
stable or unstable mass transfer (for the second mass transfer we follow Hurley et al. (2002), and we use the same prescription for alpha as in the first CE phase, if it is unstable), thus the first stable mass transfer contributes
to the distribution of the primary CO WD masses, and the distribution in this Figure more or less differs from that of Figure 3.
The various parameters influence the primary mass
distribution to some extent, but generally a global pattern in the distribution of the primary CO WD
masses is not sensitive to these assumptions in the BPS study. The distribution of the mass ratios
between the secondary WD and the primary WD is given in Figure 7, the
distribution has double peaks for almost all the models.

\cite{san15} analyzed the observed data of the central object in the Planetary star Henize 2-428,
and claimed that Henize 2-428 has a double WD system with a mass ratio of nearly unity. They also
reported its combined mass (1.76$M_\sun$, which is well above the Chandrasekhar limit mass) and
its short orbital period (4.2 hours). Based on these values, they suggested that the system should
merge within 700 million years, being the first candidate progenitor of the super-Chandrasekhar-mass
channel in the context of the DD model of SNe Ia. On the other hand, \cite{gar15} reanalyzed the
results of \cite{san15} and suggested that the central object of Henize 2-428 is not likely
a double degenerate system. In particular, \cite{gar15} argued that the possibility of forming double
WD systems with mass ratio of $\sim 1$ is very low, in view of the stellar evolution process.
However, our results agree with \cite{san15}. In our models, the binaries initially containing
binary members with similar masses can produce double WD systems with a mass ratio close to
unity. Certain number of observed samples used in this work also have a mass ratio close to
unity. This means that there are a number of binaries which are likely to have the evolutionary path
as described in \cite{san15}.

\subsubsection{Type Ia supernovae rate}


In the DD scenario, the orbit of a double WD binary shrinks through the gravitational wave emission
and eventually the two WDs merge. The timescale of this process is given as follows \citep{la71}.
\begin{equation}
t_{\rm GW} = 8\times10^{7}\times\frac{(M_1+M_2)^{1/3}}{M_1 M_2}P^{8/3} \ {\rm year} \ ,
\end{equation}
where $P$ is the orbital period (in units of hour), $M_1$ and $M_2$ are the masses of the
primary WD and the secondary WD (in units of $M_\sun$), respectively. Recently, it has been proposed,
both theoretically and observationally, that both the SD and DD channels might have own (non-negligible)
contributions to produce SNe Ia. As for the SNe Ia birth rate, it has been suggested from BPS studies that
the prediction from the DD scenario is closer to the observed rate, but there is still some difference between
theory and obervation \citep{cl14}. Hereafter, we focus on the DD scenario in this paper, and a consistent modeling
of the SD and DD scenarios is beyond the scope of this paper.

The criteria for a DD system to explode as an SN Ia have not been clarified yet, except for the requirement
that the system should merger within the Hubble time. Adding to the total mass of the system, a mass ratio of
the two WDs ($q  = {M_2}/{M_1}$) could also be an important factor, but it depends on yet-unclarified explosion
mechanism. Different criteria are adopted in different BPS studies. In this section, we investigate how the BPS
parameters affect the predicted SN Ia rate, by adopting the same criteria as \citet{ch12}. We consider three
conditions as follow: (1) the combined mass $\geq$ the  Chandrasekhar limit mass ($1.38 M_\sun$ in this work),
(2)$q={M_2}/{M_1}\geq{2/3}$, (3) the two WDs must merge within the Hubble time. Note that while \citet{ch12}
obtained the SN Ia rate in their conservative model as consistent with the observation, the adopted criterion
on $q$ might indeed be optimistic. Also, \citet{ch12} found that the evolution of the SN Ia rate with time did
not fit the observation. The aim of this section is to clarify how these conclusions would be affected by the
different choice of the BPS parameters.

We compute the expected SN Ia rate under the DD scenario, including those from mergers of two CO WDs and mergers
of a CO WDs with a He WD. Galaxies have complicated star formation histories, but in this paper we adopt two simple
 models for demonstration; (1) a constant star formation rate (SFR) over the past 13.7 Gyr, and (2)  a single star
 burst (i.e., a delta function). For the case of the single star burst, we assume that the burst produce the stellar
 mass of $10^{11} M_\sun$. In the case of the constant SFR, we assume it to be $5\,{M_\sun}{\rm{yr}^{-1}}$ \citep{wi04}.

Figure 8 displays the evolution of the SN Ia birthrates with time, for the two models of the star formation history.
The colors and styles of lines are the same as those used in the previous figures. In the left panels for the single
star burst, we also show a fit with the uncertainty of the delay-time distribution (DTD) inferred from observations
\cite[the three black thin-solid lines,][see]{ma11,ma12}. Specifically, the middle line represents the one obtained
with the formula for the DTD,
$\phi(t)=4\times10^{-13}{\rm SN {yr}^{-1}} {\rm M_\odot}^{-1}{(\frac{t}{1 \rm Gyr})^{-1}}$ \citep{mao12},
and the upper and lower lines are for the $\pm50\%$ uncertainties.
As a general prediction from our BPS simulations, we note that the DD model could result is a power-law like distribution,
but it is not described as simple as a single power law as frequently attributed to the DD scenario. The peak in the DTD
results from the peak in the orbital periods in the DD systems (Figure 5). Hereafter in comparing the observed DTD (for
the single burst case), we mainly focus on the peak in the predicted rate.
In Figure 8, the upper two panels show the results for the models with $\alpha_{\rm CE}=1$
and solar metallicity (models 2-4). For a constant SFR case, the predicted birthrates of SNe Ia are lower than that of
the standard model ($\sim 1.4\times10^{-3}\,\rm{yr}^{-1}$). The lowest birthrate
among these models ($\sim 7.4\times10^{-4}\,\rm{yr}^{-1}$) corresponds to the case with $\lambda=\lambda_{\rm g}$ and
$q_{\rm cr} = q_{\rm cr2}$. The birthrate is the highest ($\sim 1.7\times10^{-3}\,\rm{yr}^{-1}$) when
$\lambda=\lambda_{\rm e}$ and $q_{\rm cr} = q_{\rm cr1}$. As shown in the left panel for the single star burst case,
the evolution of the delay time marginally fits the observational lower limit. The second two panels of Figure 8
(models 5-7, in which $\alpha_{\rm CE}$ changes with $q$) show that there is no obvious change in both the evolution
of the birthrate and the DTD, except that the peaks are just slightly lower and a little change happens in the evolution
of birthrate when $\lambda=\lambda_{\rm b}$.

The third two panels of Figure 8 (models 8-10) show the results with  IMRD $n(q)\propto q$. In the constant SFR case,
the birthrates are higher than the standard model when $\lambda = \lambda_{\rm e}$, but comparable or lower for
$\lambda=\lambda_{\rm g}$ and
$\lambda_{\rm b}$. The range of the predicted birthrate for these models is between
$\sim 1.2\times10^{-3}\,\rm{yr}^{-1}$ and $\sim  2.8\times10^{-3}\,\rm{yr}^{-1}$. For the single star burst case,
the delay time evolution is marginally consistent with the observationally derived rate, considering an uncertainty
of $\sim 50$\%.
In the lowest panels (models 11-13) only the metallicity is different ($Z=0.001$) from the corresponding models 2-4.
The birthrates are typically not as high as in the standard model. The low metallicity results in higher rates than
the solar metallicity, especially when $\lambda=\lambda_{\rm e}$. \cite{to12} estimated that the low metallicity has
no effect on the delay-time evolution. However, our results show that the low metallicity changes the delay-time
evolution to some extent.

In summary, our DD models generally under-predict the SN Ia rate, except for the model with $\lambda=\lambda_{\rm e}$
and $q_{\rm cr} = q_{\rm cr1}$. The criteria of $q={M_2}/{M_1}\geq{2/3}$ and combined mass $> 1.38M_\sun$ implies that
the mass of secondary WDs must be at least 0.55$M_\sun$,
so the secondary He WDs donot significantly contribute to the SN Ia rate. This motivates us to further investigate
a condition to increase the SN Ia rate under
the DD scenario. Another key parameter to describe the nature and outcome of the DD systems is the distribution of the
initial orbital separation. Figure 9 shows the predicted SN Ia birthrate when we adopt the range of the initial orbital
separations as $3\leq a_{\rm i}/R_\odot \leq 10^4$ \citep{hu02}, instead of $3\leq a_{\rm i}/R_\odot \leq 10^6$ in our
reference models. The predicted SN Ia rate increases, ranging between
$\sim 1.3\times10^{-3}$ and $\sim 4\times10^{-3}\,\rm{yr}^{-1}$ depending on the BPS parameters. This brings a large
fraction of our BPS moldes to the values as high as observationally derived, under the particular criteria we assumed
for a DD system to explode as an SN Ia.

\subsubsection{A link between the explosion scenarios and the BPS SN Ia rate}

We note that the results shown in the previous section does not necessarily provide a fair comparison
to the {\em absolute} values of the predicted SN Ia rates to the observed rate. The rate is highly dependent on
the criteria for a DD system to become an SN Ia, where different scenarios have different criteria (and frequently
the criteria are not accurately determined from the first principle). In the previous section, we adopted the same
criteria as those adopted by \citet{ch12}, so that we can focus on dependence of the SN Ia rate on different BPS
parameters by calibrating our BPS models with previous study. In this section, we study the SN Ia rate based on
physically-motivated criteria, taking into account recent development in the explosion simulations.

Given a lack of the detailed knowledge on the {\em real} explosion mechanisms, it is not possible to cover all the
possible models. However, investigating the following two scenarios provides useful insight, as these models likely
represent the lower and upper limits on SN Ia rate within the DD scenario.

\begin{enumerate}
\item {\bf Carbon-Ignited Violent Merger Model:} When two CO WD merge, the high accretion rate would create hot
spots on the primary WD's surface where the temperature is so high that carbon burning would be ignited explosively
and produce a detonation wave \citep{pa13}. While there is still a numerical convergence issue (see, e.g.,
Tanikawa et al. 2015), the numerical simulations agree that this mode is likely a result
for a merging DD system if a total mass well exceeds the Chandrasekhar limiting mass and the mass ratio is
nearly unity. Since this prediction is robust, this will give a lower limit for the SN Ia rate from the DD system.
We adopt the following criteria: (1) $0.8M_\sun< M_2 < M_1$ \citep{sa15} (the combined mass of two CO WDs is at
least $1.6 M_\sun$), (2) $q > 0.8$ and (3) the system must merger within the Hubble time.
\item {\bf Chandrasekhar Mass Model:} If a DD system avoid a prompt detonation at the merging, the system is then
represented by a massive CO WD that accretes materials from a thick accretion torus and a hot envelope. Given the
high accretion rate, the WD is suggested to become a ONeMg WD by a carbon deflagration and then the system would not
explode as an SN Ia \citep{Sa85}. However, it could still avoid the deflagration \citep{yo07}, and in this case the
primary WD is expected to evolve into a Chandrasekhar mass WD. Thus, adopting the following criteria, we should obtain
an upper limit for the SN rate under the DD scenario: (1) the combined mass of the two WD $\geq$ $1.38 M_\sun$, and
(2) the merging of two WDs within the Hubble time.
\end{enumerate}

The upper two panels of Figure 10 show the case of the `Chandrasekhar mass model'. Depending on the BPS parameters,
the range of SNe Ia birthrate is between $\sim 8.0\times10^{-4}$ and $\sim 2.24\times10^{-3}\,\rm{yr}^{-1}$. From
these results, we can see that the mass ratio criterion to become SNe Ia has some effects on SNe rates but not that
much in this context (i.e., either the criterion is set at $q \sim 2/3$ or not). With this model, the secondary He WDs contribute
to the SN Ia rate in some extent as well. To have a combined mass exceeding
$\sim 1.38 M_{\odot}$, the systems should in any case have the mass ratio of $q > 0.5$.

However, this modest dependence on the critical mass ratio is not necessarily the case if we consider the criterion
much tighter than $q \sim 0.5$. The simulated SNe rates with models 1-4 for the `Carbon-Ignited Violent Merger Model' are
shown in the lower two panels of Figure 10. Compared to the results with the modest criterion ($q > 2/3$), the
birthrates of the SNe Ia are much lower. Also, virtually one peak is seen for the DTD at $\sim 10^8$ yeas, unlike what
is observationally derived. This also results in a rapid rise of SN Ia rate in the case of the constant SFR. With this
criterion, a large fraction of systems having the combined mass exceeding $1.38 M_{\odot}$ are now rejected as SN Ia
progenitors. The peak in the DTD corresponds to the peak in the distribution of the mass ratio at $q > 0.8$ in the DD
systems (Figure 7). The birthrate of model 1 is $\sim 1.4\times10^{-4}\,\rm{yr}^{-1}$, and those in models 2-4 fall in
the range between $\sim 1.69\times10^{-4}$ and $\sim 8.2\times10^{-5}\,\rm{yr}^{-1}$. The model with
$\lambda=\lambda_{\rm e}$ (model 2) and  $q_{\rm cr} = q_{\rm cr1}$ gives the highest birthrate, and model 4
gives the lowest rate under this tight criterion.

\section{Discussion and conclusions}

With the Monte Carlo method, we have performed detailed BPS simulations of the evolution of WD+MS binaries
and double degenerate systems for 37 models with different recipes for the key binary evolution processes.
The final systems in our simulations come from binaries that have experienced CE evolution at least once, and
the effects of the following key processes/conditions have been investigated: the CE efficiency ($\alpha_{\rm CE}$),
the binding energy parameter ($\lambda$), the critical mass ratio ($q_{\rm cr}$), and the initial mass ratio distribution,
and metallicity.

\cite{de1993} performed pioneering simulations for the formation of PCEBs by adopting $\lambda=0.5$. Later,
\cite{de00} demonstrated that the binding energy parameter $\lambda$ changes with the evolutionary phases.
The works on the PCEB simulations were further updated \citep{wi04}. However, these previous results were different
from observations in some degree. Recently, a number of researchers performed BPS for PCEBs with different assumptions,
and altogether these studies would lead to comprehensive investigation of the formation and evolution of PCEBs.
For example, \cite{po07} used very low values for the CE efficiency that changes with the mass of the secondary star.
\cite{de10,de12} relaxed the assumption that the binding energy parameter is a constant, and considered both
$\lambda_{\rm g}$ and $\lambda_{\rm b}$ to describe $\lambda$. Similar comprehensive studies have also been made
by \cite{z10} and \cite{to13}. It seems that either the modeling assumptions did not treat the parameters in the
realistic ways or the model results were not comparable with observations. More recently, \cite{c14} presented detailed
analysis of the selection effects that affect the sample of observed PCEBs obtained through SDSS. They also provided a
thorough comparison between their BPS results and the observed sample of these systems. However, their sample was very
limited. \cite{z14} presented a systematic investigation that includes the contribution from the recombination energy to
the energy budget of the CE evolution. They found that the recombination energy leads to a large number of PCEBs with
long orbital period (longer than 10 days), considering only three different input parameters in BPS simulations. In our
present work, we have further updated and extended the previous works by systematically considering different recipes for
the key binary evolution processes (e.g., by using three different prescriptions for $\lambda$, three different recipes for
the critical mass ratios).

In our BPS simulations for PCEBs, the main features that characterize the distributions of resulting binary
parameters for the different models can be summarized as follows:
\begin{enumerate}
\item The three different prescriptions for the binding energy parameter ($\lambda$) influence the distributions
clearly. Binaries can eject the CE more easily and result in shorter orbits (for both PCEBs and double WD systems),
for a higher value of $\lambda$ (i.e., $\lambda = \lambda_{\rm e}$). The orbital period of PCEBs can be as short as 0.008
day when $\lambda=\lambda_{\rm e}$. If $\alpha_{\rm CE}$ is treated to change with $q$, a larger number of systems survive
at the CE phase than in the case where $\alpha_{\rm CE}=1$.
\item The effect of different initial mass ratio distributions are mainly reflected by the resulting mass distributions.
\item For binaries with a low metallicity, a larger number of systems can dispel the CE and tend to have shorter orbits.
A choice of the prescription for the critical mass ratios turn out to have only little effect on the distributions of
binary parameters of the resulting PCEBs.
\end{enumerate}

Double WD systems considered to be possible progenitors of SNe Ia are descendants of WD/MS binaries.
There have been a number of BPS works on the SD and DD scenarios and observational constraints
\citep[for recent reviews]{ma14,ru14}. The SNe Ia rate is one of the important constraints to understand
the natures and progenitors of SNe Ia. In our BPS study of the SN Ia rate, we tested different prescriptions for
five main input physical parameters in the binary evolution. By considering several conditions for the systems to
explode as an SN Ia, our findings are summarized as follows:
\begin{enumerate}
\item A larger number of double WD systems can survive the CE phase when $\lambda$ changes with the evolution than
a case where it is set as a constant. A number of double WD systems have
mass ratios around 1. We have shown that $\alpha_{\rm CE}$, $\lambda$, $n(q)$ and Z affect the distributions of the
resulting double WD parameters.
\item Considering a merging of CO WD with a CO WD or a He WD, the simulated SNe Ia birthrate ranges between
$\sim 7.4\times10^{-4}$ to $\sim 2.8\times10^{-3}\,\rm{yr}^{-1}$ with different parameters. With our
first criteria, He WDs donot have significant contribution to the SN Ia rate. This range applies
when the criterion for the mass ratio to become an SNe Ia is not larger than $q \sim 0.8$. The simulated birthrates
are marginally comparable with the observed SNe Ia rate if  $\lambda=\lambda_{\rm e}$ and $q_{\rm } = q_{\rm cr1}$.
\item If we adopt the initial orbital separations of $3\leq a_{\rm i}/R_\odot \leq 10^4$ instead of
$3\leq a_{\rm i}/R_\odot \leq 10^6$, then we obtain the higher birthrate ranging between
$\sim 1.3\times10^{-3}$ and $\sim 4\times10^{-3}\,\rm{yr}^{-1}$ (see the Figure 9). The
observed SNe Ia rate is well explained by this model if $\lambda=\lambda_{\rm e}$.
\end{enumerate}

To summarize, we conclude that the CE parameters, the metallicity and other parameters affect the evolution
of SNe Ia rate with time. While even our optimistic models still show a discrepancy in the evolution of the SN Ia
rate as compared to the observational inferred power law distribution (i.e., the ratio of the `young' population
to the `old population' is quite high in our BPS models), the simulated SNe Ia rate can be comparable to the observations,
depending on the treatment of the parameters. Especially, a combination of $\lambda=\lambda_{\rm e}$ and
$q_{\rm cr} = q_{\rm cr1}$ results in the highest SNe Ia rate as being compatible to the observations.

While the above argument should be correct for the dependence of the resulting SNe Ia rate to the BPS parameters, the
absolute rate should depend on the particular criteria for a DD system to explode as an SNe Ia. The criteria are
different for different explosion models. To see the effect and to link the present BPS study to the existing explosion
models, we also calculate the simulated SNe Ia rate under different assumptions on the criterion of the mass ratio for
the SN Ia progenitors.
\begin{enumerate}
\item Even if we totally remove the criterion of the mass ratio as compared to our reference model (where $q > 2/3$),
the resulting SN Ia rate is not that significantly affected. This results in a range of the SNe Ia birthrate between
$\sim 8.0\times10^{-4}$ and $\sim 2.24\times10^{-3}\,\rm{yr}^{-1}$. This prescription corresponds to the classical
Chandrasekhar mass scenario.
\item The high value of the mass ratio is an essential ingredient in the so-called violent merger scenario.  Setting
the critical mass ratio higher than $\sim 0.8$, the resulting SNe Ia rate is decreased substantially. By
adopting $0.8M_\sun< M_2 < M_1$ \citep{sa15} and $q > 0.8$, the SN Ia birthrates are $\sim 1.4\times10^{-4}\,\rm{yr}^{-1}$
(model 1), and $\sim 1.69\times10^{-4}$ -- $\sim 8.2\times10^{-5}\,\rm{yr}^{-1}$ (models 2--4 with different prescriptions
for $q_{\rm cr}$). Even for the optimistic model with $\lambda=\lambda_{\rm e}$ (model 2) and $q_{\rm cr} = q_{\rm cr1}$,
the simulated SNe Ia rate is far below the observationally derived SNe Ia rate.
\end{enumerate}

\acknowledgments This work was funded by China Postdoctoral Science Foundation, the Natural Science
Foundation of China under grant numbers 11390371, 11133001 \& 11333004,
the Strategic Priority Research Program of CAS under grant No.
XDB09000000. The work also was partly supported by JSPS KAKENHI (No. 26800100) from MEXT and by WPI Initiative,
MEXT, Japan.



\begin{table}

\begin{center}
\caption{Different models used in our calculation }

\begin{tabular}{l l l l l}
 \hline\hline
Model & $\alpha_{\rm CE}$ & $\lambda$ & $n(q)$ & Z\\
\hline
 1 & 1 & 1 & 1 & 0.02\\
 2 & 1 & $\lambda_{\rm e}$ & 1 & 0.02\\
 3 & 1 & $\lambda_{\rm b}$ & 1& 0.02\\
 4 & 1 & $\lambda_{\rm g}$ & 1& 0.02\\
 5 & Eq.(6) & $\lambda_{\rm e}$ & 1& 0.02\\
 6 &  Eq.(6)  & $\lambda_{\rm b}$ & 1& 0.02\\
 7 &  Eq.(6)  & $\lambda_{\rm g}$ & 1& 0.02\\
 8 & 1 & $\lambda_{\rm e}$ & $\propto q$& 0.02\\
 9 & 1 & $\lambda_{\rm b}$ & $\propto q$& 0.02\\
 10 & 1 & $\lambda_{\rm g}$ & $\propto q$& 0.02\\
 11 & 1 & $\lambda_{\rm e}$ & 1& 0.001\\
 12 & 1 & $\lambda_{\rm b}$ & 1& 0.001\\
 13 & 1 & $\lambda_{\rm g}$ & 1& 0.001\\
\hline
\end{tabular}
\end{center}
\end{table}

\clearpage

\begin{figure}
\centering
\includegraphics[totalheight=2.5in,width=3.0in]{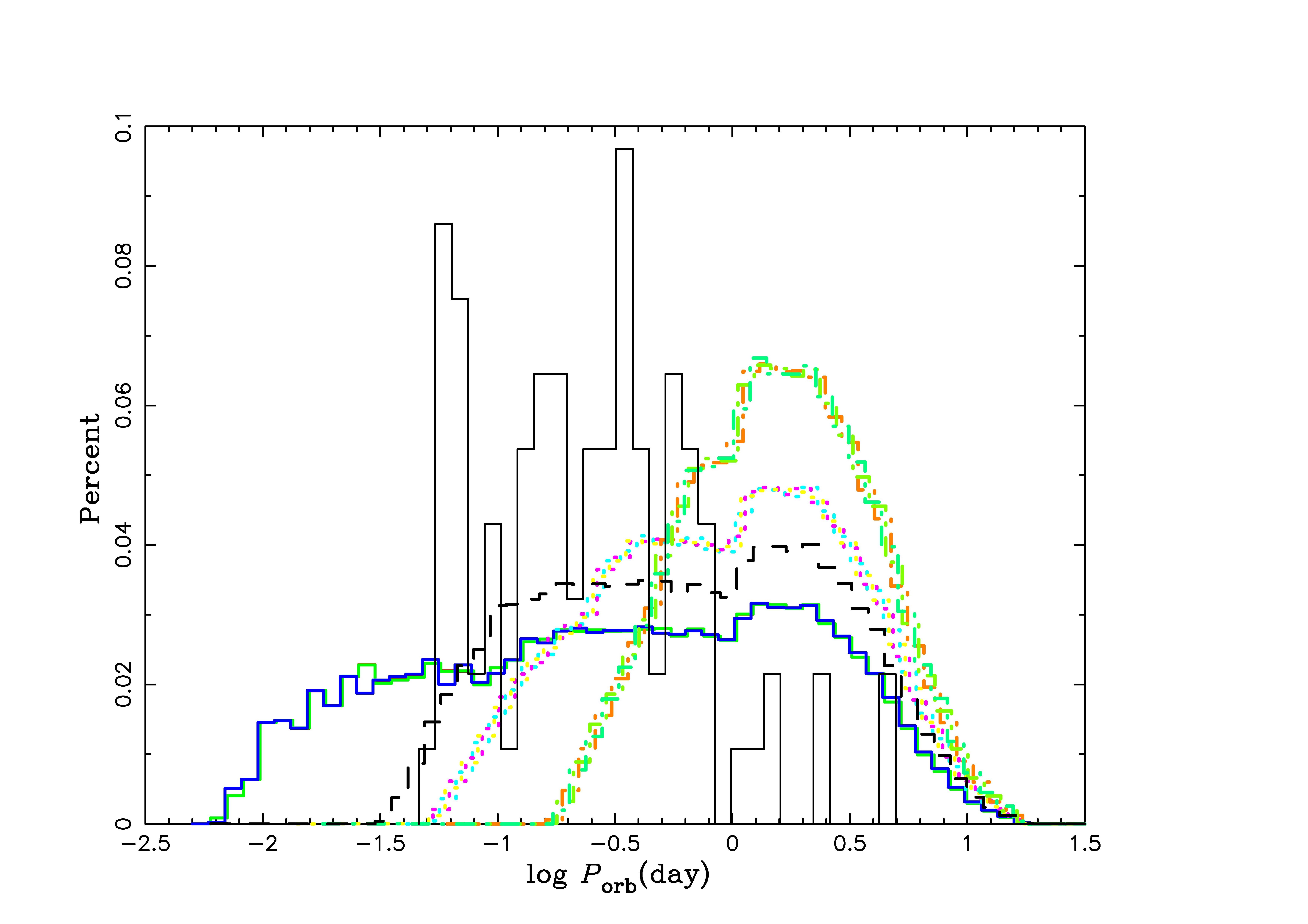}
\includegraphics[totalheight=2.5in,width=3.0in]{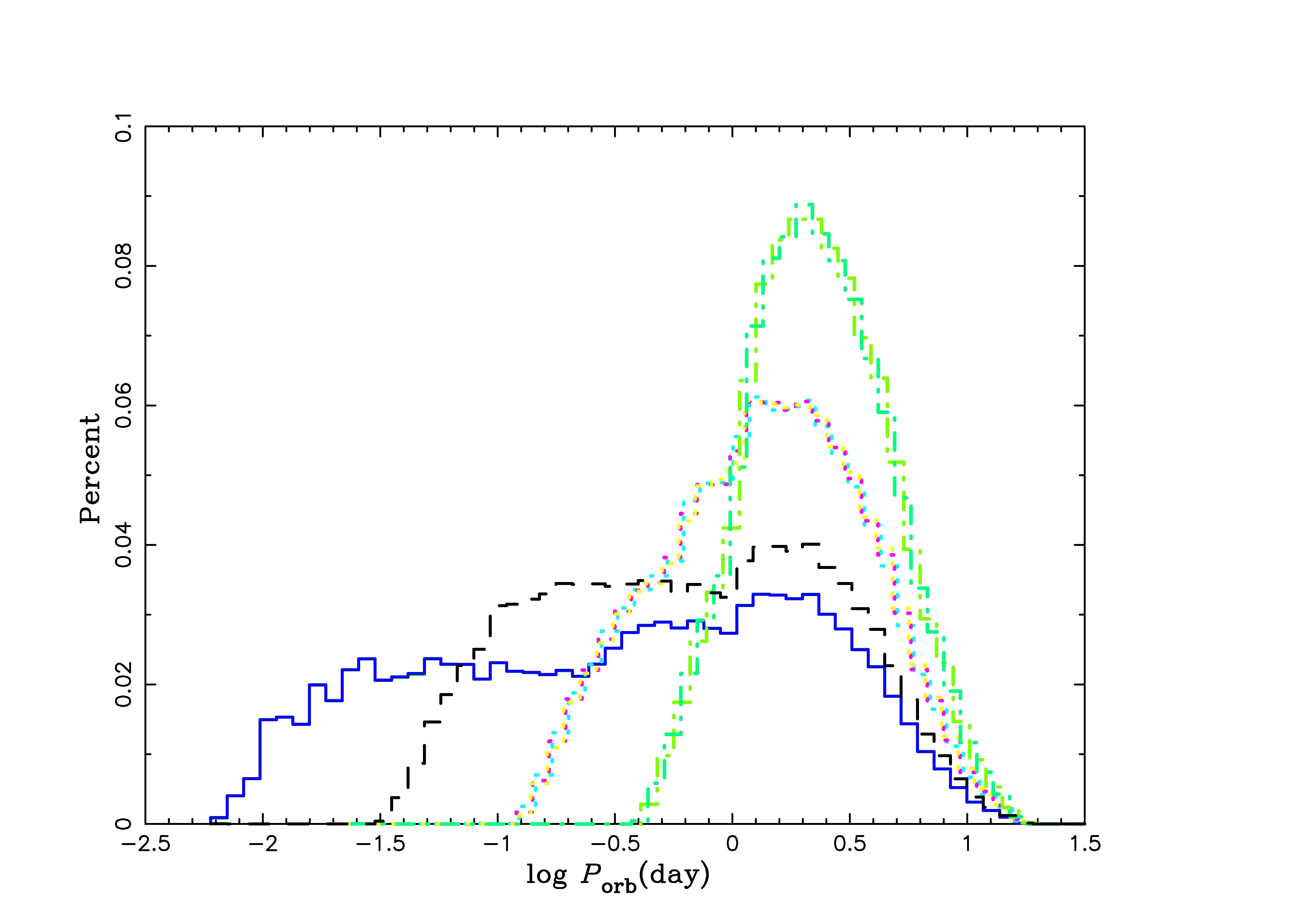}
\includegraphics[totalheight=2.5in,width=3.0in]{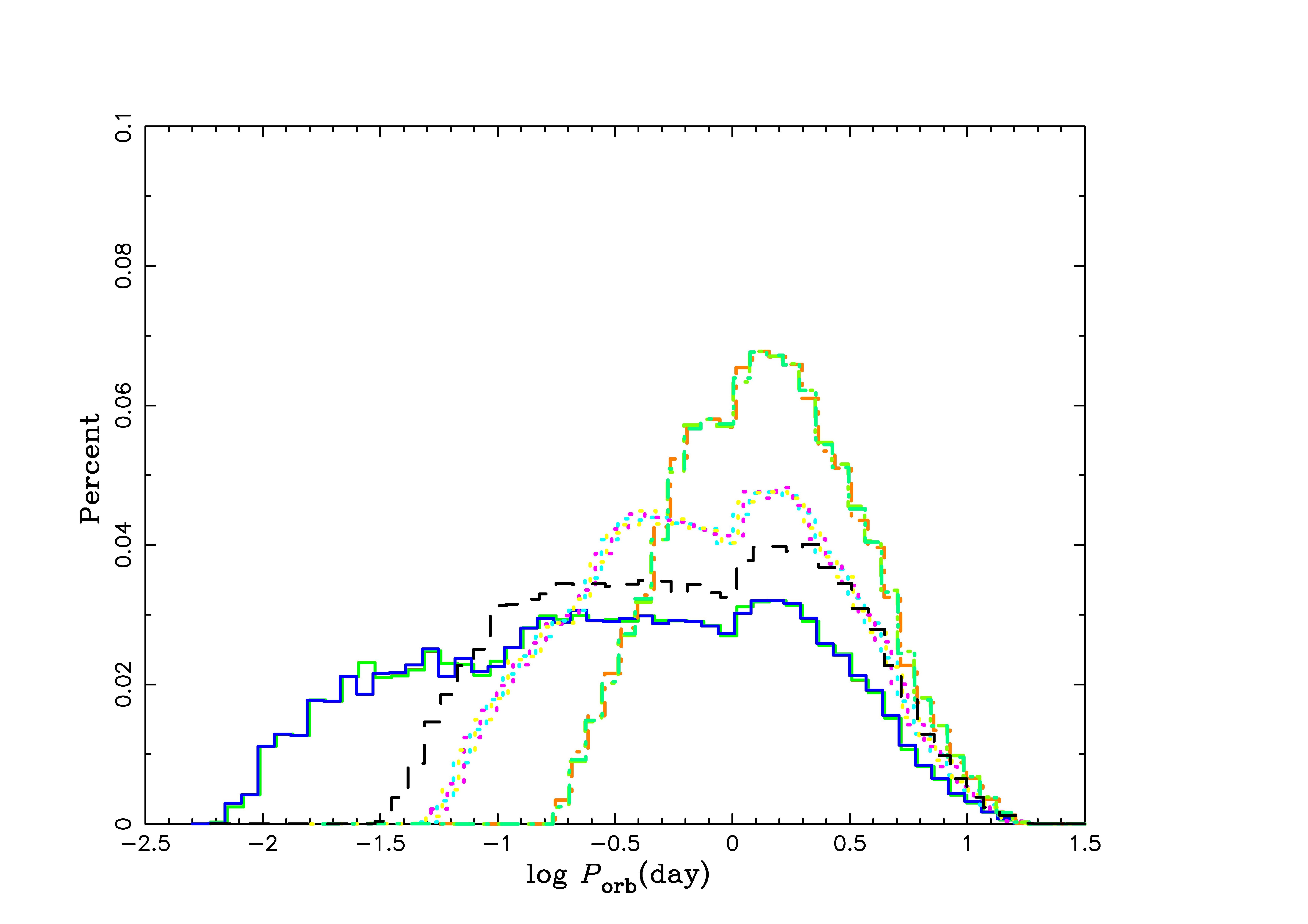}
\includegraphics[totalheight=2.5in,width=3.0in]{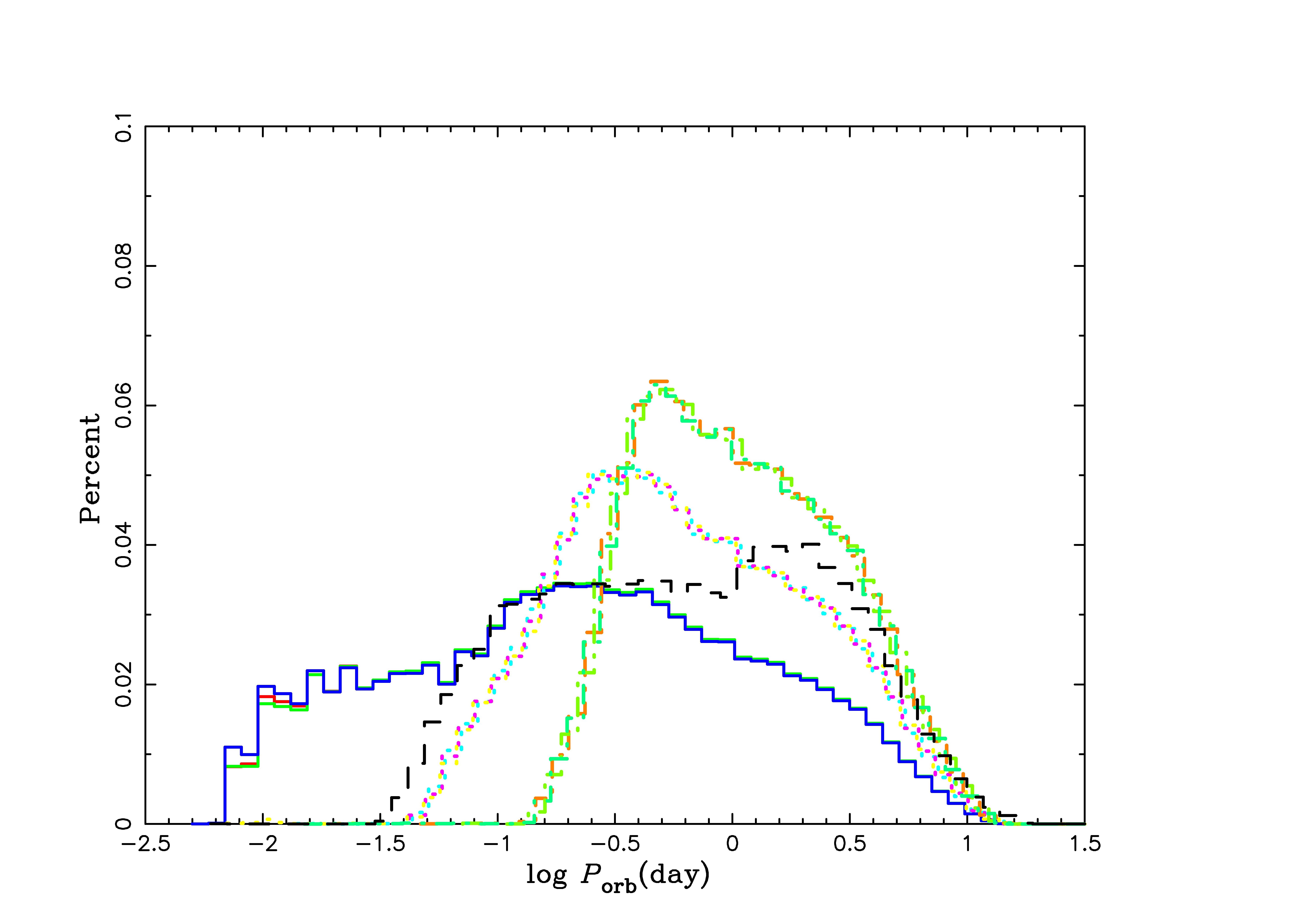}
\caption{Distributions of the orbital periods of PCEBs, for models 2--4
(upper left panel), models 5--7 (upper right panel), models 8--10 (lower left panel),
and models 11--13 (lower right panel). In all the panels, the result from our standard model
is shown by the black dashed line. Taking into account different prescriptions for the binding
energy parameters ($\lambda$, which specifies a model) and the critical mass ratios ($q_{\rm cr}$
for which the three cases are considered for each model), there are three lines (except for the standard model)
shown in each panel. The models are indicated as follows: The solid red,  green and blue lines for
$\lambda=\lambda_{\rm e}$ with $q_{\rm cr1}$, $q_{\rm cr2}$ and $q_{\rm cr3}$, respectively.
The dotted cyan, magenta and yellow lines are for  $\lambda=\lambda_{\rm b}$ with $q_{\rm cr1}$,
$q_{\rm cr2}$ and $q_{\rm cr3}$, respectively. The dash dotted orange, green$+$yellow and green$+$cyan
lines for $\lambda=\lambda_{\rm g}$ with $q_{\rm cr1}$, $q_{\rm cr2}$, and $q_{\rm cr3}$, respectively.
The sold black line is for the selected observed samples.}
\label{fig:1}
\end{figure}

\clearpage

\begin{figure}
\centering
\includegraphics[totalheight=2.5in,width=3.0in]{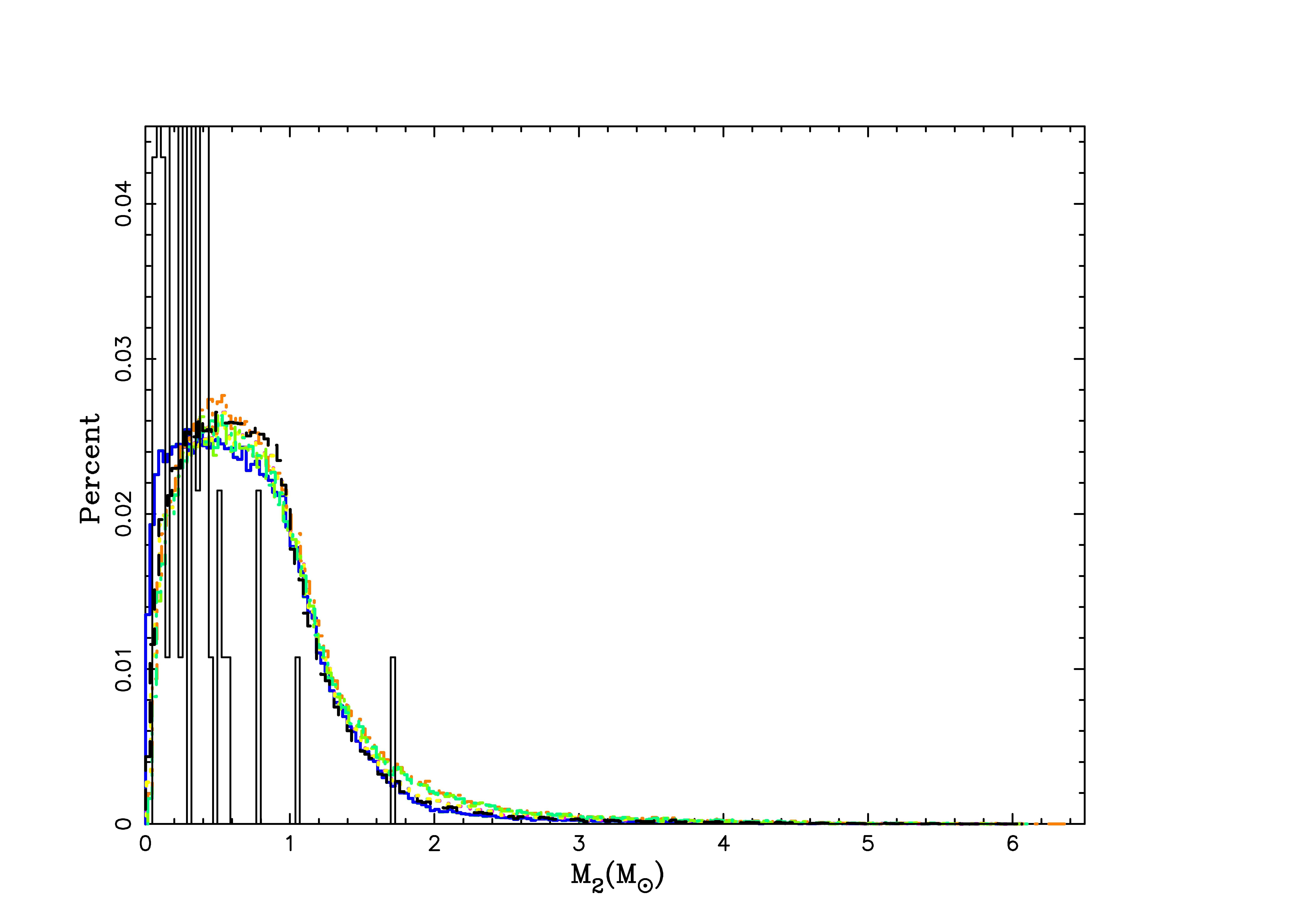}
\includegraphics[totalheight=2.5in,width=3.0in]{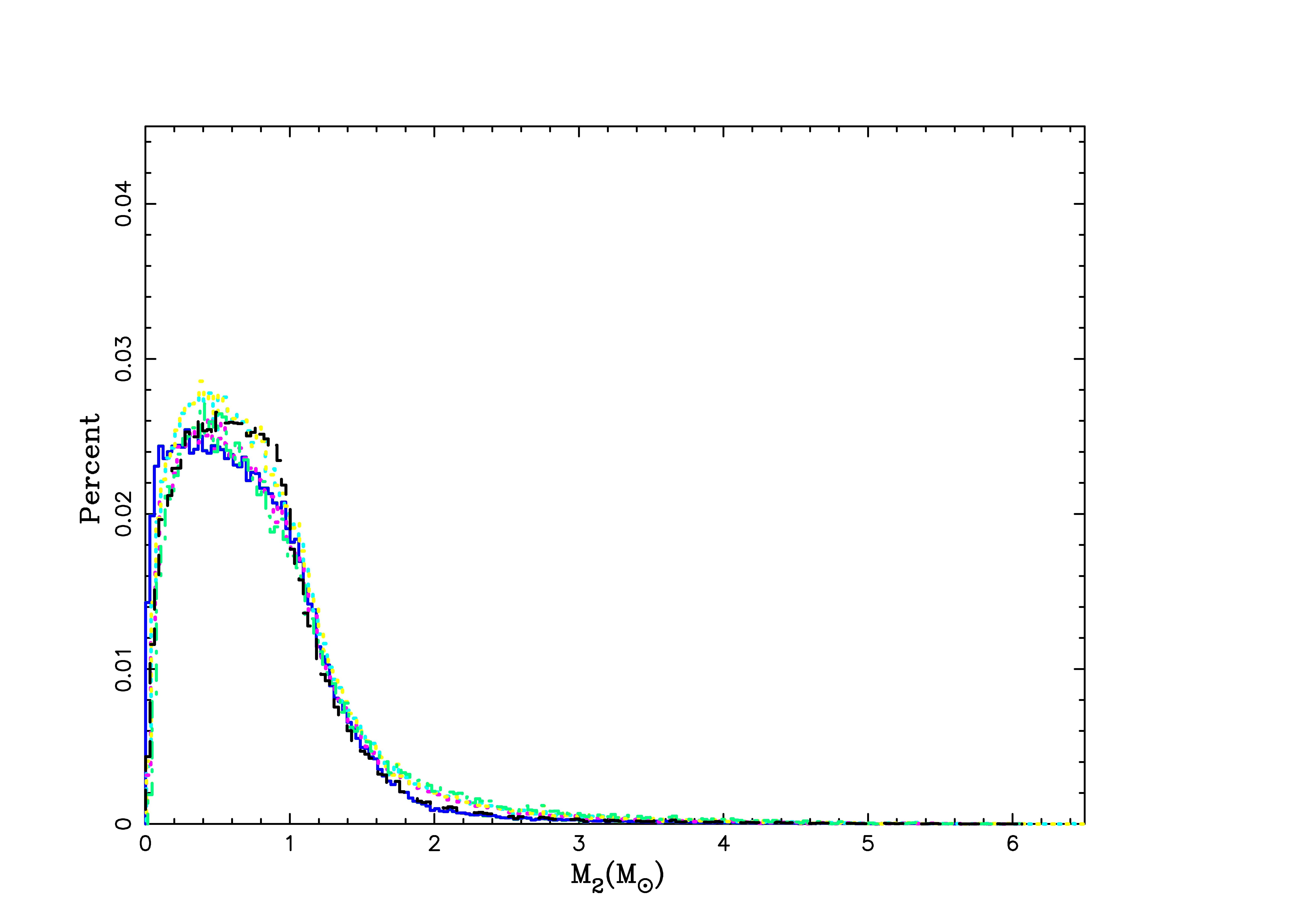}
\includegraphics[totalheight=2.5in,width=3.0in]{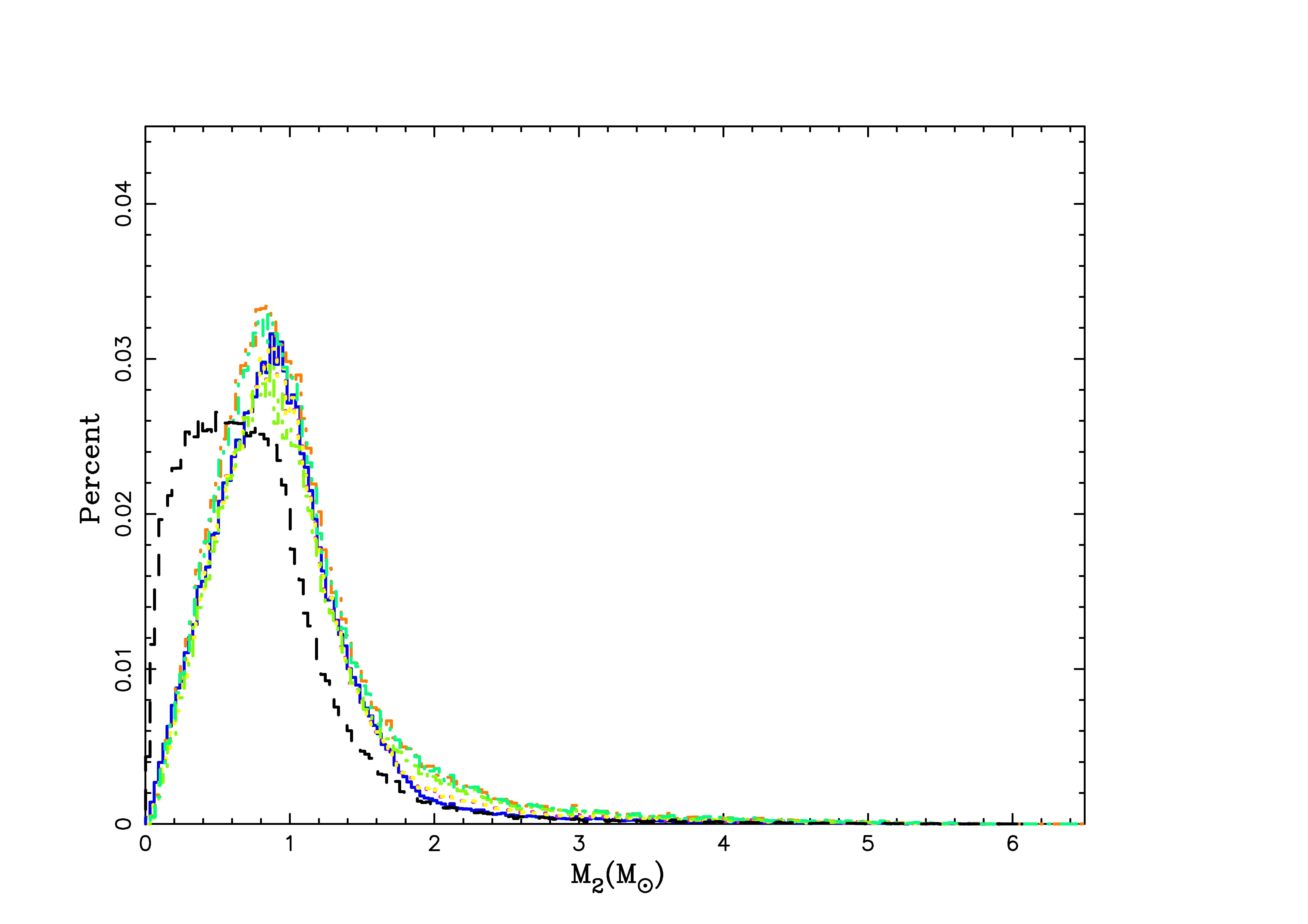}
\includegraphics[totalheight=2.5in,width=3.0in]{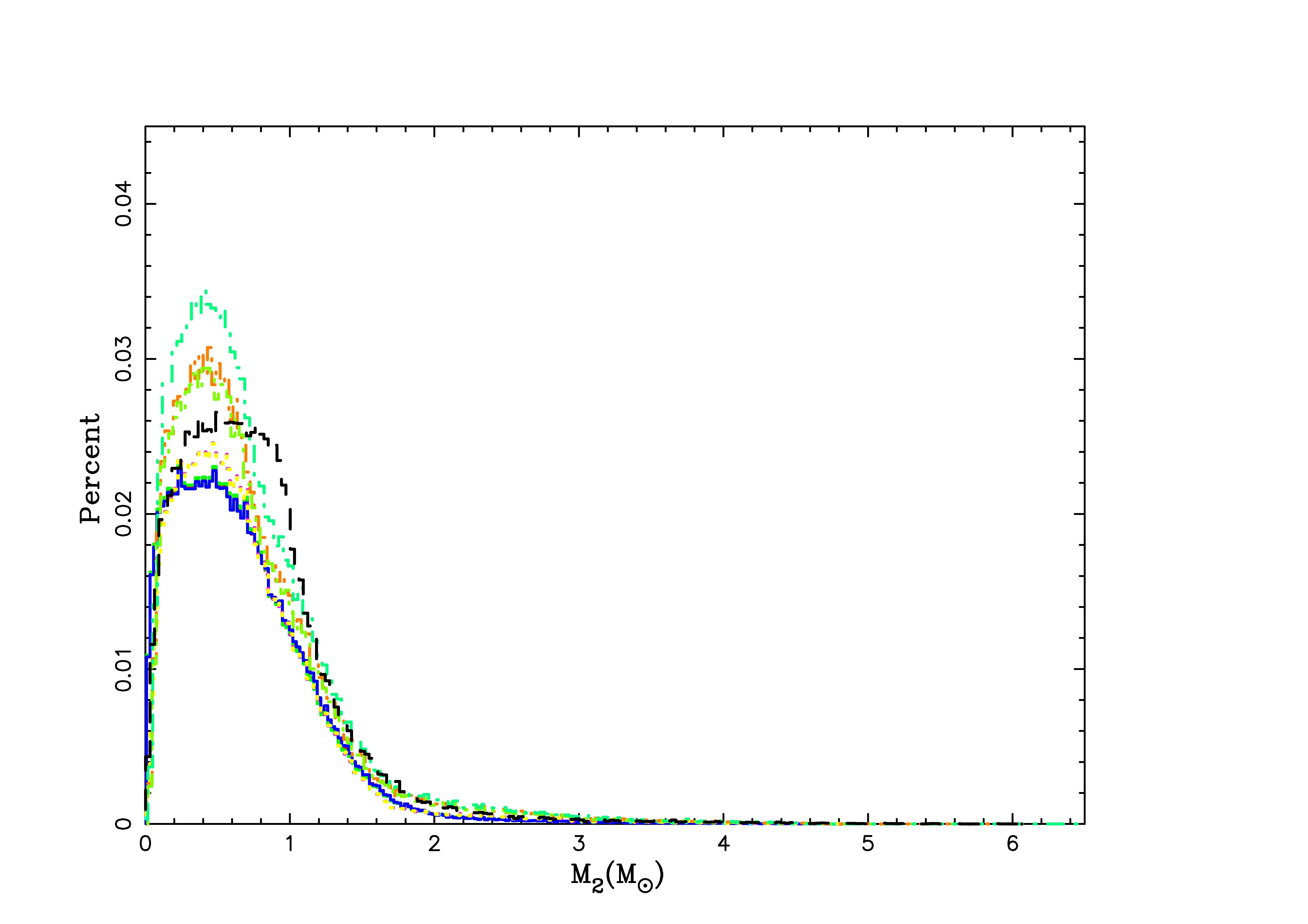}
\caption{Distribution of the secondary MS masses. See the caption of Figure 1 for the model description.}
\label{fig:1}
\end{figure}

\clearpage

\begin{figure}
\centering
\includegraphics[totalheight=2.5in,width=3.0in]{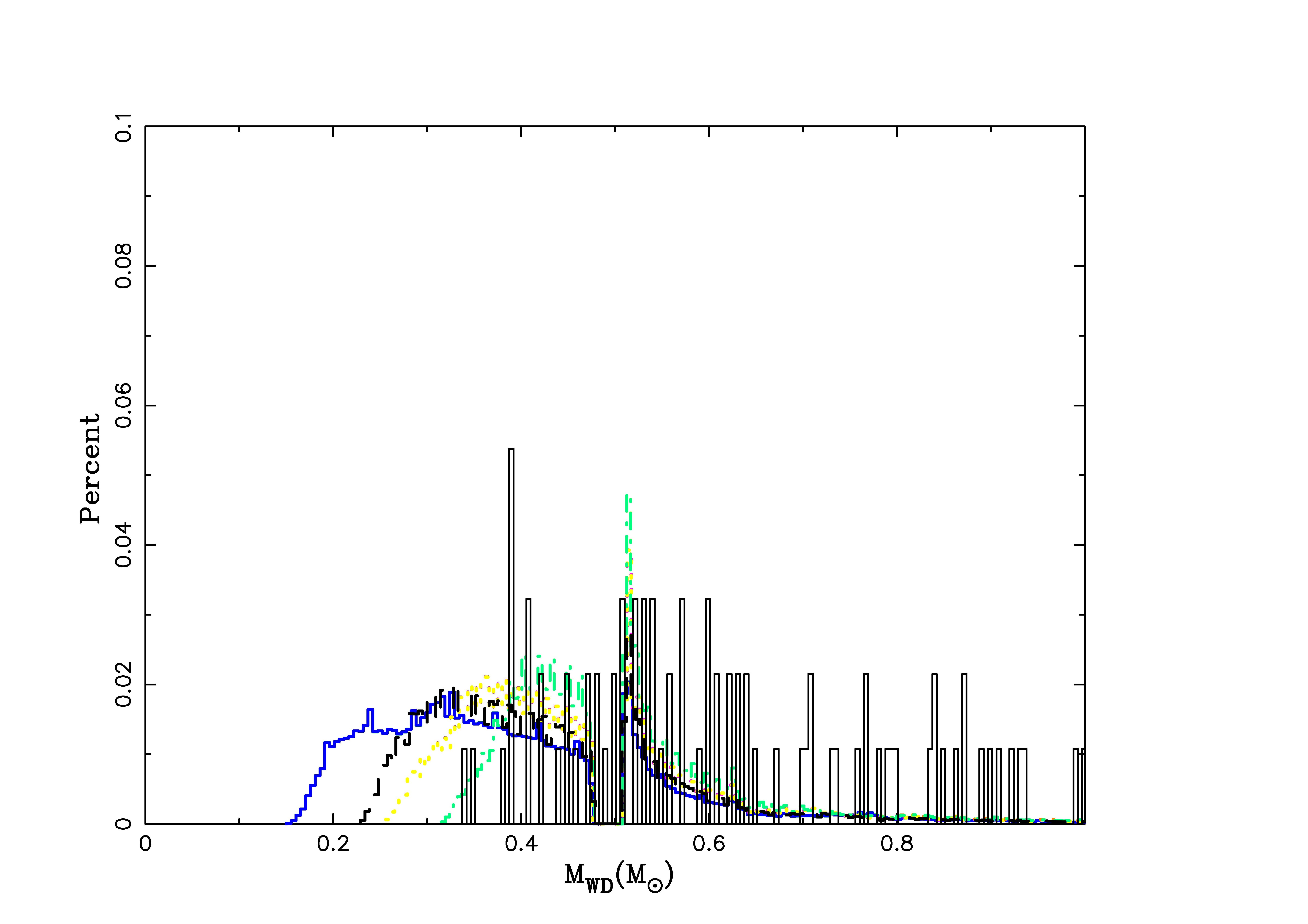}
\includegraphics[totalheight=2.5in,width=3.0in]{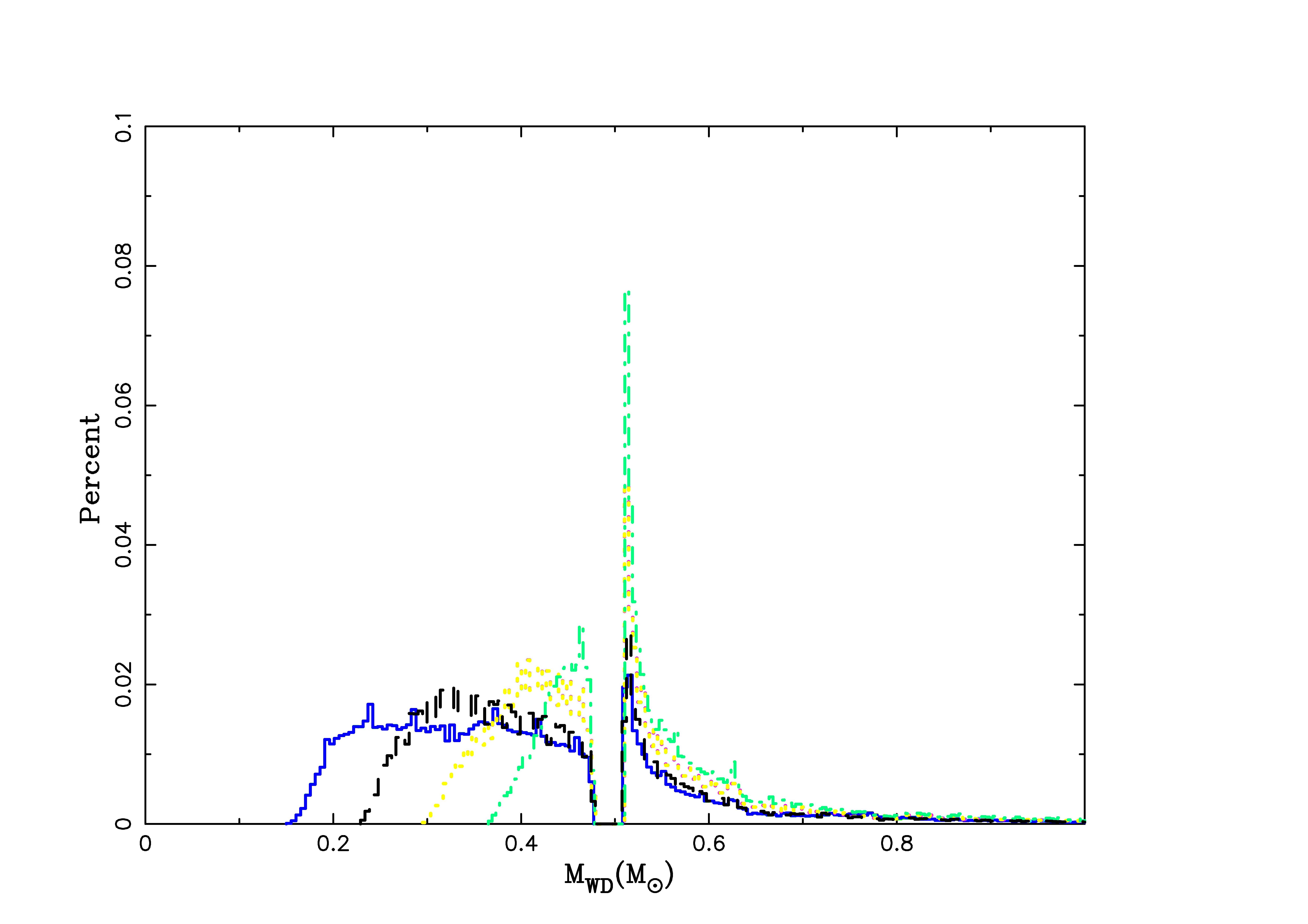}
\includegraphics[totalheight=2.5in,width=3.0in]{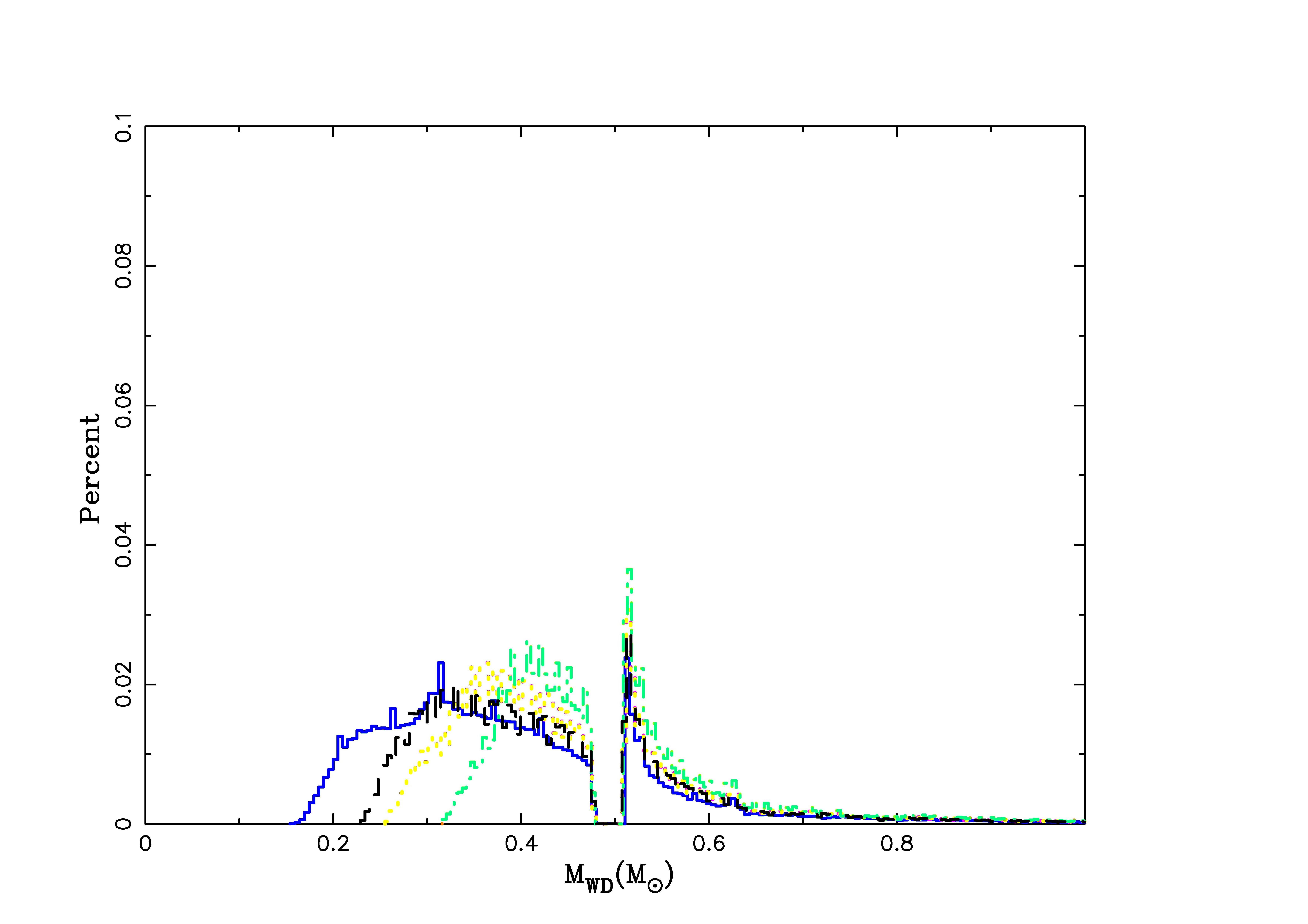}
\includegraphics[totalheight=2.5in,width=3.0in]{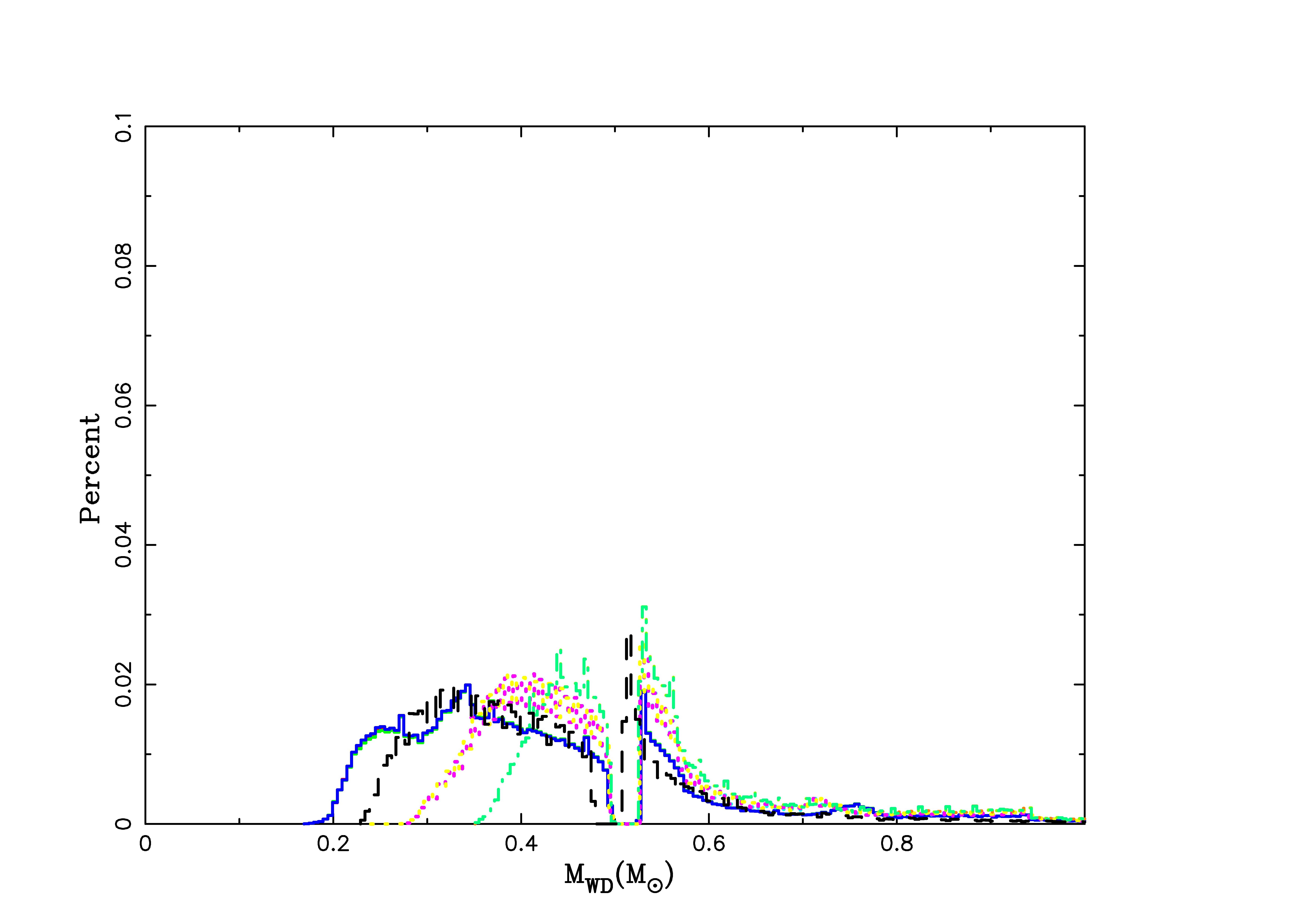}
\caption{Distribution of the WD masses. See the caption of Figure 1 for the model description.}
\label{fig:1}
\end{figure}

\clearpage

\begin{figure}
\centering
\includegraphics[totalheight=3.1in,width=3.2in]{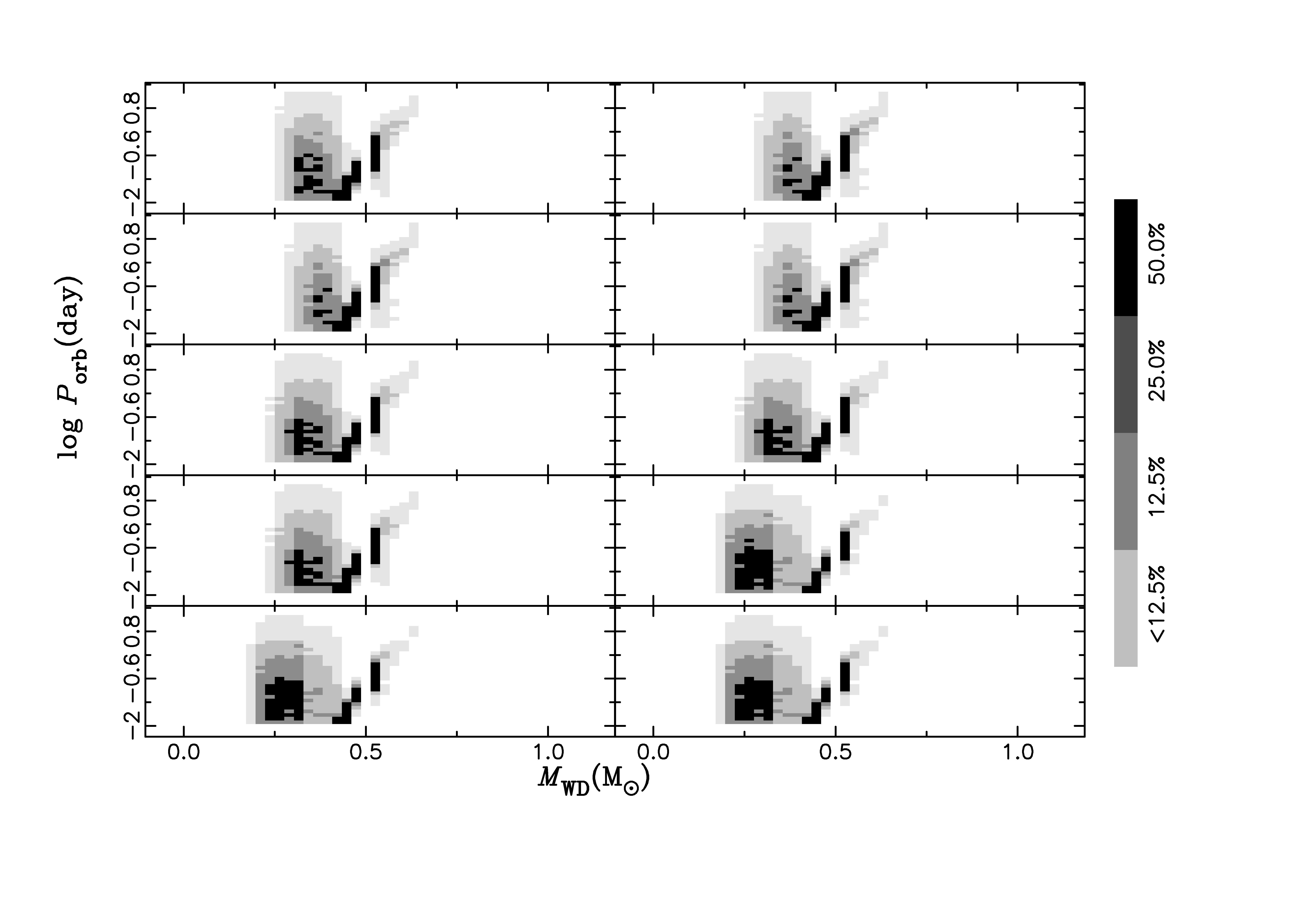}
\includegraphics[totalheight=3.1in,width=3.2in]{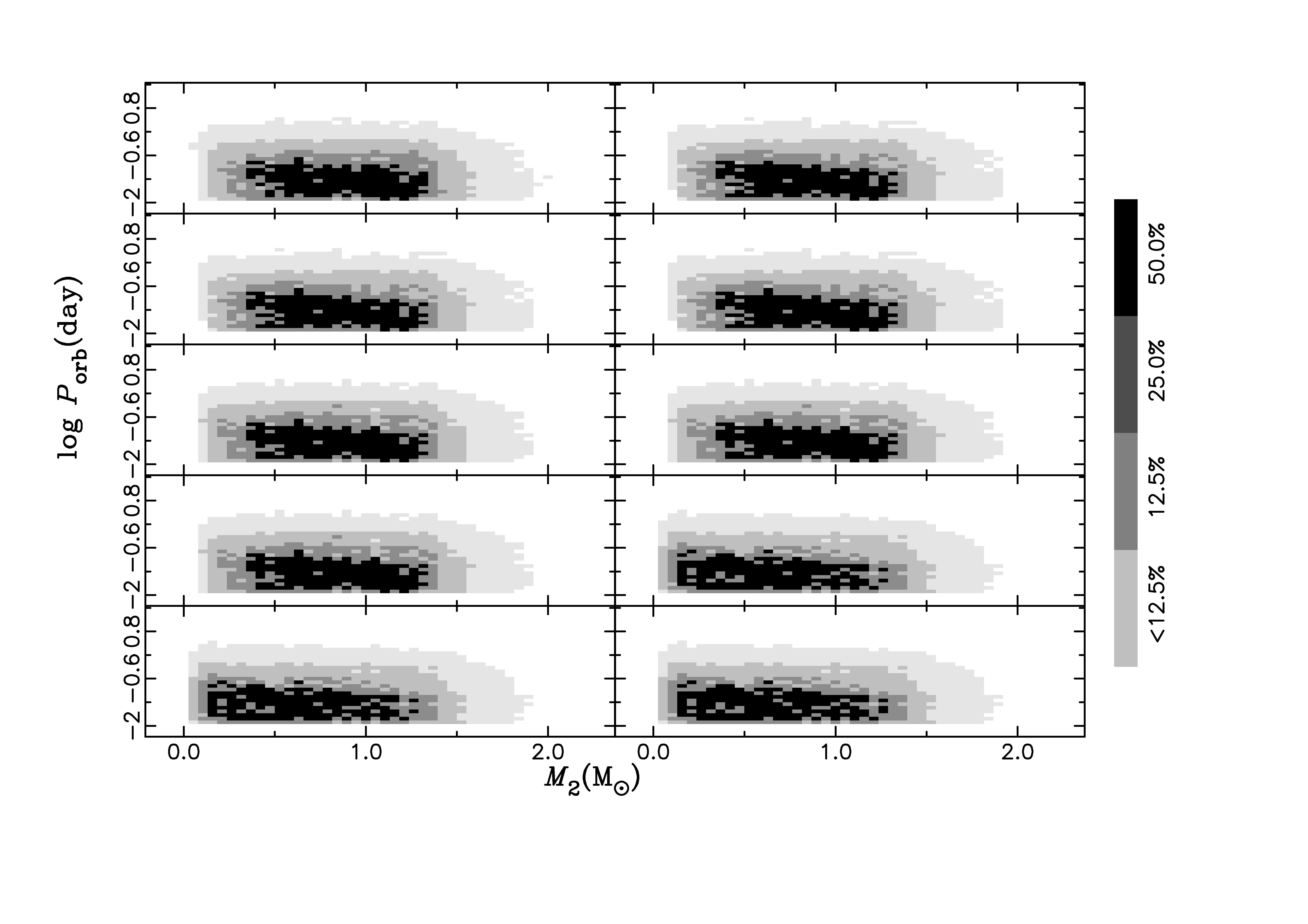}
\includegraphics[totalheight=3.1in,width=4.0in]{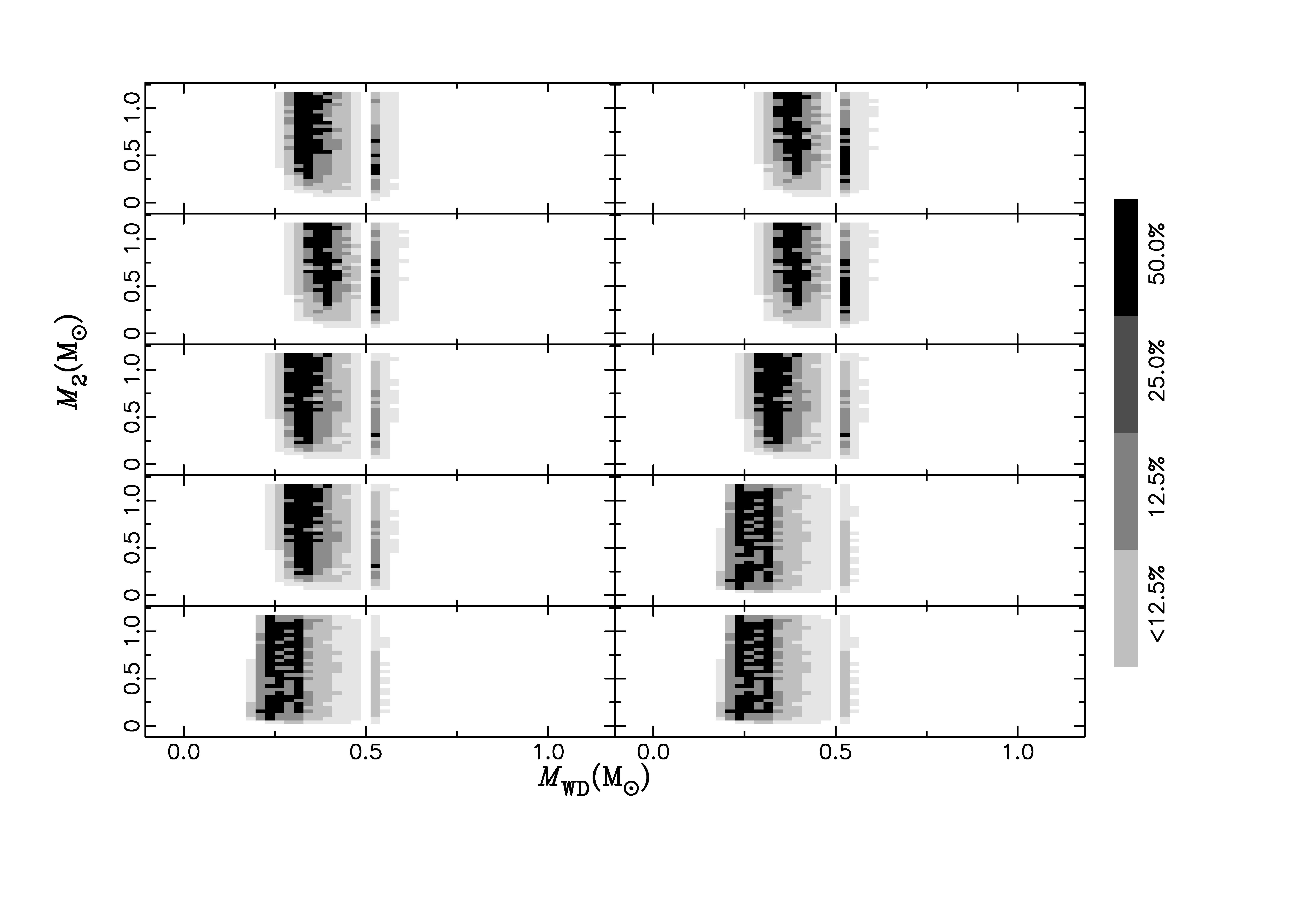}
\caption{The left(1st) panel shows the distribution of the orbital period and the WD masse. The middle(2nd)
panel shows the distribution of the orbital period and secondary mass. The right (3rd) panel
shows the distribution of the secondary mass and the WD mass all at the PCEB phase. In each panel,
10 different models are shown, including model 1 (the first panel), model 4 with $q_{\rm cr1,\,2,\,3}$
(the 2nd--4th panels), model 3 with $q_{\rm cr1,\,2,\,3}$ (the 5th--7th panels) and model 2 with
$q_{\rm cr1,\,2,\,3}$ (the 8th--10th panels), respectively. Here, the panel numbers are coordinated
as follows: 1st (the first row, left), 2nd (the first row, right), 3rd (the second row, left), 4th
(the second row, right), and so forth.}
\label{fig:1}
\end{figure}

\clearpage

\begin{figure}
\centering
\includegraphics[totalheight=2.5in,width=3.0in]{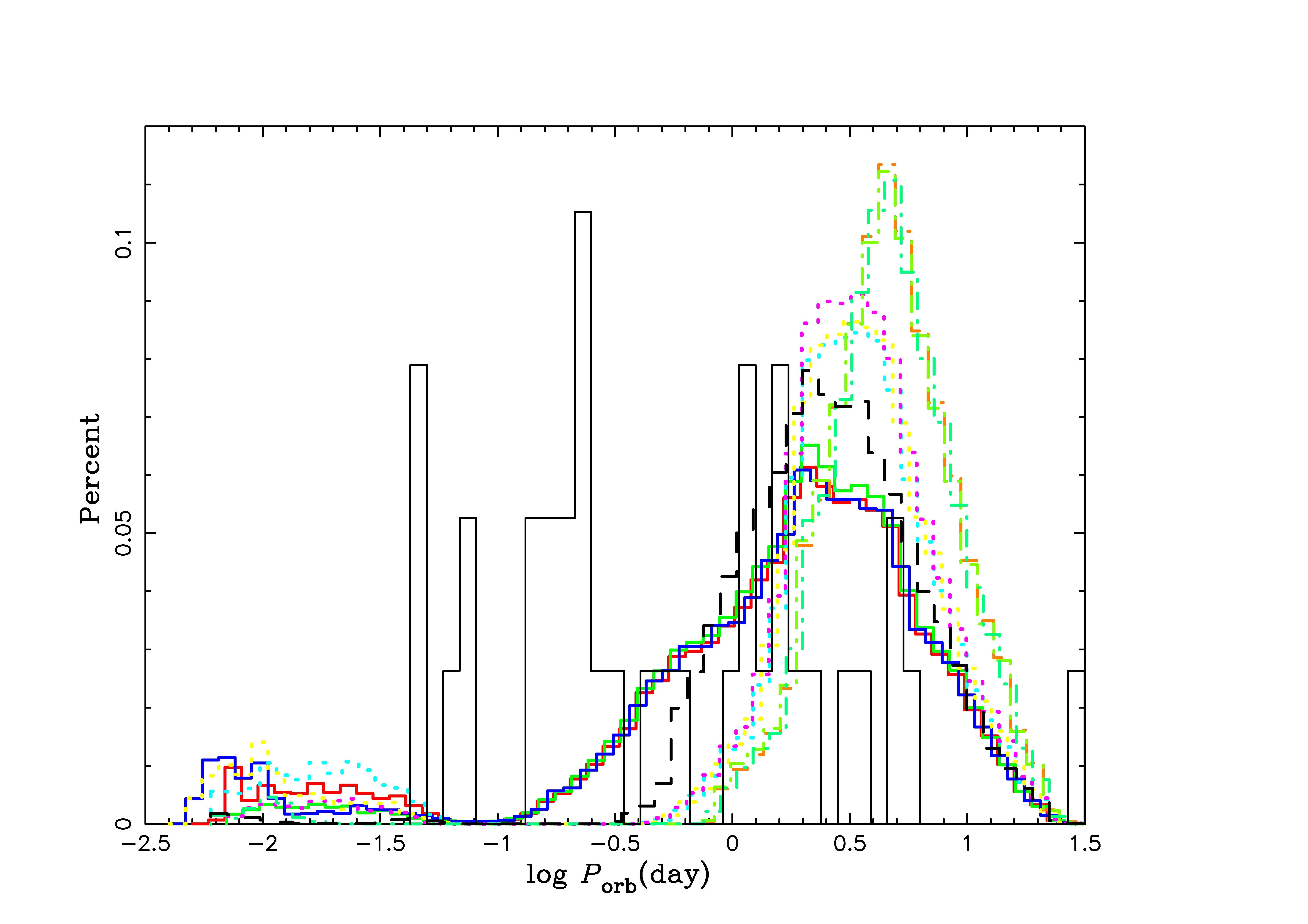}
\includegraphics[totalheight=2.5in,width=3.0in]{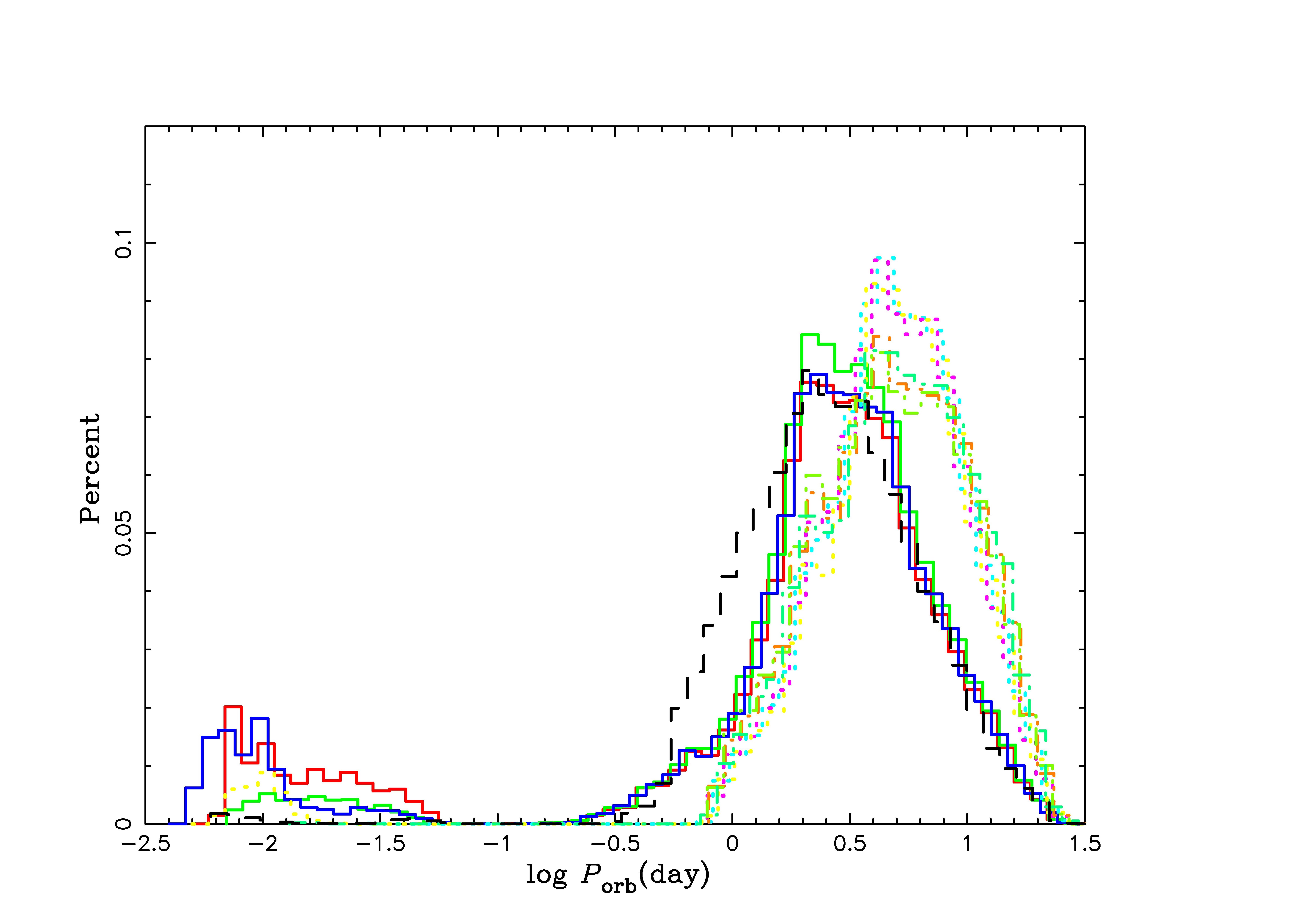}
\includegraphics[totalheight=2.5in,width=3.0in]{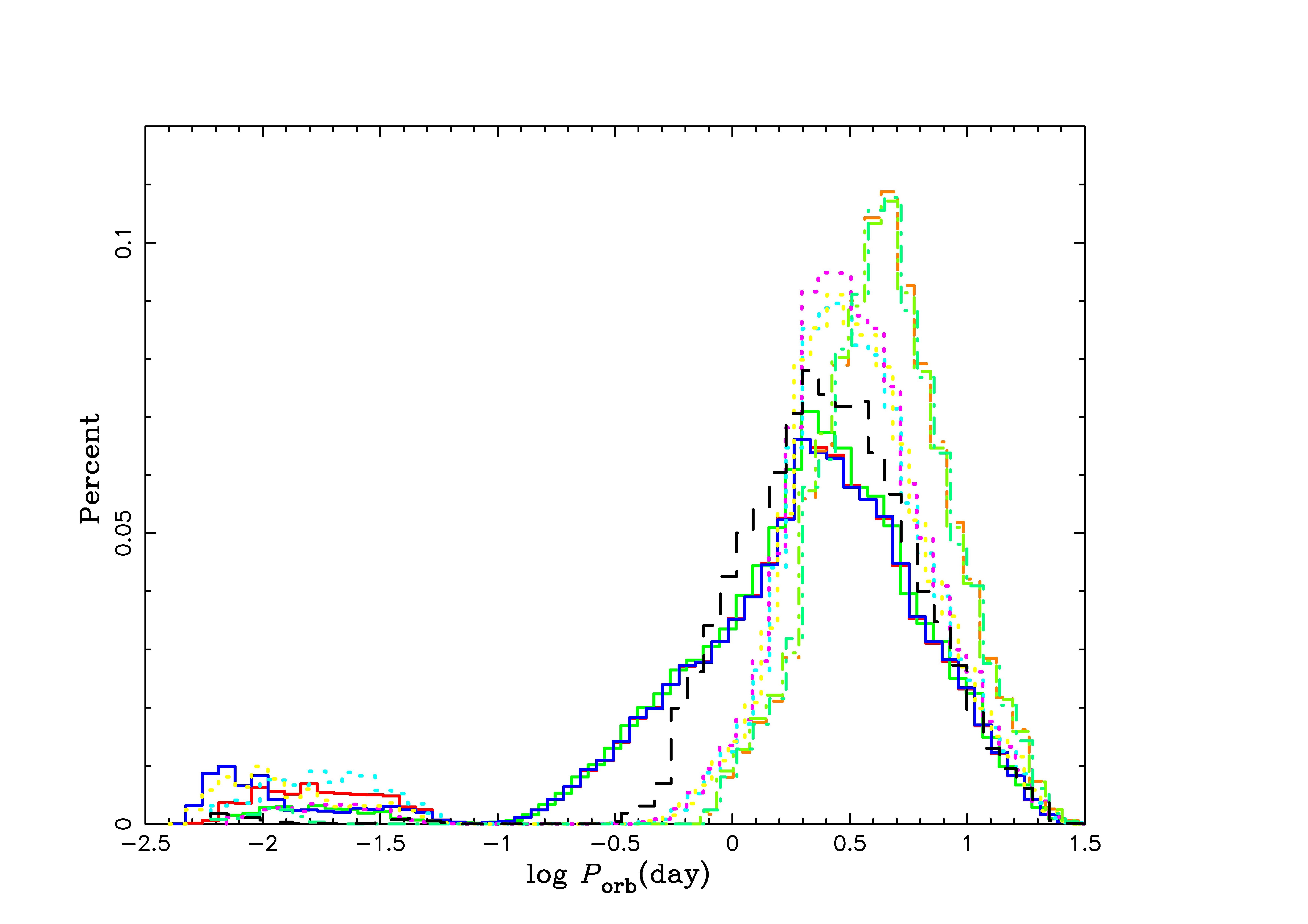}
\includegraphics[totalheight=2.5in,width=3.0in]{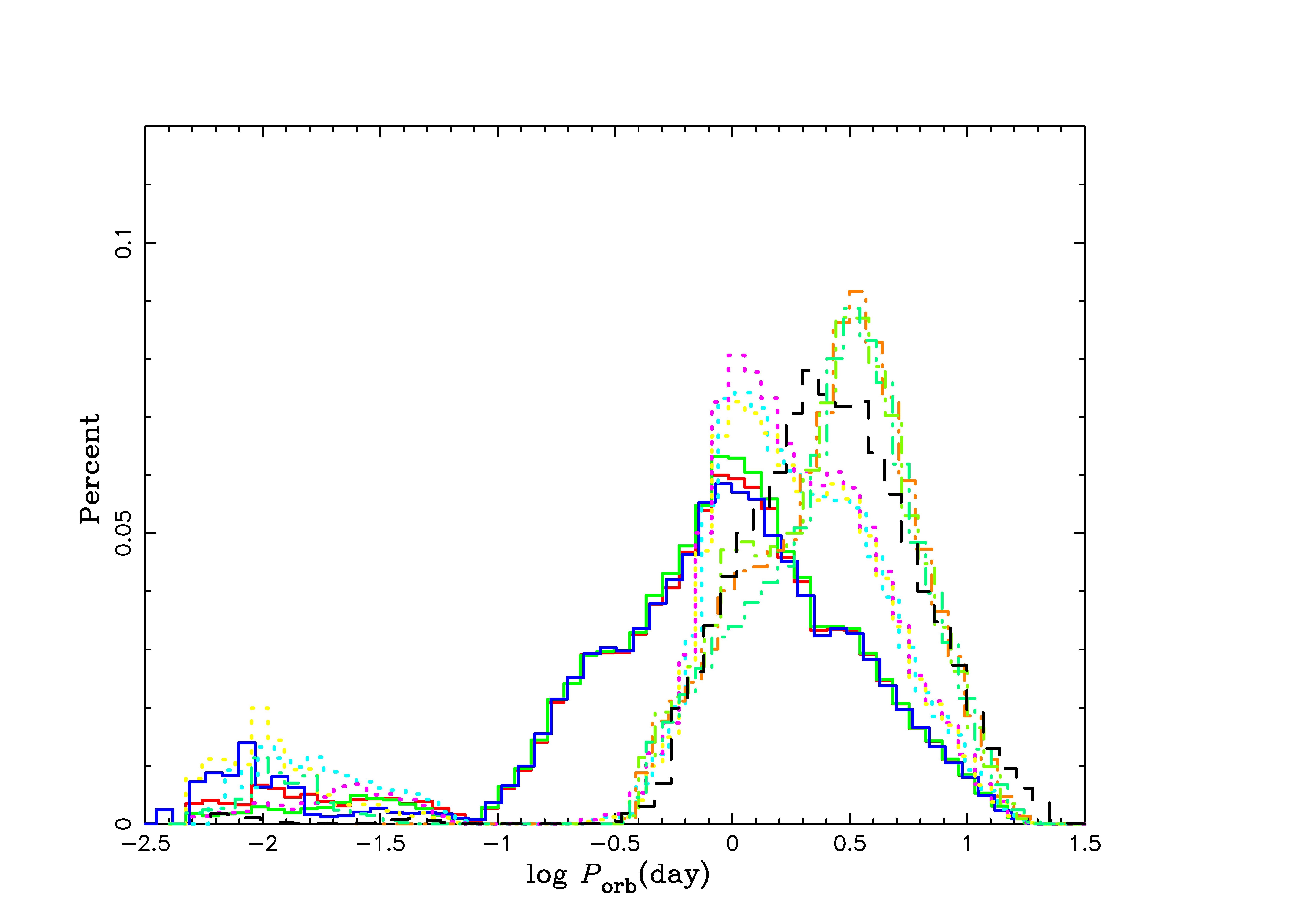}
\caption{Distributoin of the orbital periods of double WD systems. See the caption
of Figure 1 for the model description. he sold black line is for the selected observed
samples (for the details see the text).}\label{fig:1}
\end{figure}

\clearpage

\begin{figure}
\centering
\includegraphics[totalheight=2.5in,width=3.0in]{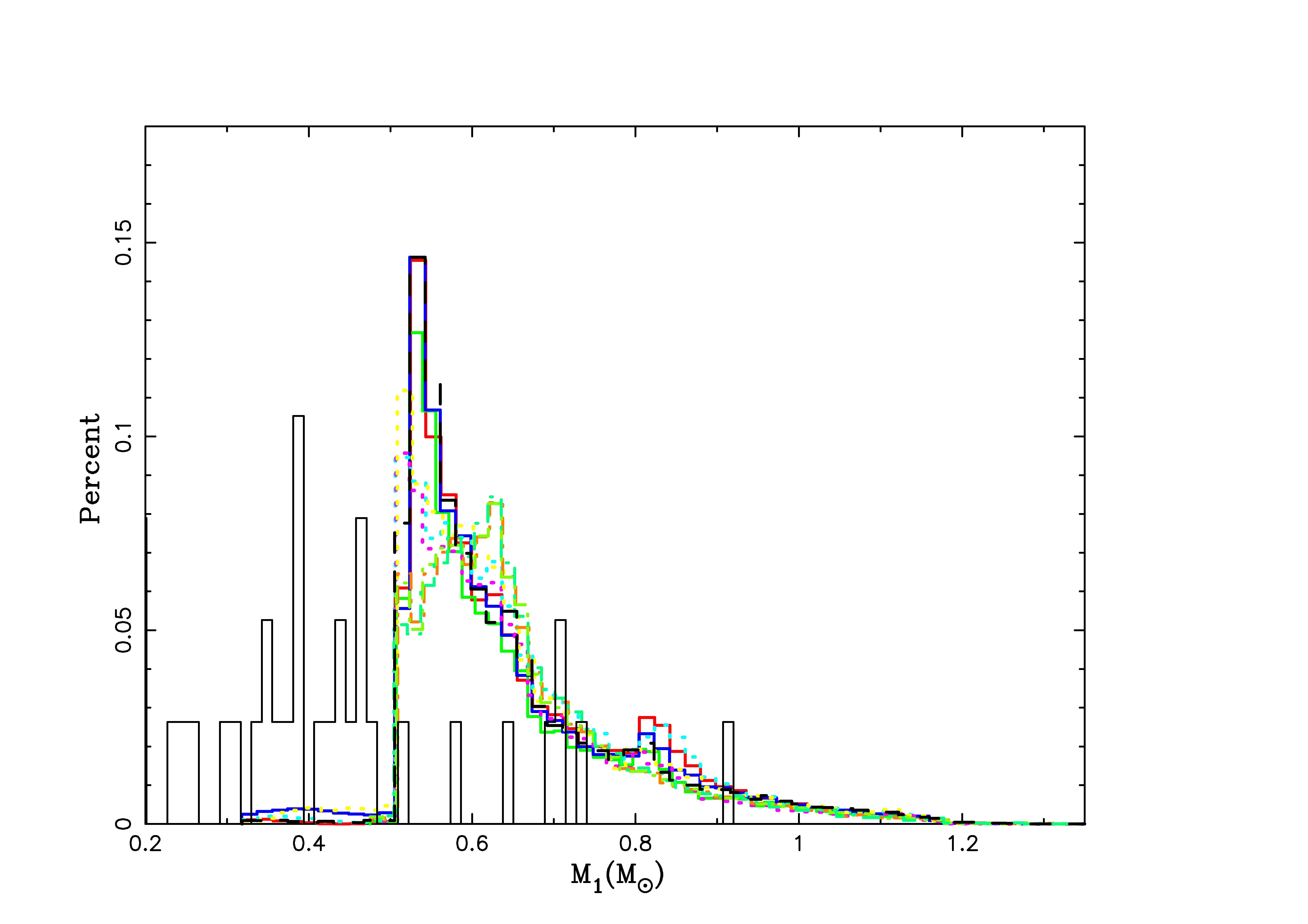}
\includegraphics[totalheight=2.5in,width=3.0in]{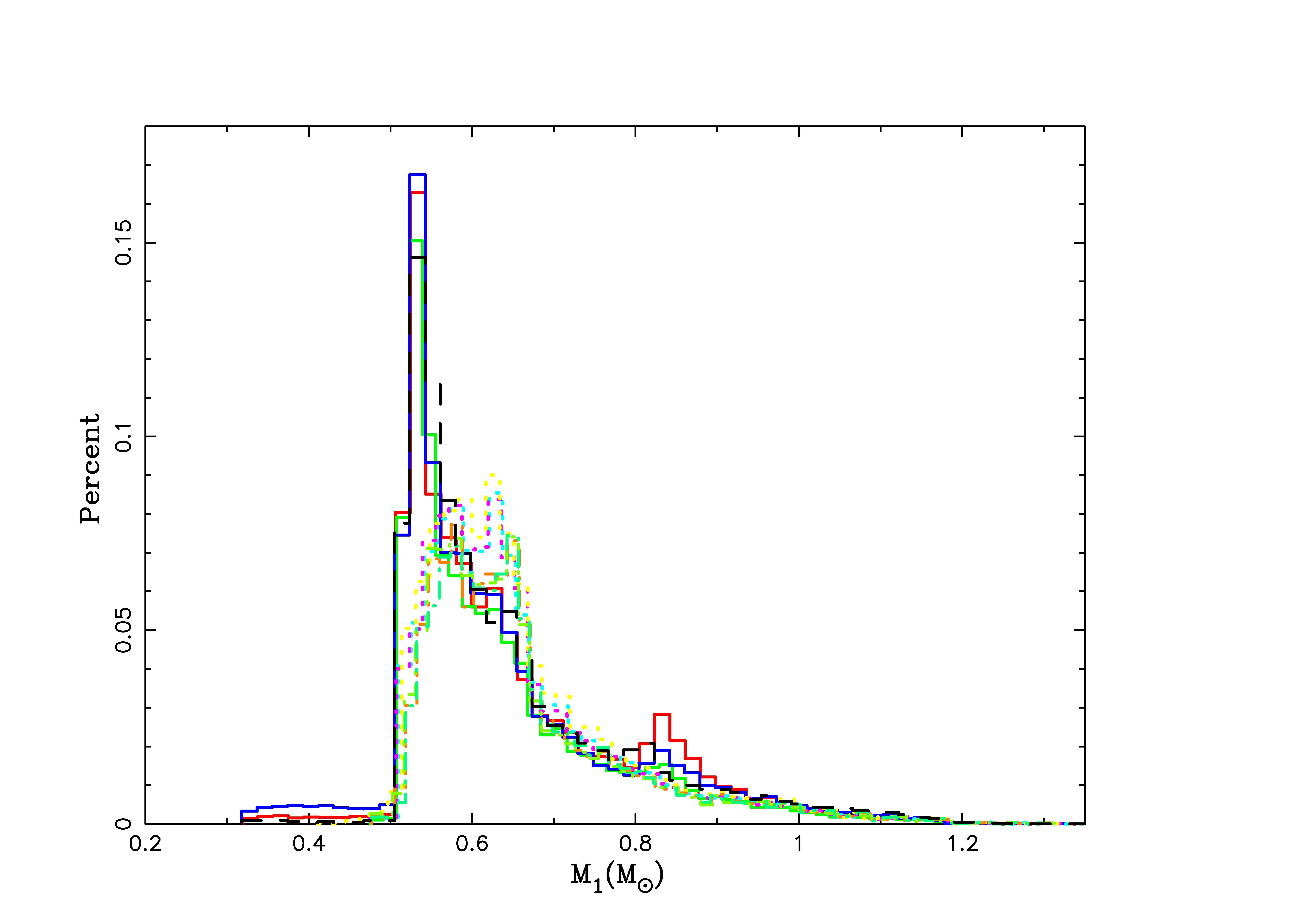}
\includegraphics[totalheight=2.5in,width=3.0in]{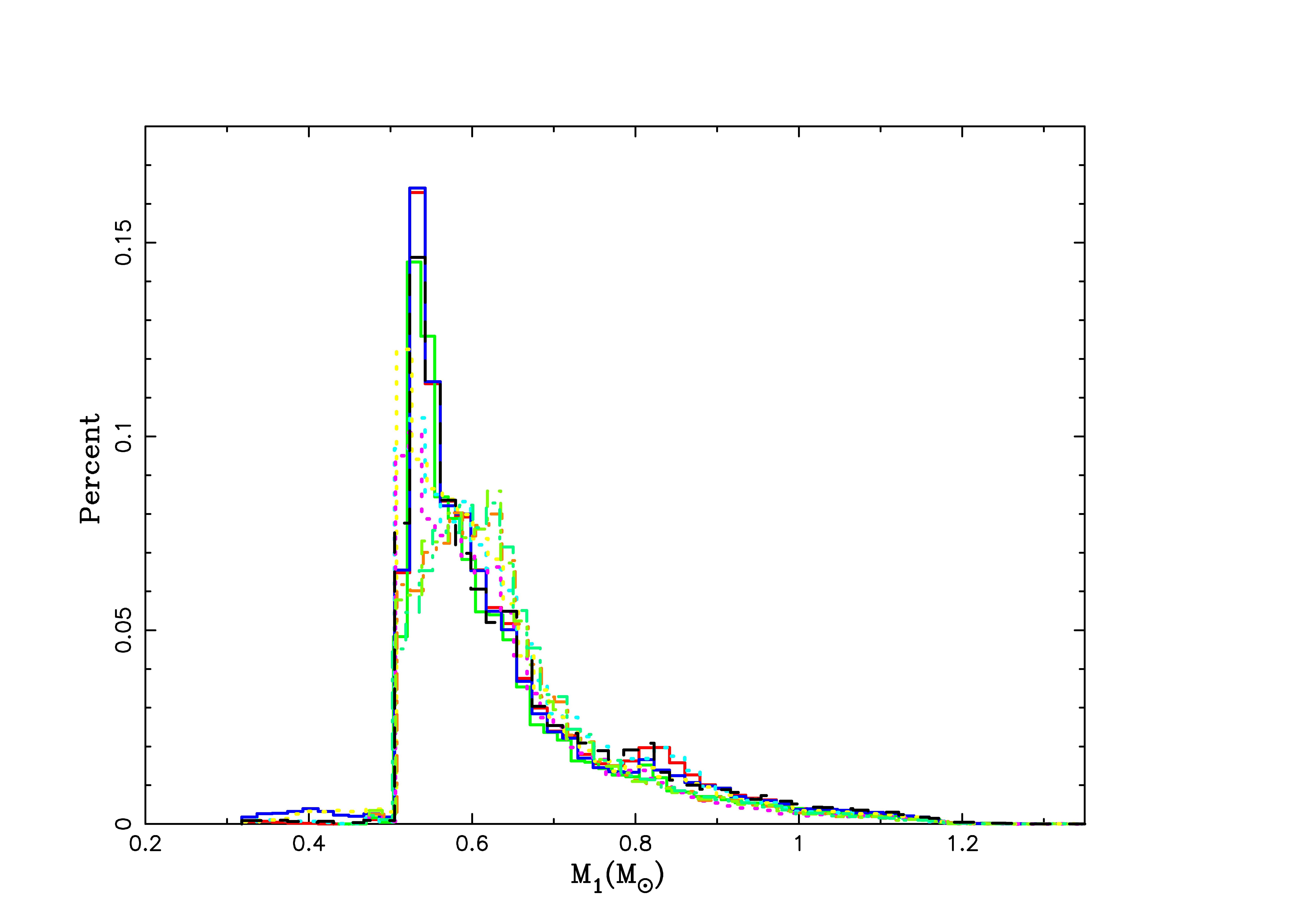}
\includegraphics[totalheight=2.5in,width=3.0in]{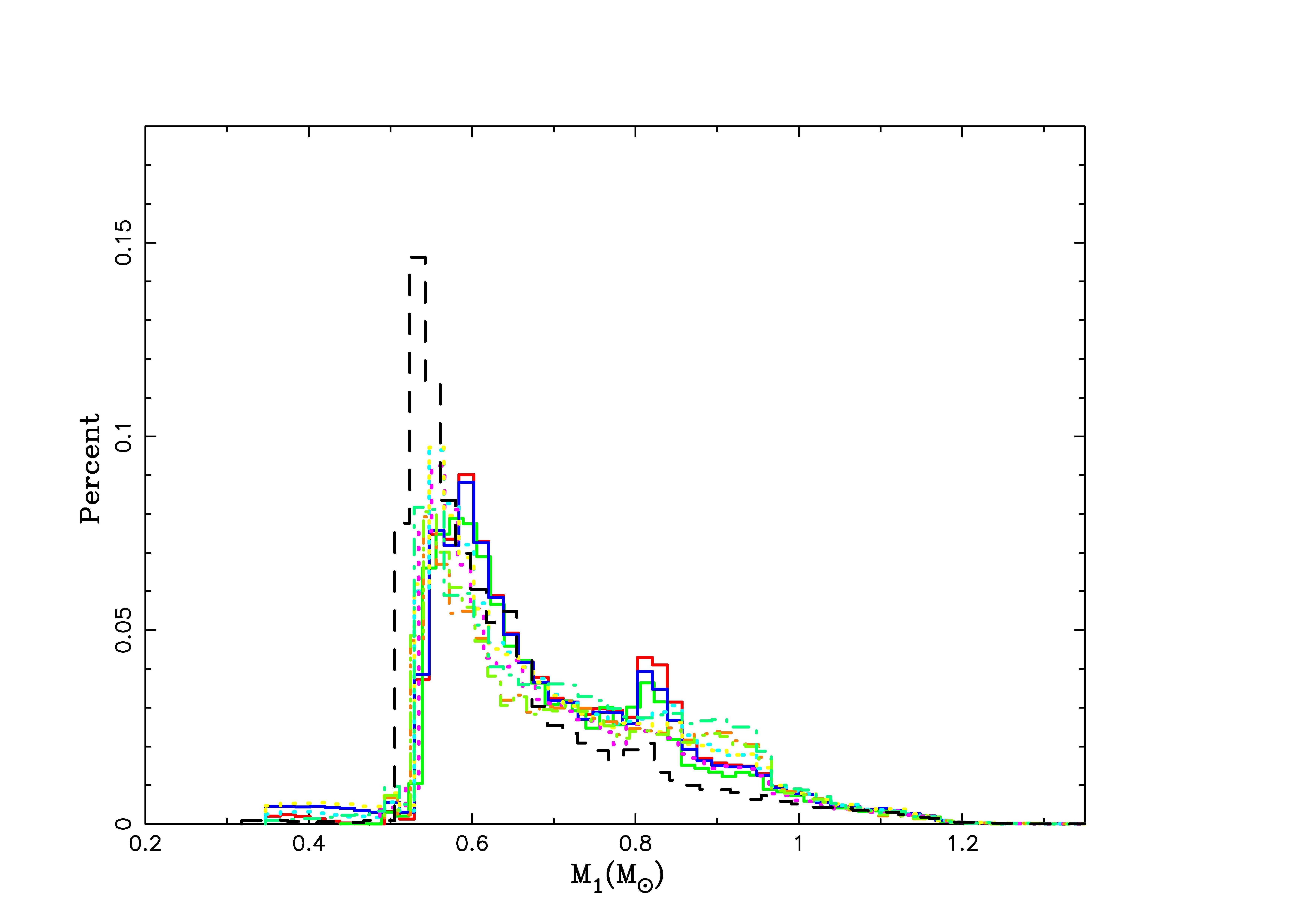}
\caption{Distribution of the primary CO WD masses. See the caption of Figure 1 for the model description.}\label{fig:1}
\end{figure}

\clearpage

\begin{figure}
\centering
\includegraphics[totalheight=2.5in,width=3.0in]{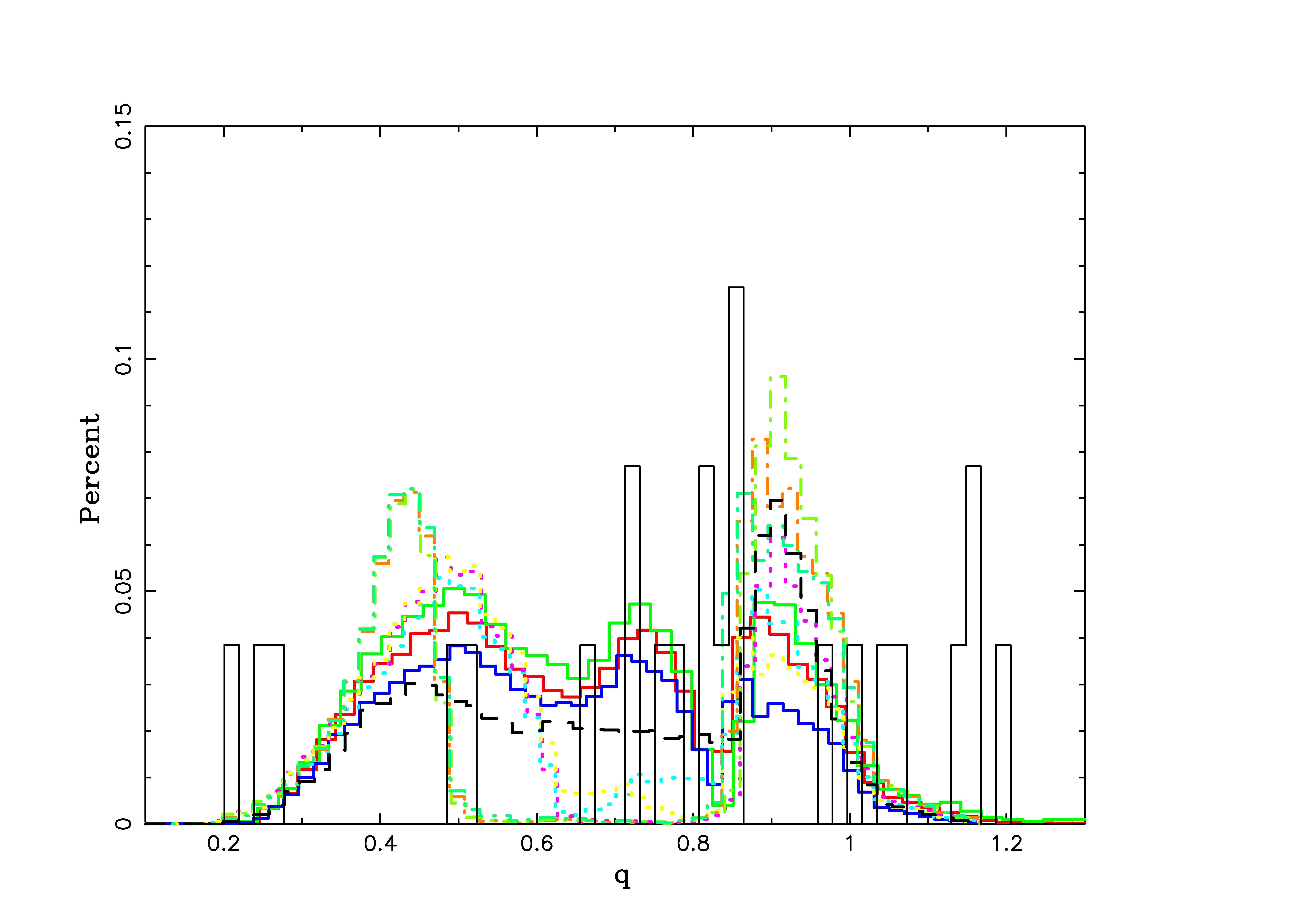}
\includegraphics[totalheight=2.5in,width=3.0in]{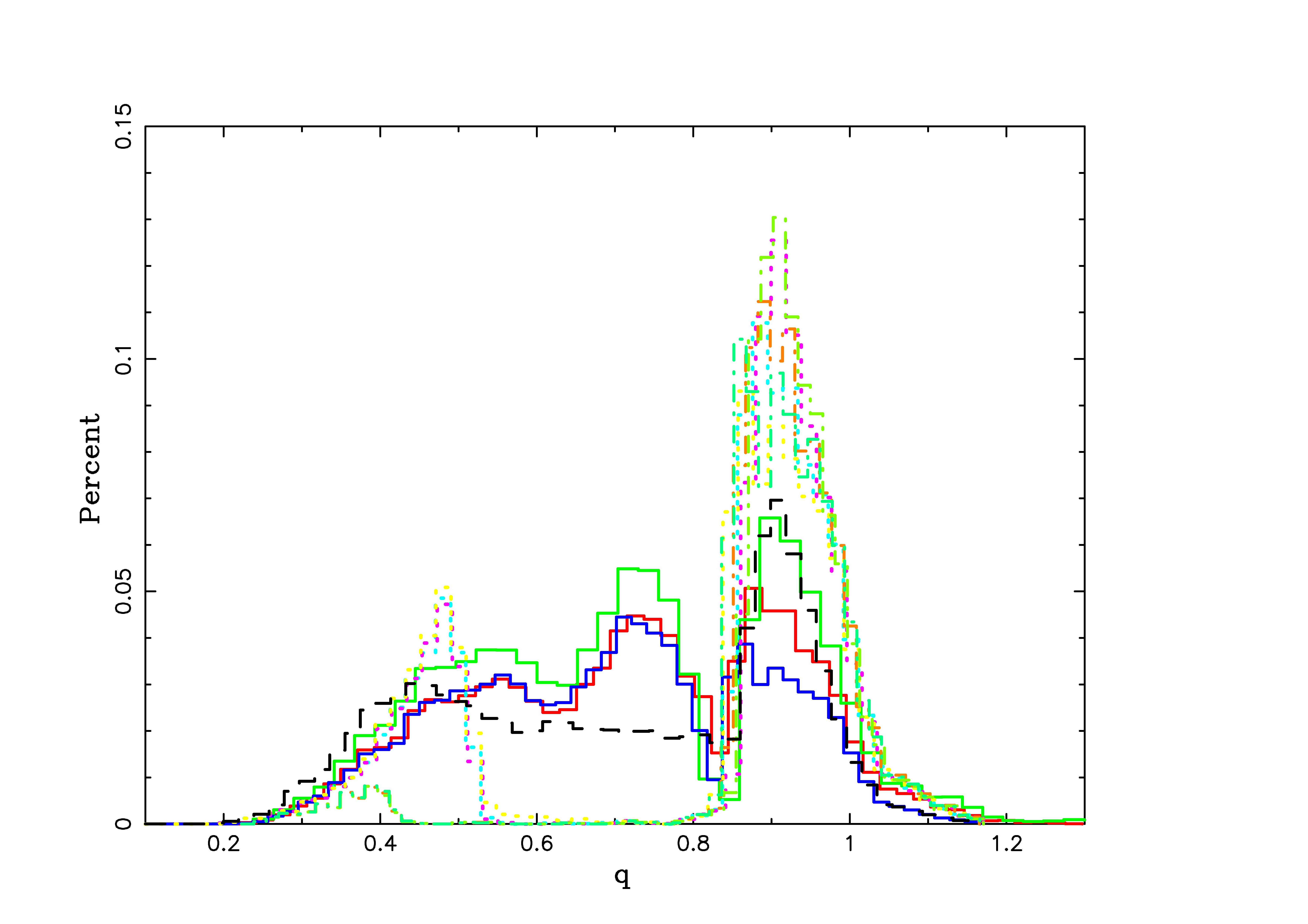}
\includegraphics[totalheight=2.5in,width=3.0in]{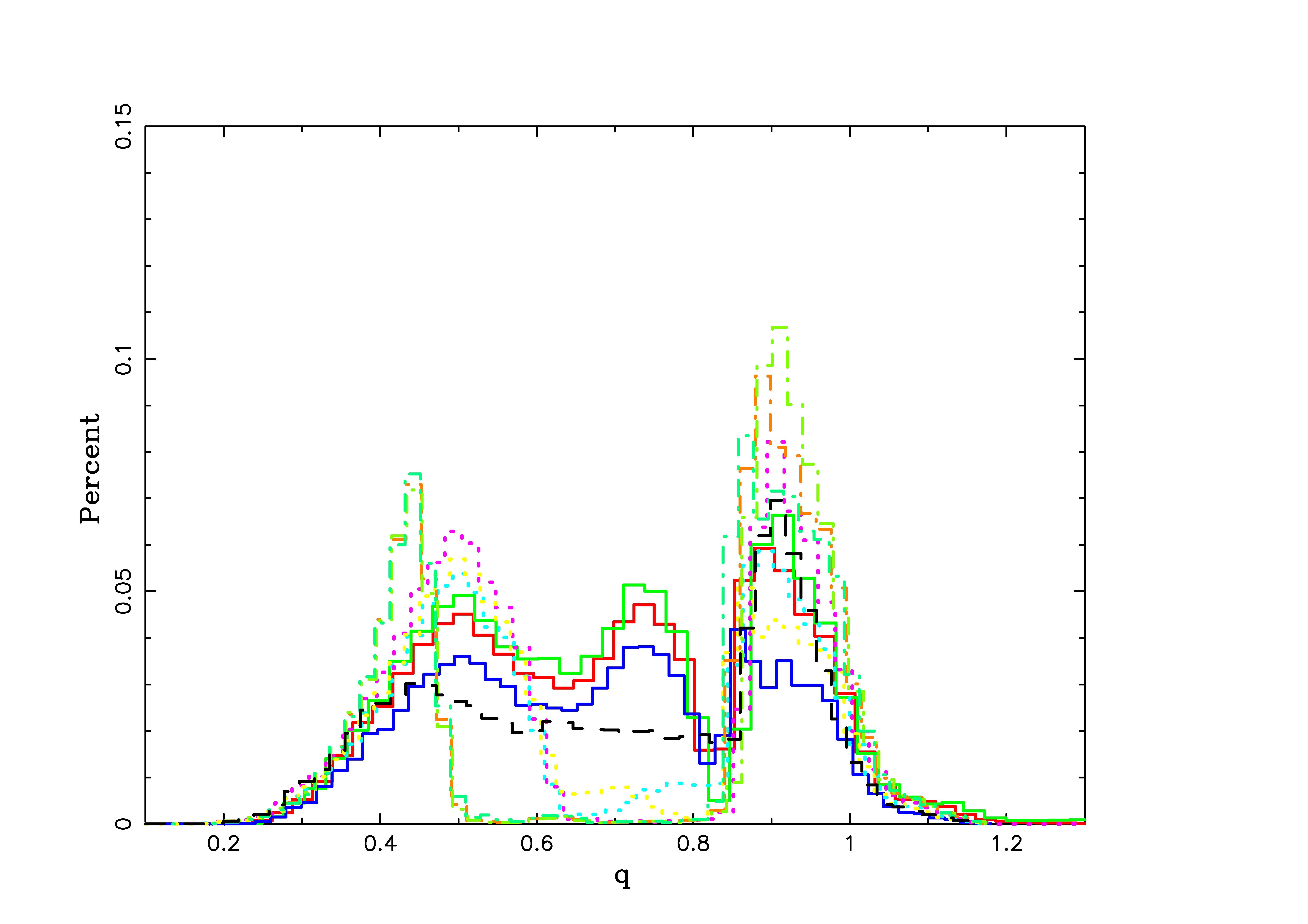}
\includegraphics[totalheight=2.5in,width=3.0in]{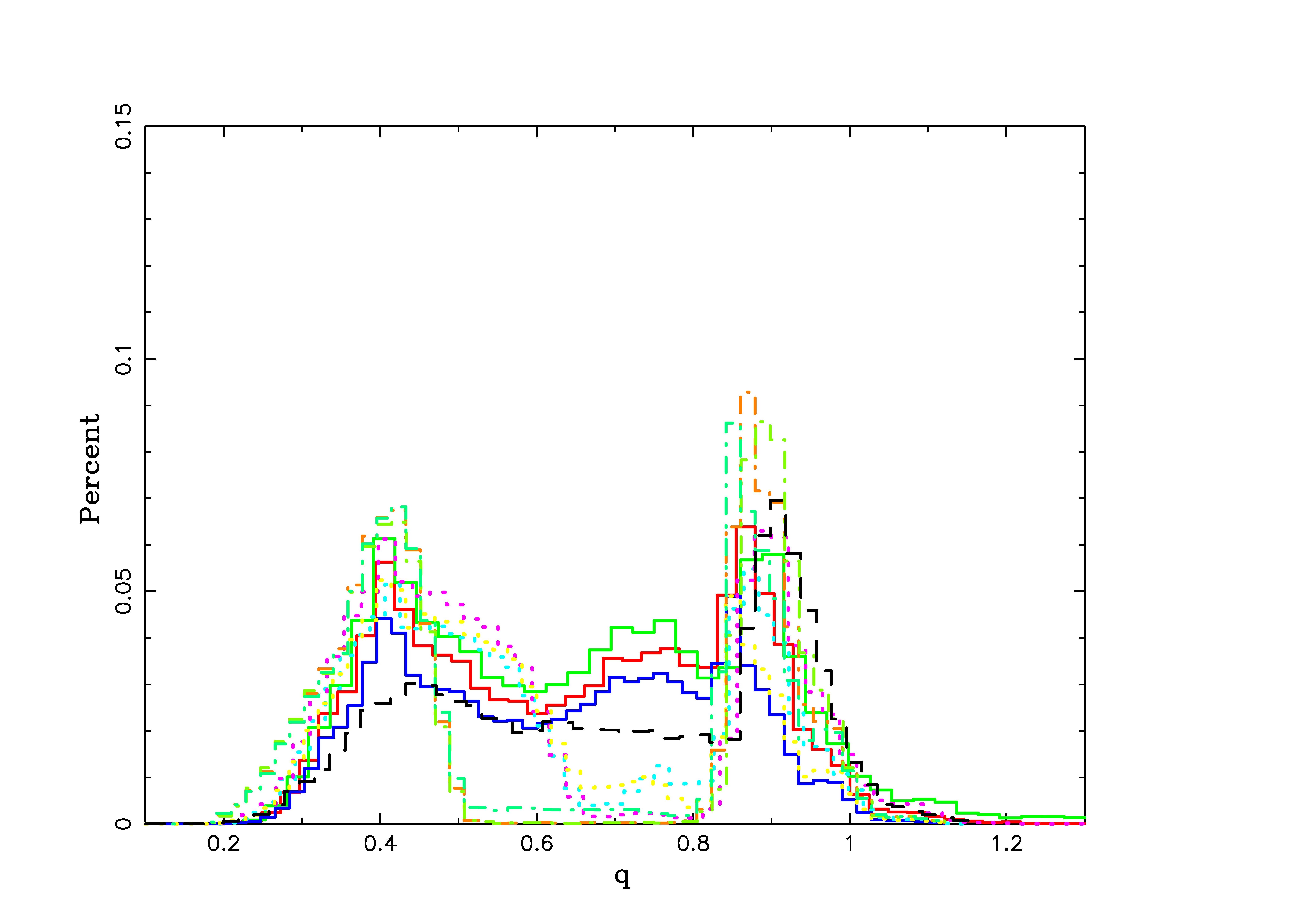}
\caption{Distribution of the mass ratios between secondary WDs and primary CO WDs.
See the caption of Figure 1 for the model description.}\label{fig:1}
\end{figure}

\clearpage

\begin{figure}
\centering
\includegraphics[totalheight=2.1in,width=2.5in]{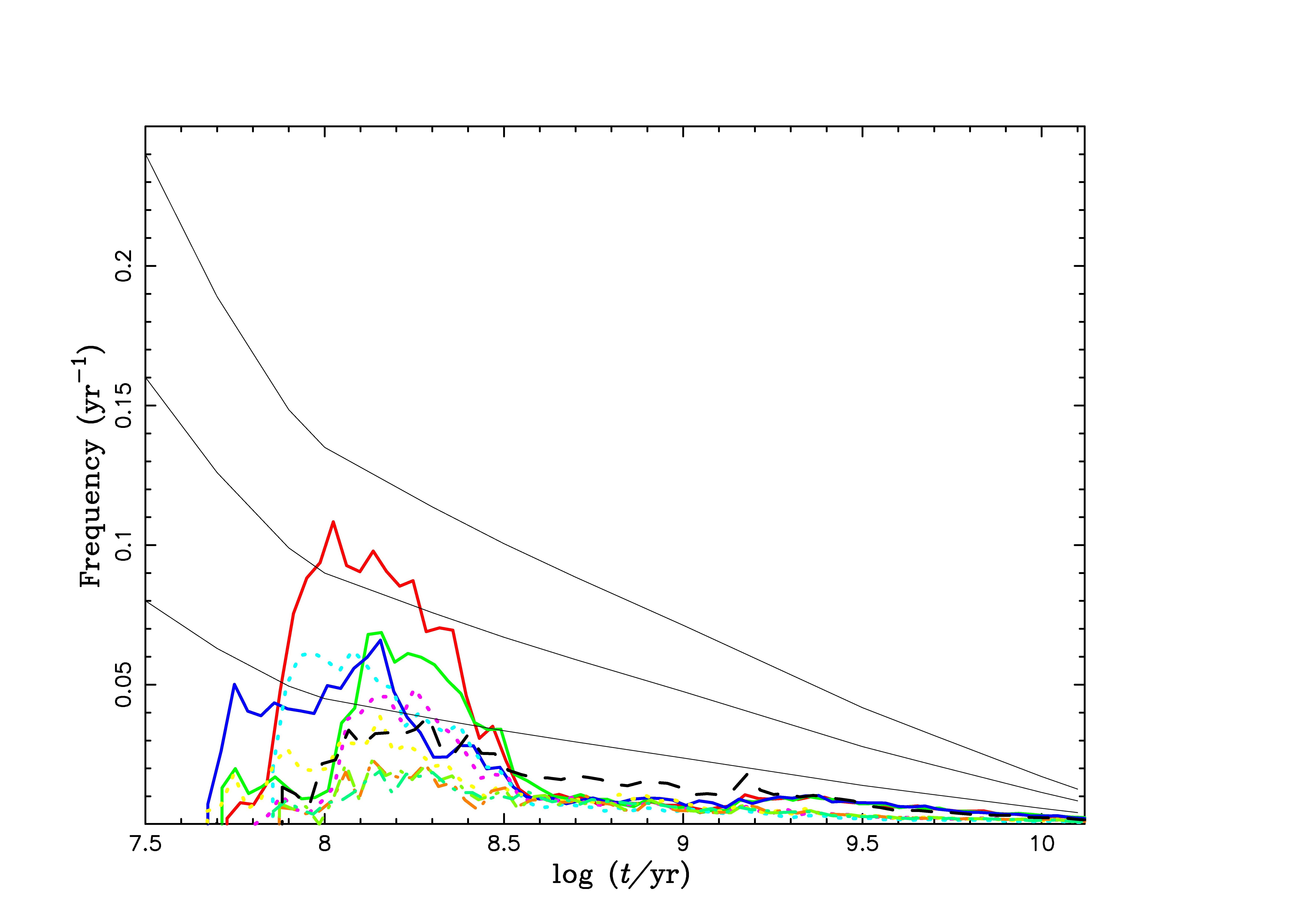}
\includegraphics[totalheight=2.1in,width=2.5in]{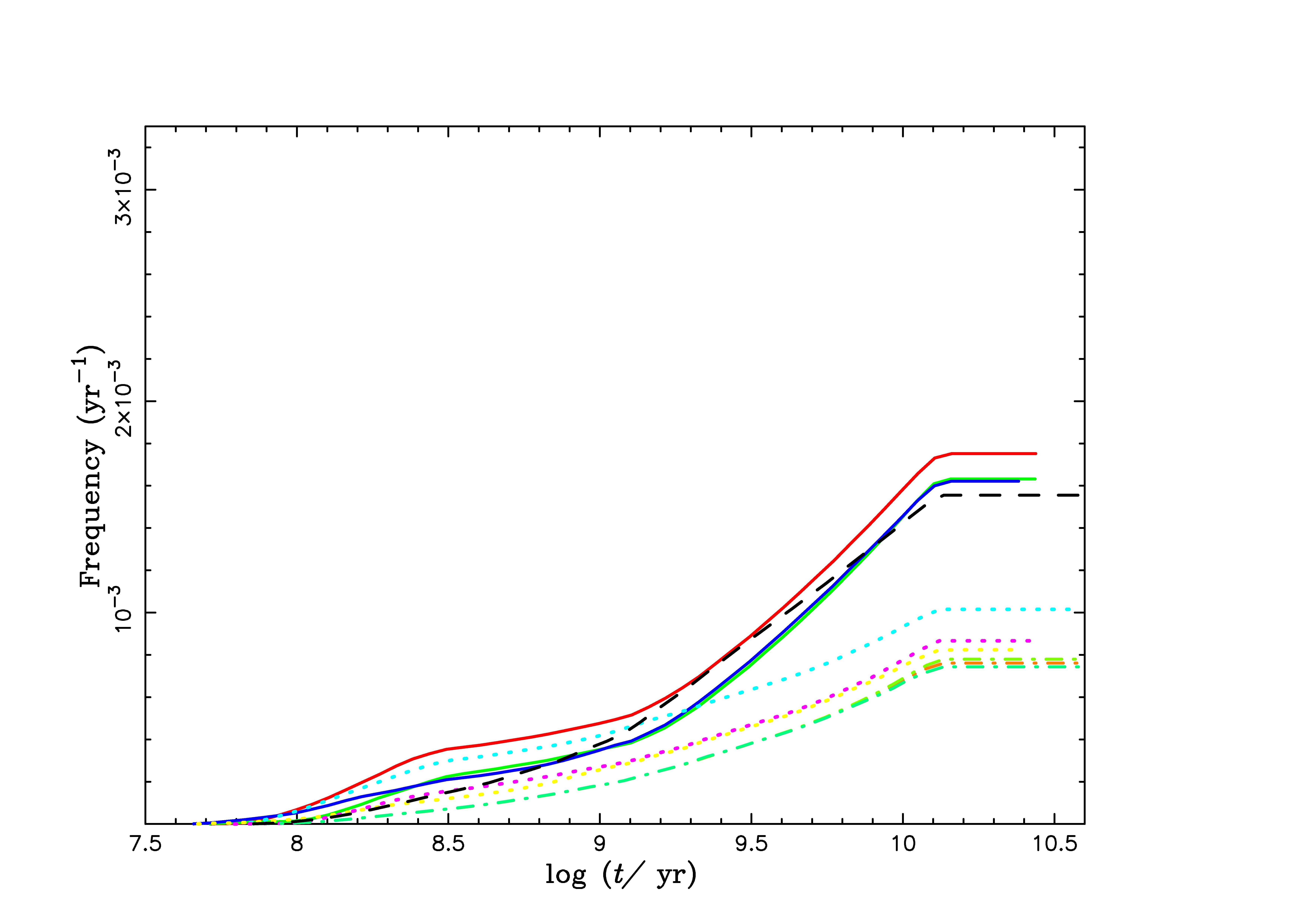}
\includegraphics[totalheight=2.1in,width=2.5in]{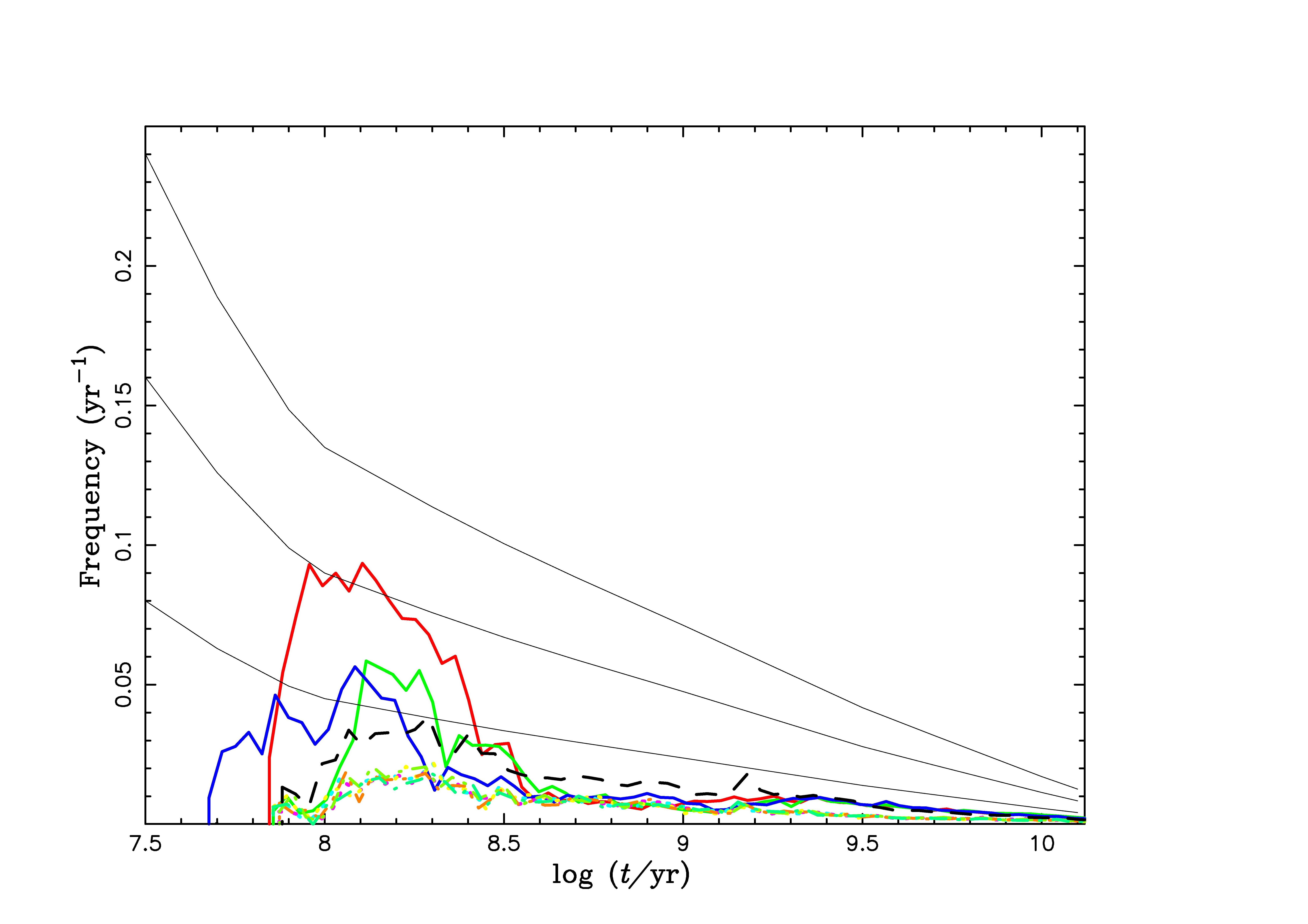}
\includegraphics[totalheight=2.1in,width=2.5in]{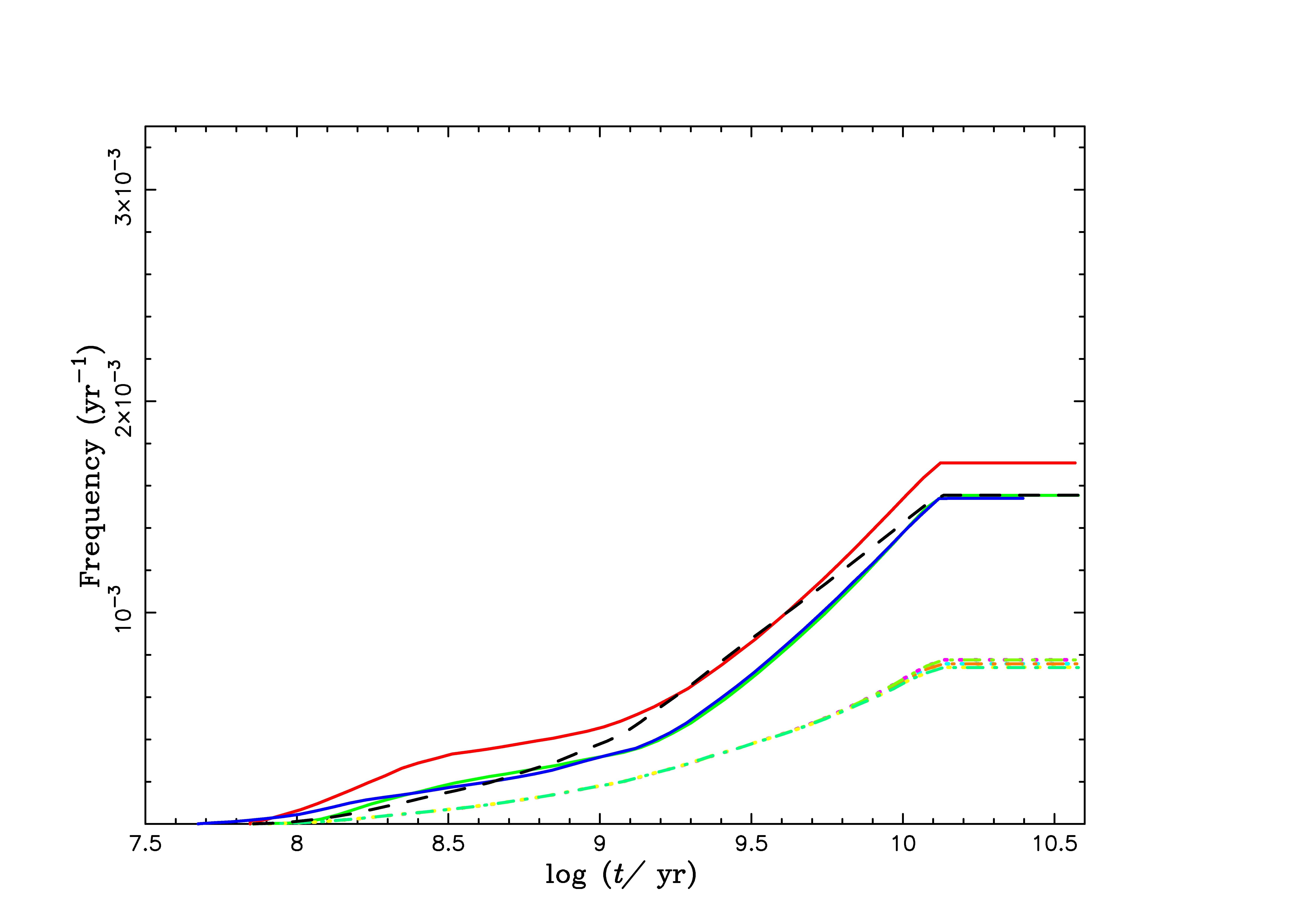}
\includegraphics[totalheight=2.1in,width=2.5in]{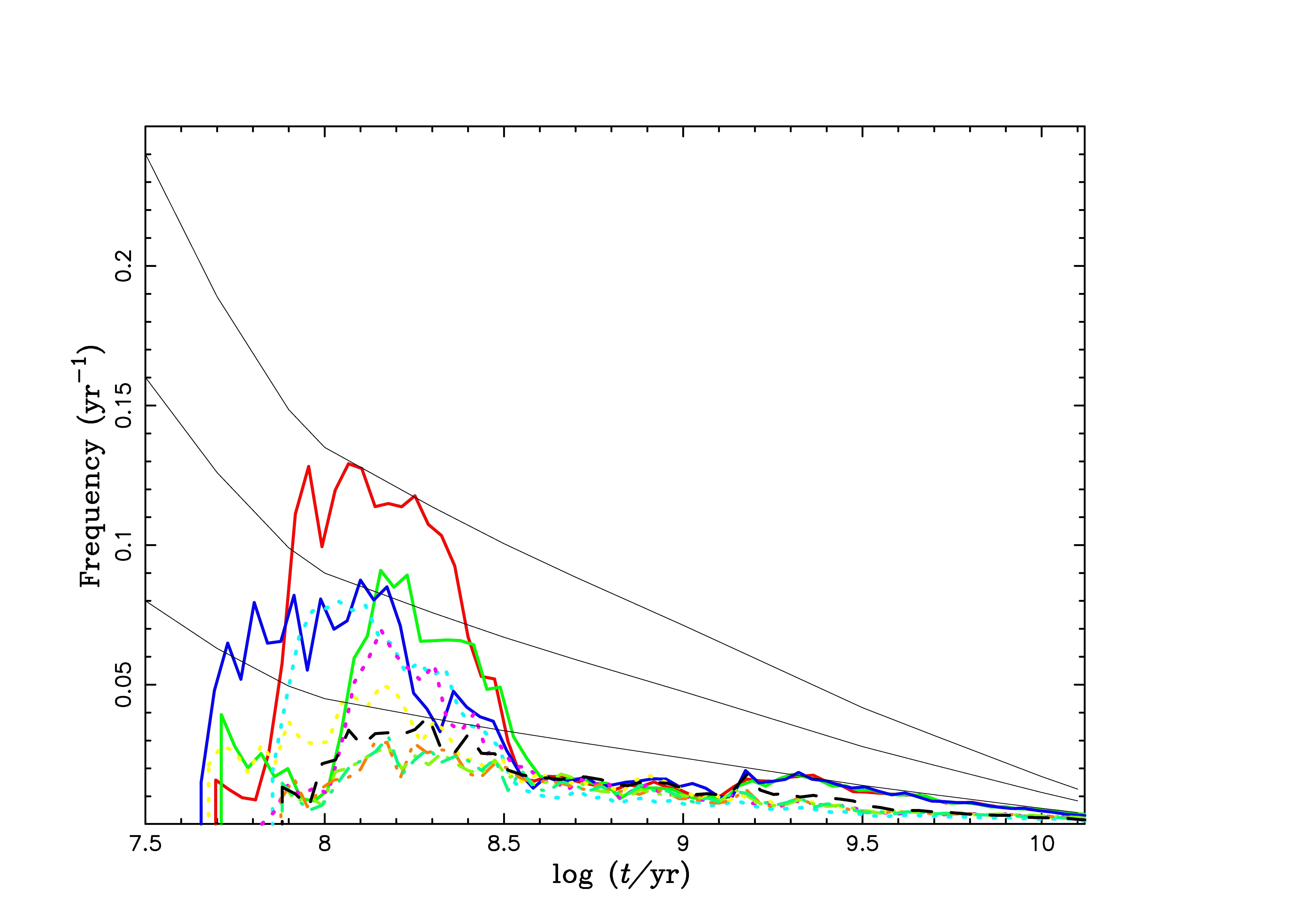}
\includegraphics[totalheight=2.1in,width=2.5in]{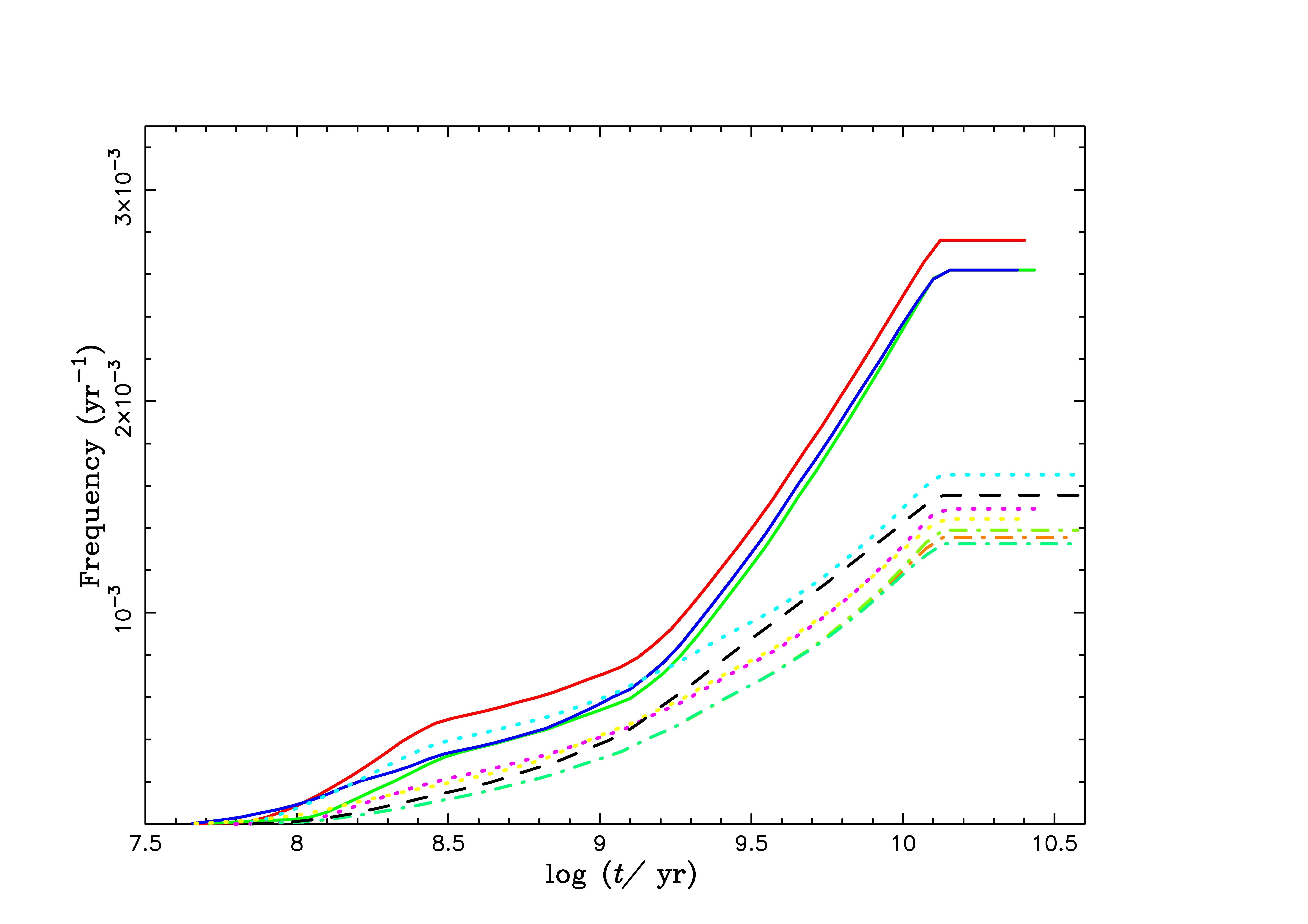}
\includegraphics[totalheight=2.1in,width=2.5in]{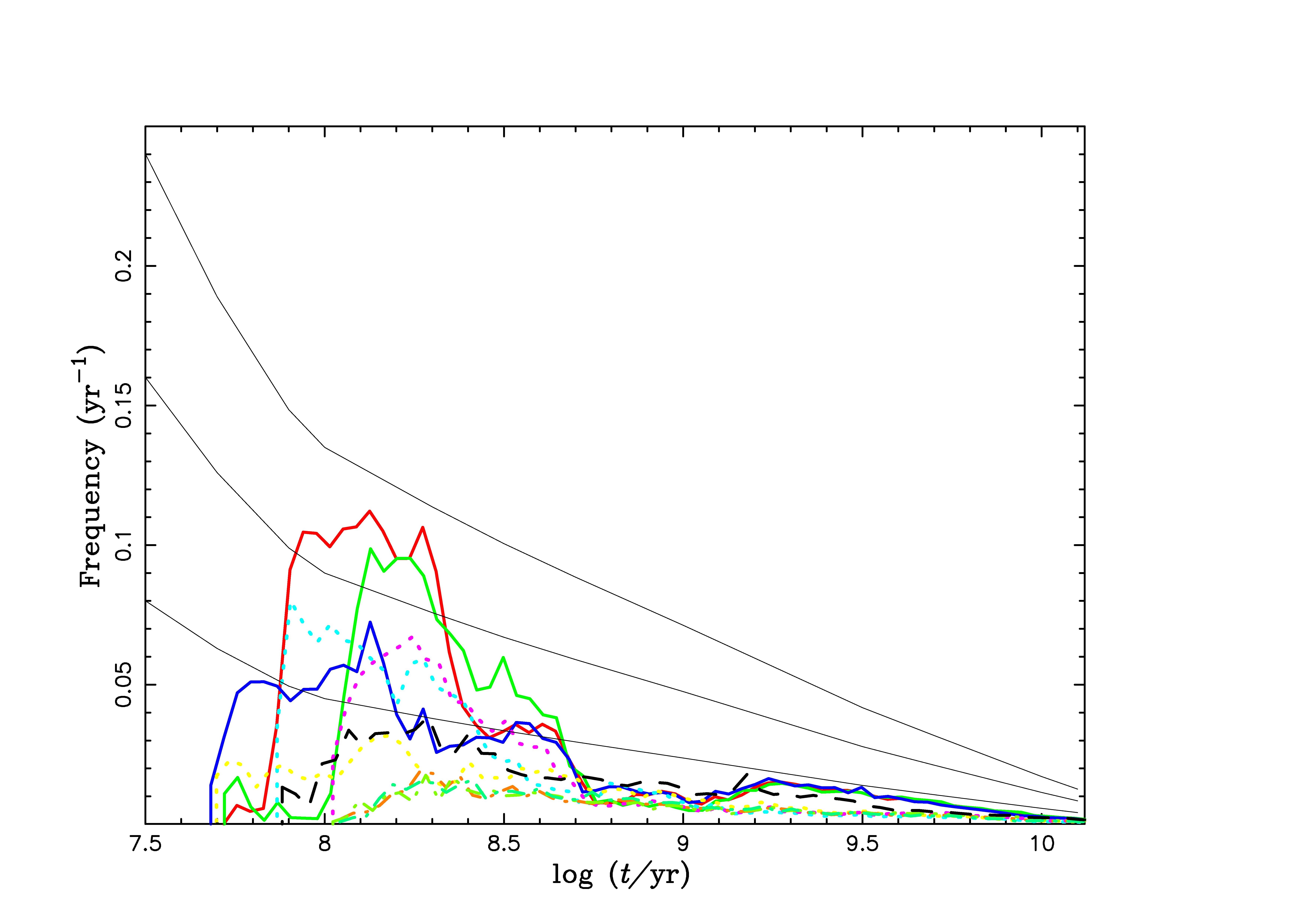}
\includegraphics[totalheight=2.1in,width=2.5in]{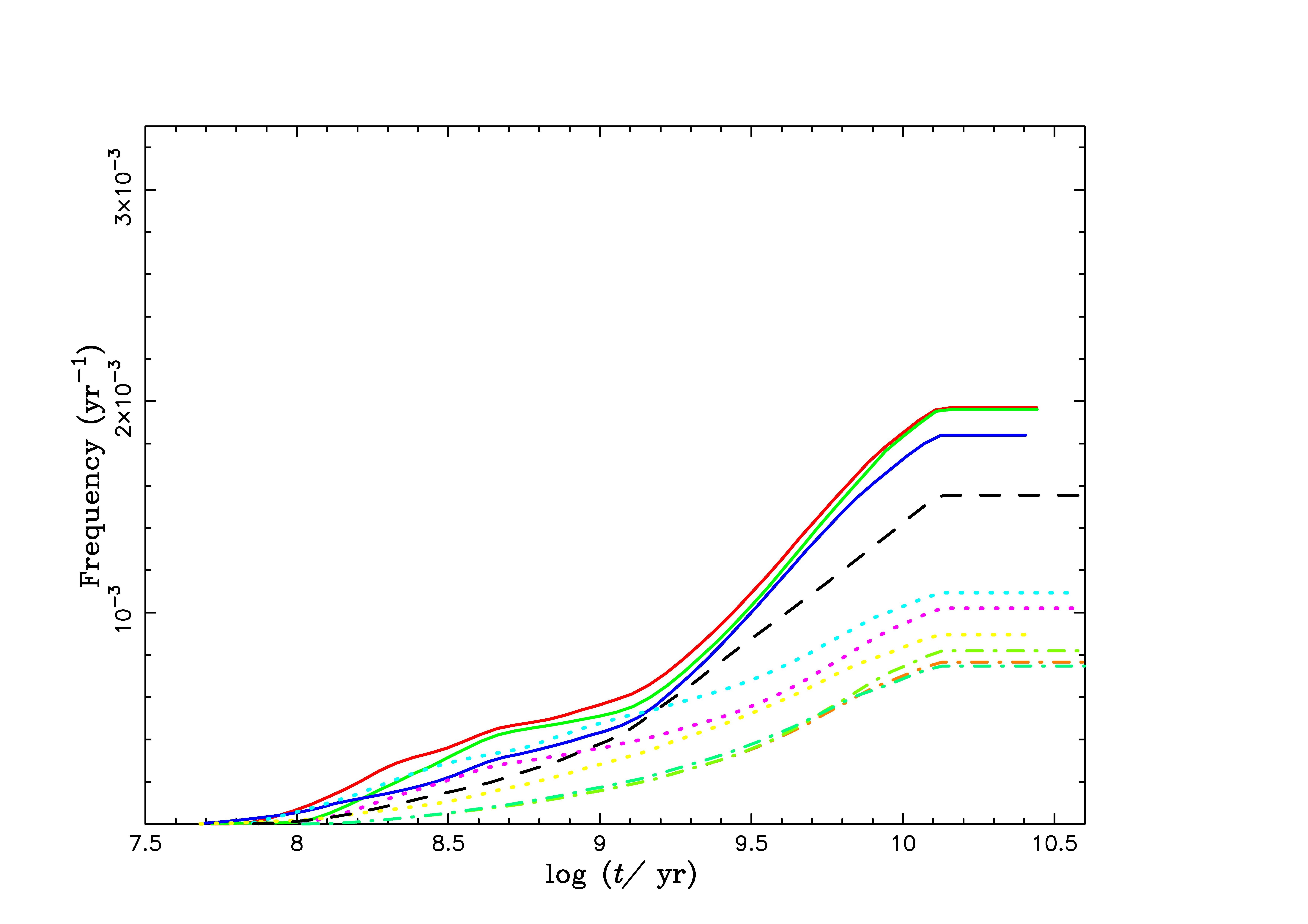}
\caption{The SN Ia birth rates calculated for various models, for a single starburst
of $10^{11} M_\sun$(left panels) and for a constant star formation rate of
$5\,{M_\sun}{\rm{yr}^{-1}}$ over the past 13.7 Gyr(right panels). The letter represents
the Milky Way.
For the panels(models) \& lines(same as Fig. 1), and for the three sold black lines of observational results, see the text.}\label{fig:1}
\end{figure}

\clearpage

\begin{figure}
\centering
\includegraphics[totalheight=2.1in,width=2.5in]{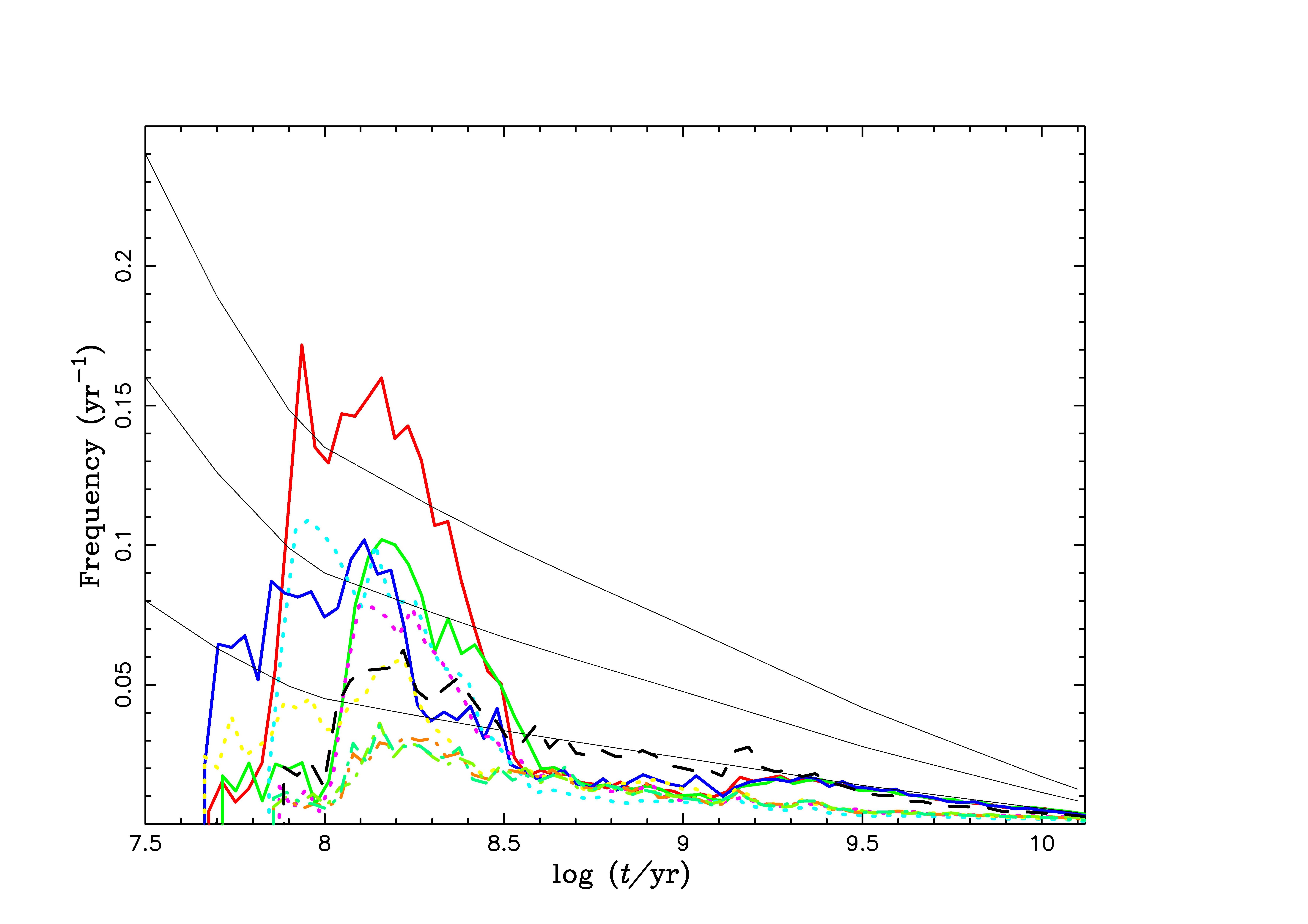}
\includegraphics[totalheight=2.1in,width=2.5in]{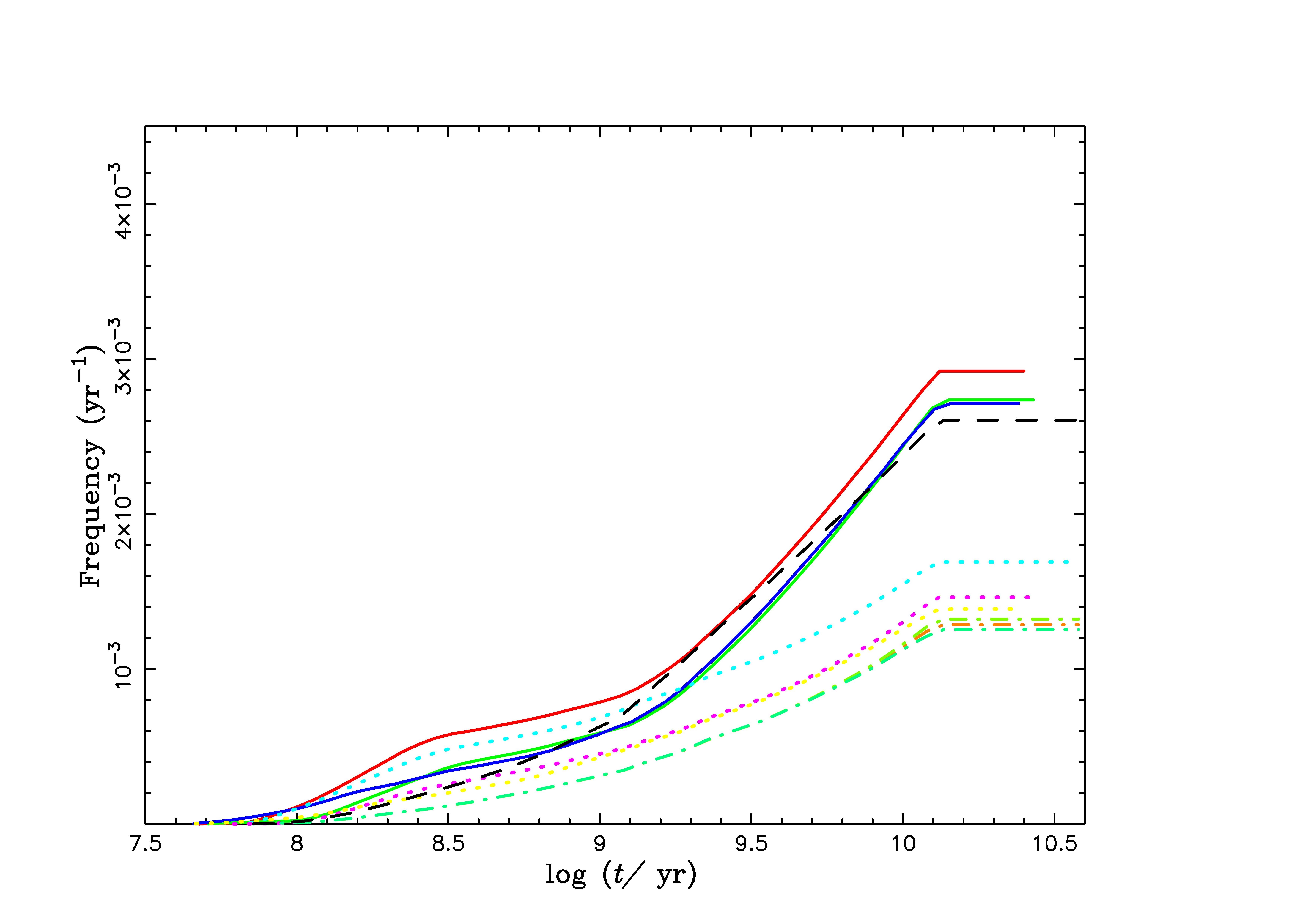}
\includegraphics[totalheight=2.1in,width=2.5in]{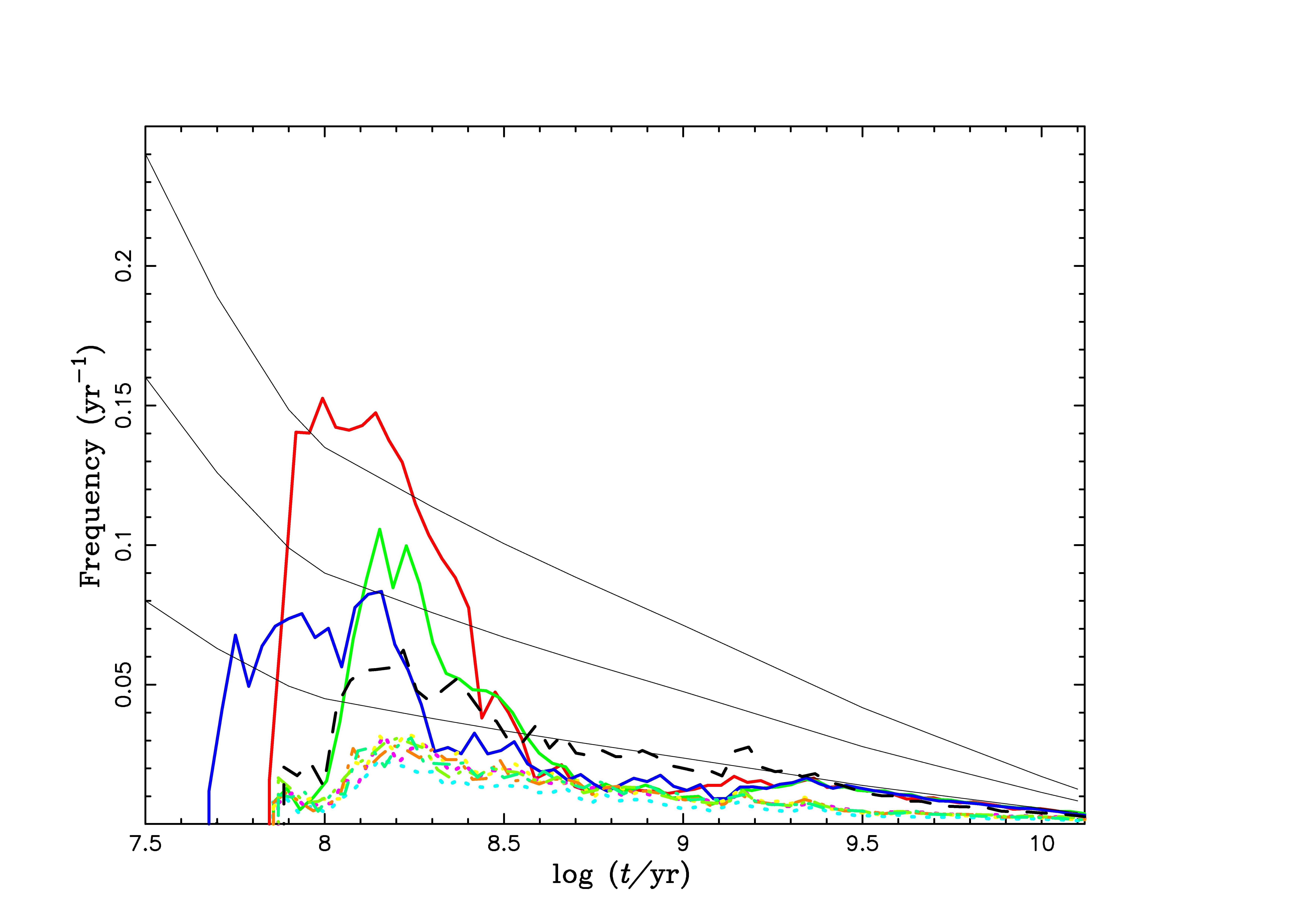}
\includegraphics[totalheight=2.1in,width=2.5in]{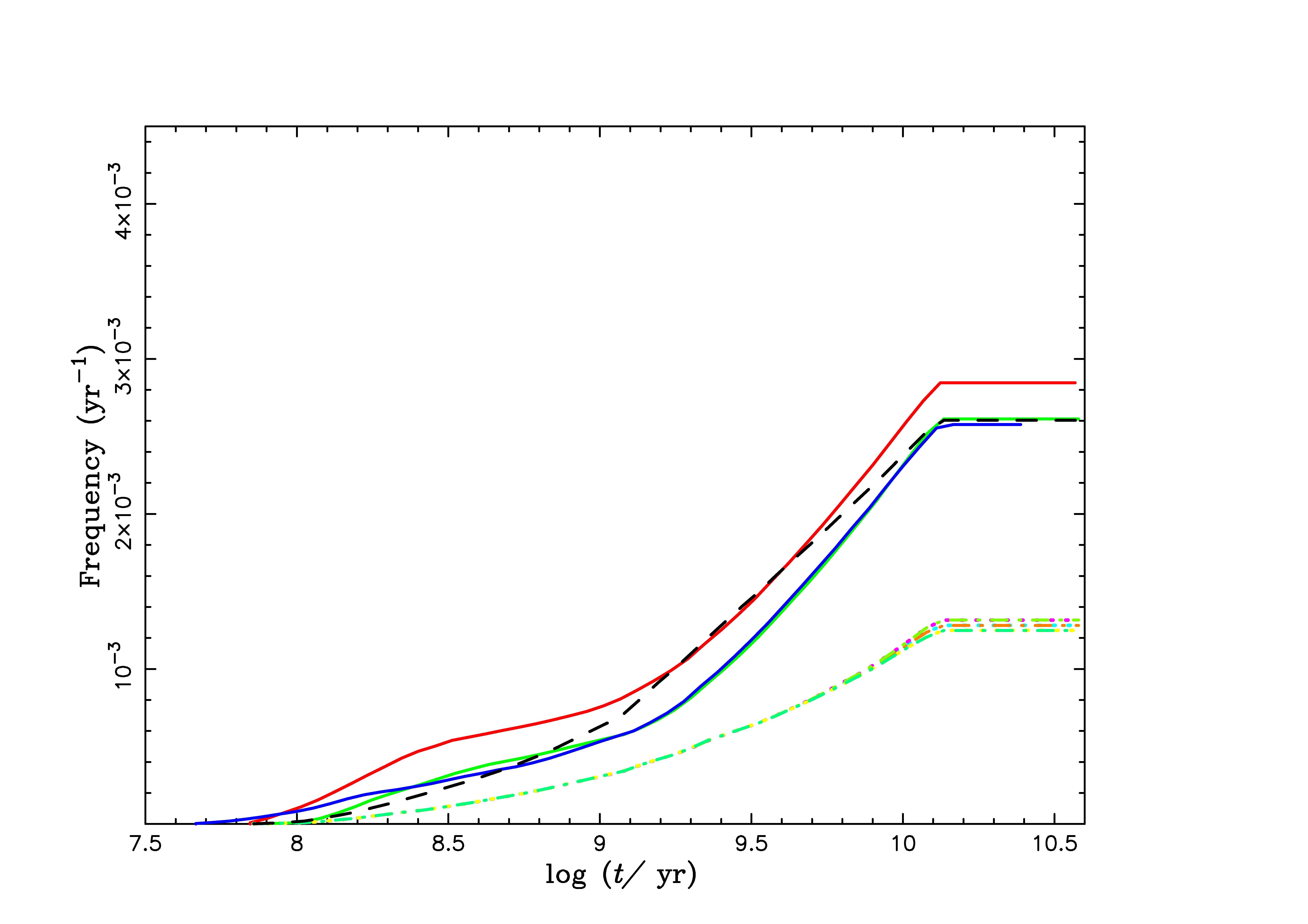}
\includegraphics[totalheight=2.1in,width=2.5in]{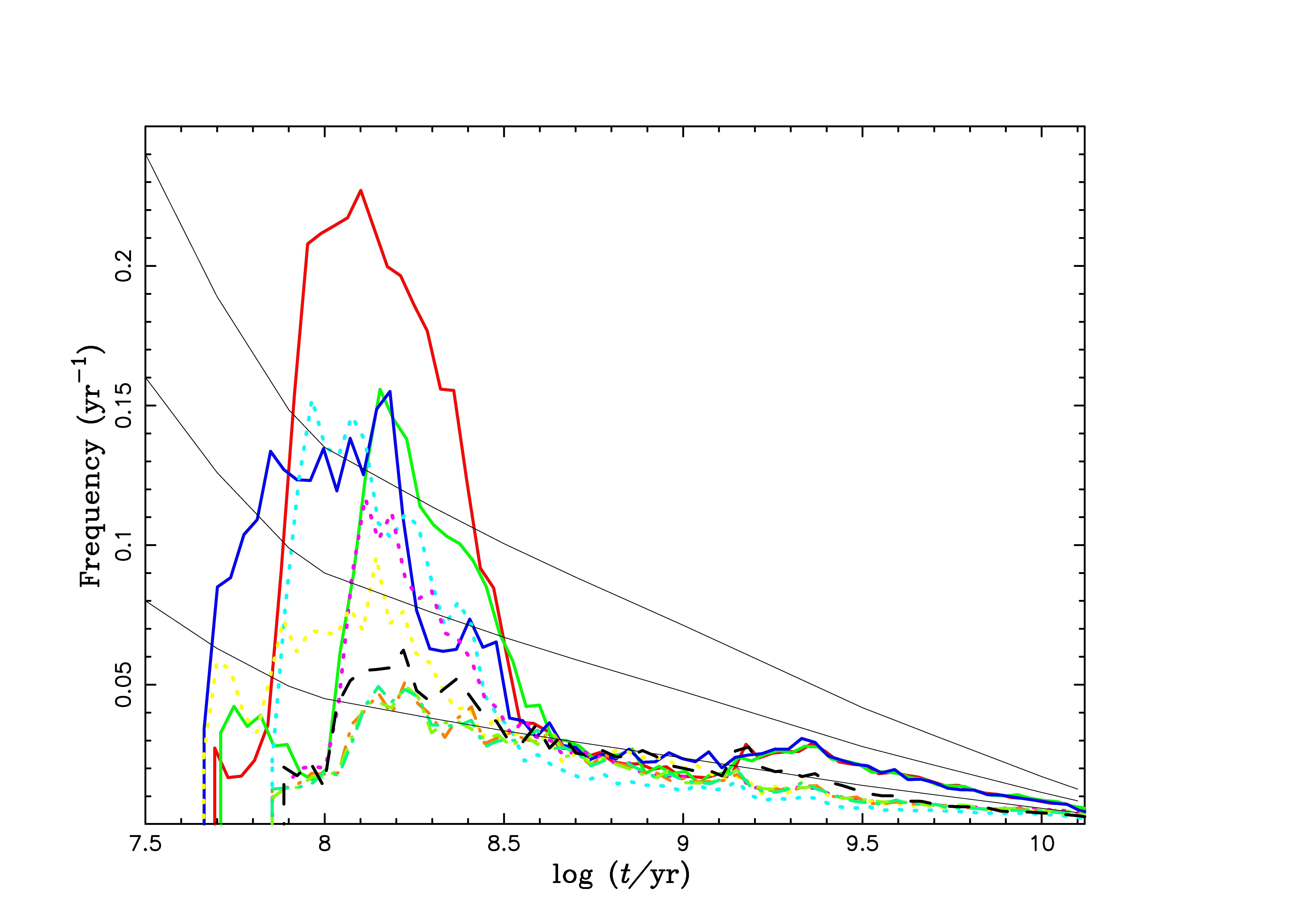}
\includegraphics[totalheight=2.1in,width=2.5in]{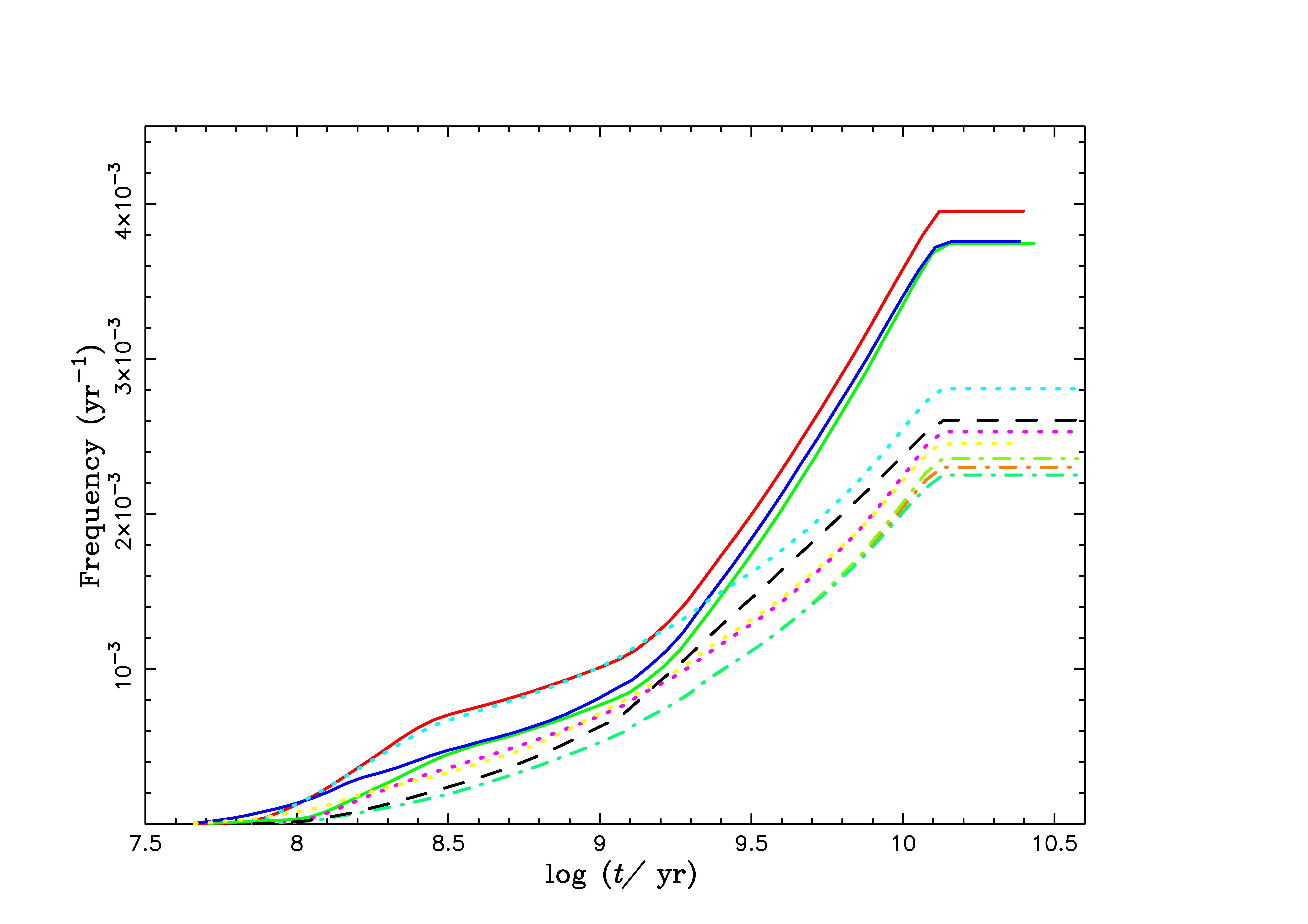}
\includegraphics[totalheight=2.1in,width=2.5in]{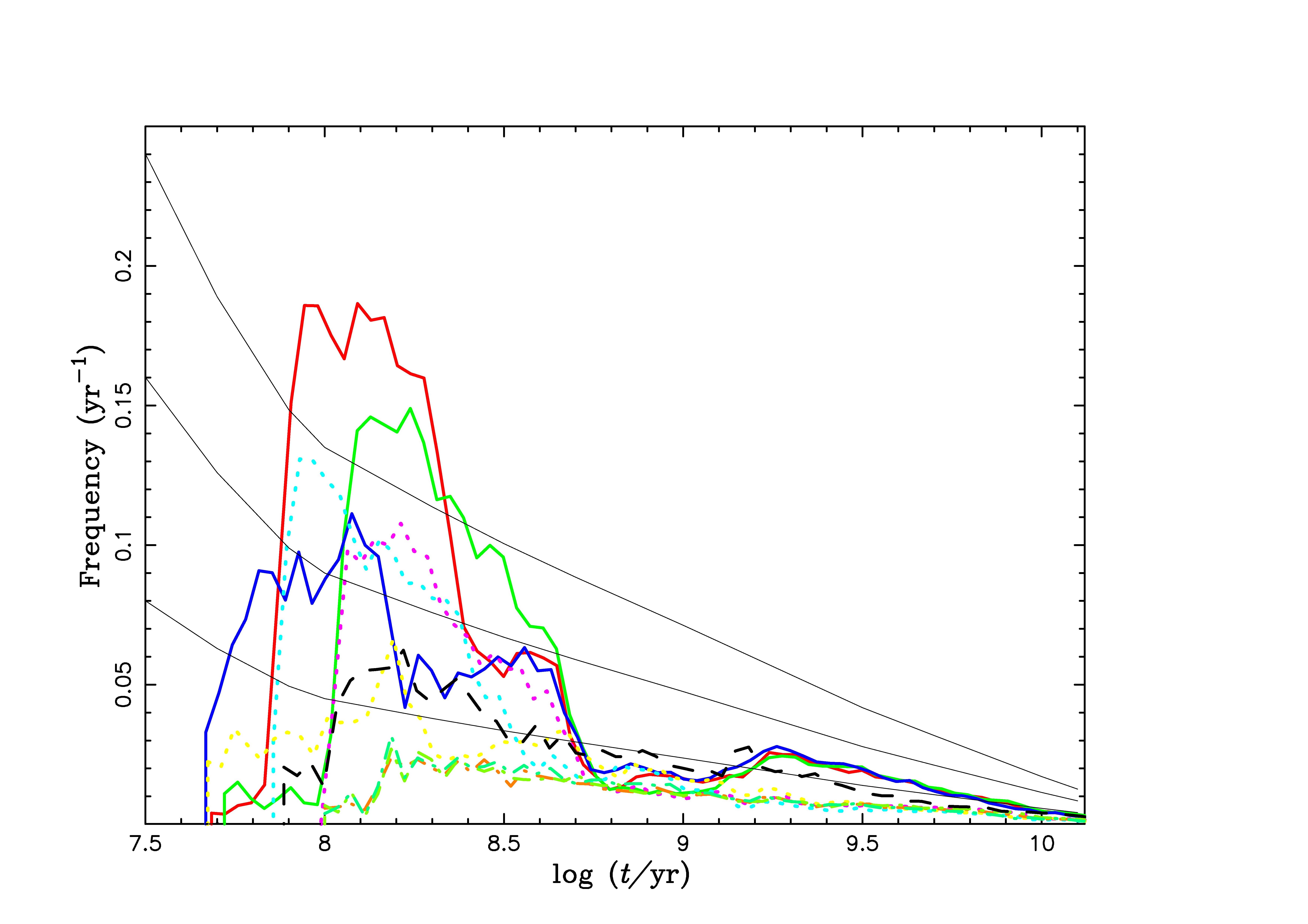}
\includegraphics[totalheight=2.1in,width=2.5in]{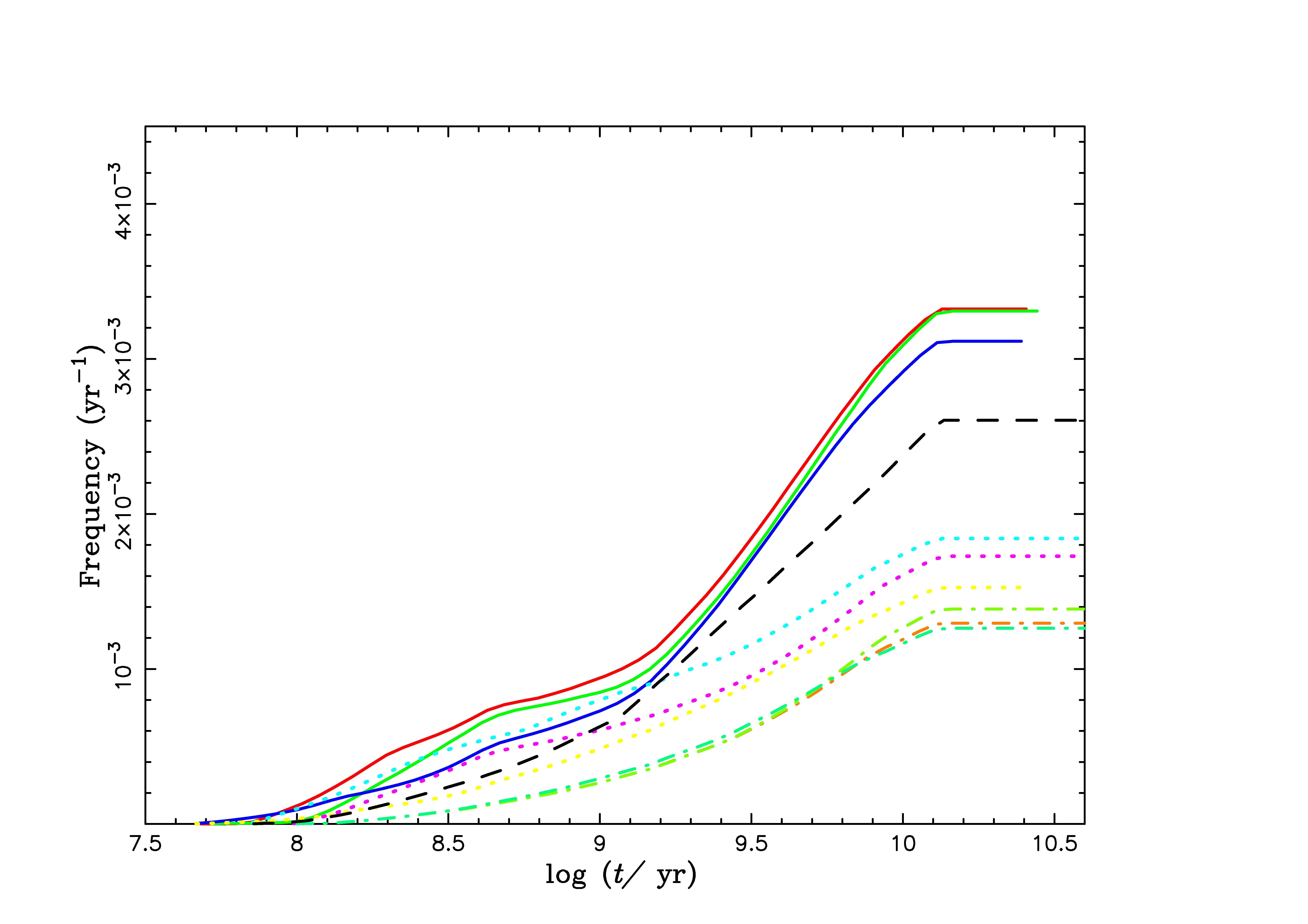}
\caption{In this figure, the range of the initial orbital separation is set to be
$3\leq a_{\rm i}/R_\odot \leq 10^4$, models are the same as in Figure 8.}
\label{fig:1}
\end{figure}

\clearpage

\begin{figure}
\centering
\includegraphics[totalheight=2.65in,width=3.2in]{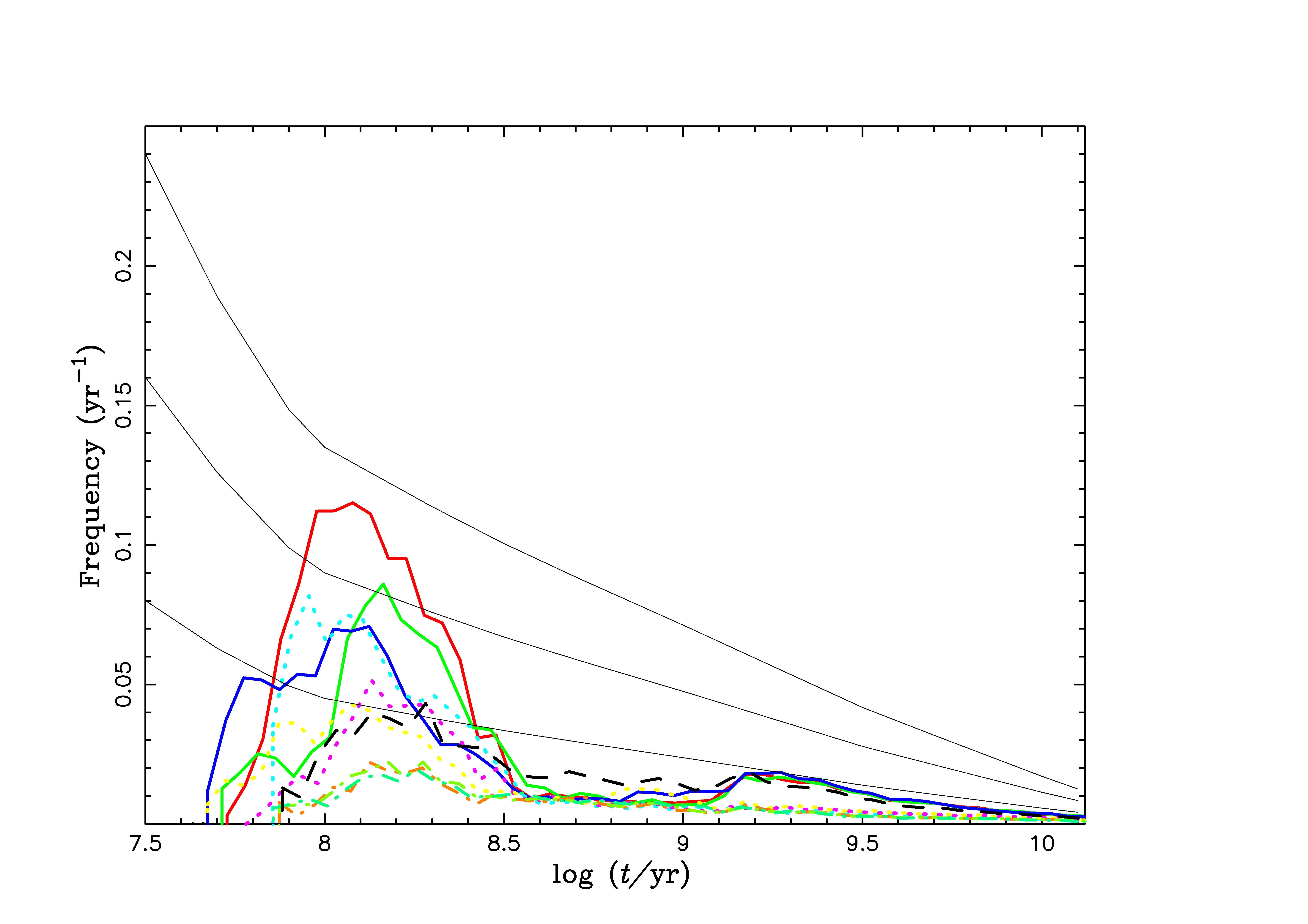}
\includegraphics[totalheight=2.65in,width=3.2in]{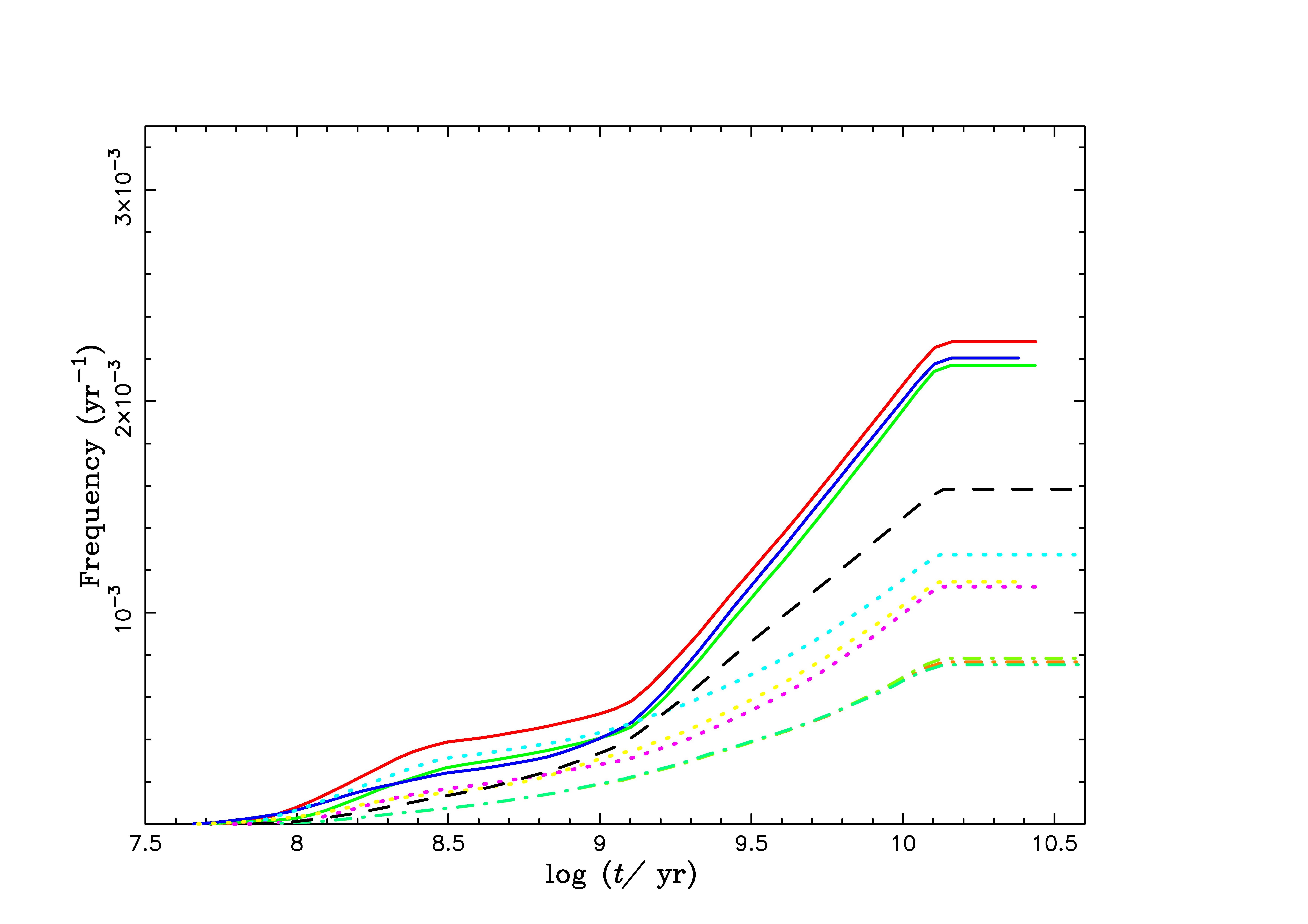}
\includegraphics[totalheight=2.65in,width=3.2in]{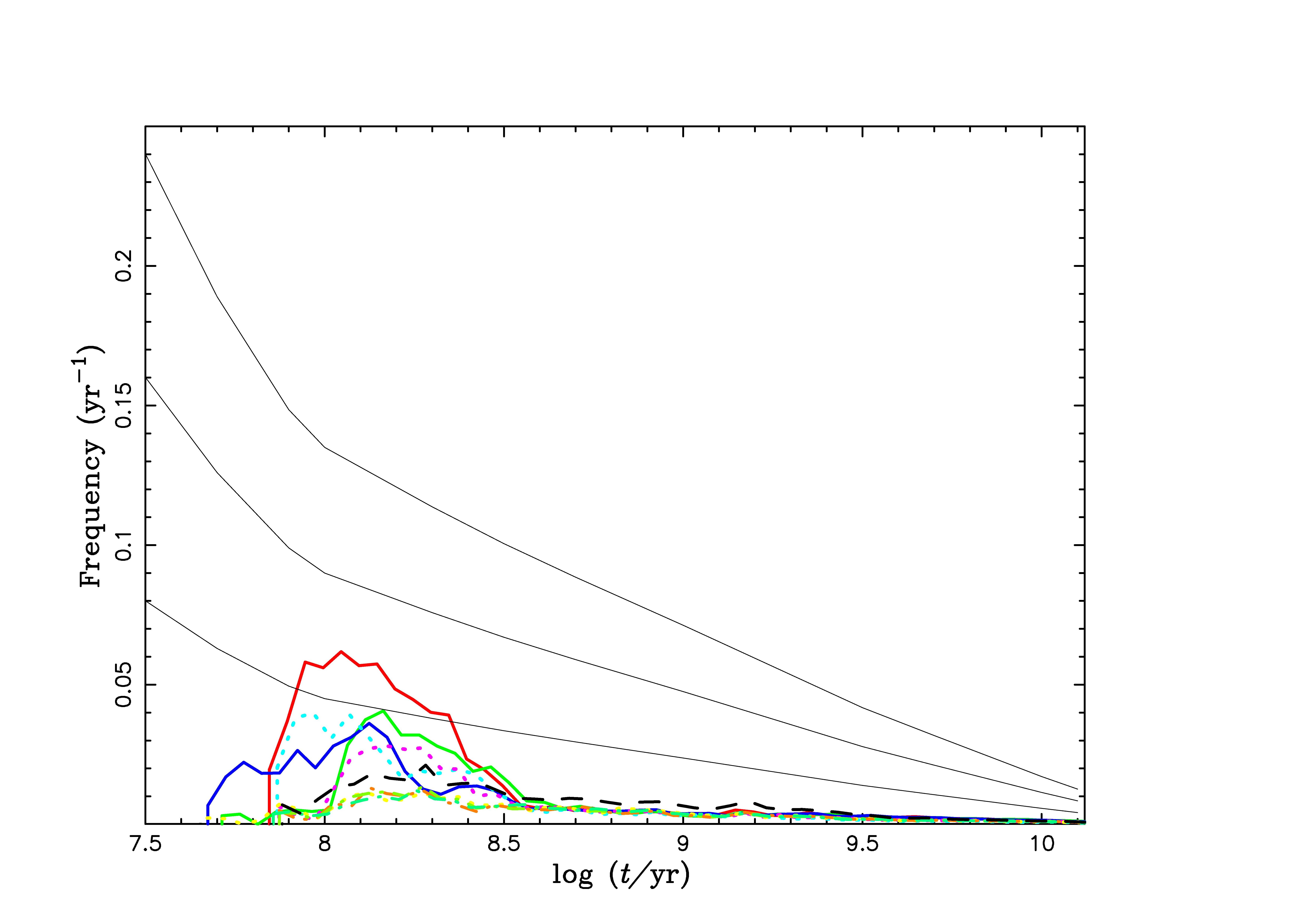}
\includegraphics[totalheight=2.65in,width=3.2in]{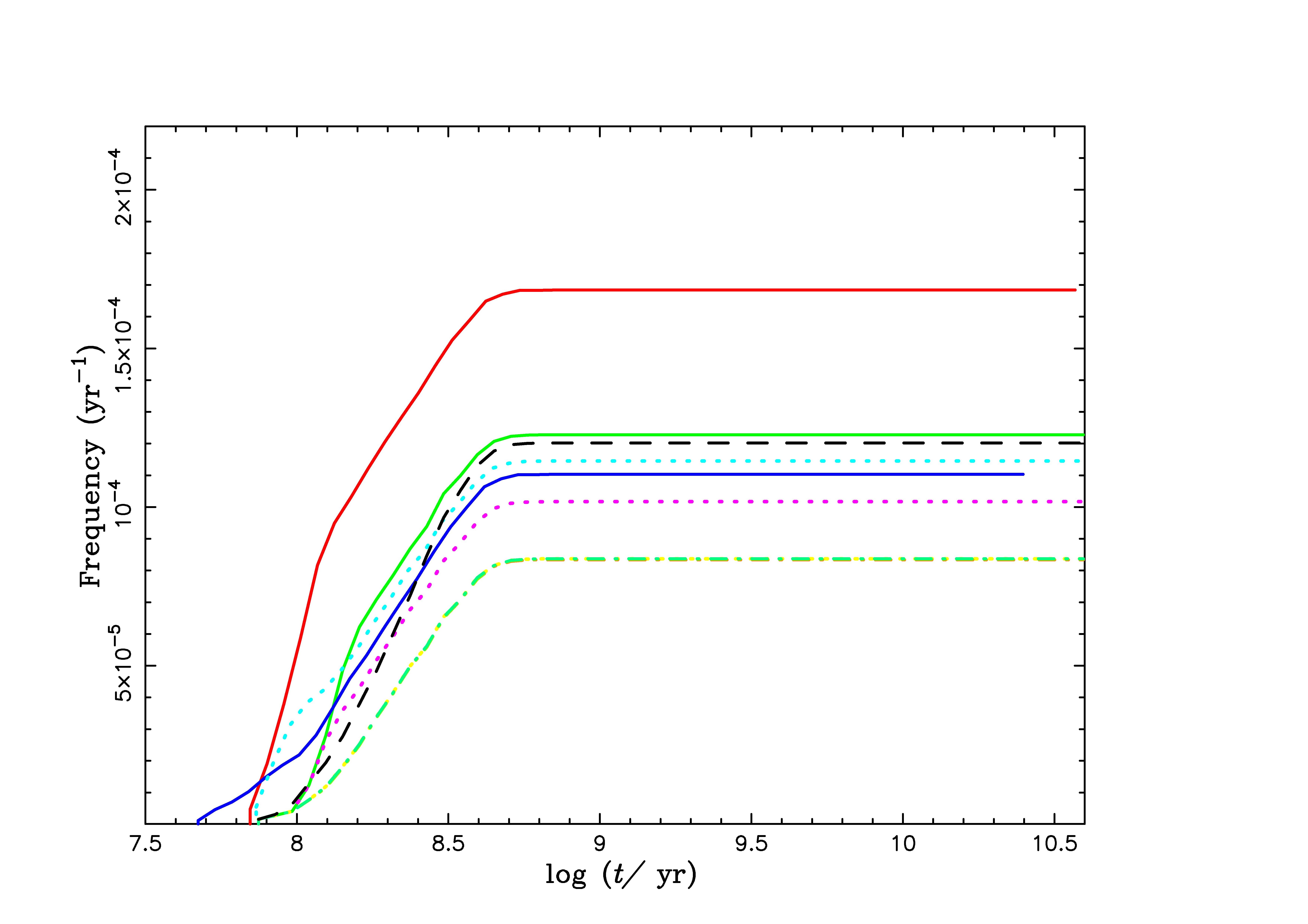}
\caption{This Figure shows the results of model 1 and models 2--4 with three
prescriptions for $q_{\rm cr}$, under our different criteria for a DD system
to explode as an SN Ia. The upper panels show the results from the `Chandrasekhar Mass Scenario'
and the lower panels show the results from the `Carbon-Ignited Violent Merger Scenario'
(see \S 3.2.3 for details).}\label{fig:1}
\end{figure}

\clearpage

\label{lastpage}

\end{document}